\documentclass[preprint]{ptephy}

\preprintnumber{KYUSHU-HET-220, KEK-TH-2294}

\usepackage{amssymb}
\usepackage{amsmath}
\usepackage{bbm}
\usepackage{mathtools}
\usepackage{hyperref}
\usepackage{url}
\usepackage{comment}
\newcommand{\be}{\begin{equation}}
\newcommand{\ee}{\end{equation}}
\newcommand{\Slash}[1]{{\ooalign{\hfil/\hfil\crcr$#1$}}}

\newcommand{\lt}{\left}
\newcommand{\rt}{\right}

\newcommand{\non}{\nonumber \\}

\newcommand{\fn}{\footnote}

\newcommand{\MSb}{\overline{\rm MS}}

\newcommand{\LMS}{\Lambda_{\overline{\rm MS}}}

\newcommand{\mO}{\mathcal{O}}
\numberwithin{equation}{section}

\graphicspath{{../Figs/}}

\begin{document}

\date{\today}

\title{$t \to 0$ extrapolation function in SF$t$X method for the energy--momentum tensor
}

\author{%
\name{\fname{Hiroshi} \surname{Suzuki}}{1} and
\name{\fname{Hiromasa} \surname{Takaura}}{2,\ast}
}

\address{%
\affil{1}{Department of Physics, Kyushu University
744 Motooka, Nishi-ku, Fukuoka, 819-0395, Japan}
\affil{2}{Theory Center, High Energy Accelerator Research Organization (KEK),
Tsukuba, Ibaraki, 305-0801, Japan}\email{htakaura@post.kek.jp}
}

\begin{abstract}
We  theoretically clarify the functional form to be used in $t \to 0$ extrapolation
in the small flow time expansion (SF$t$X) method for the energy--momentum tensor (EMT),
which facilitates lattice simulation of the EMT based on the gradient flow.
We argue that in the $t \to 0$ extrapolation analysis,
lattice data should be fitted by a power function in~$g(\mu(t))$,
the flow time dependent running coupling, where the power 
is determined by the perturbation order we consider.
From actual lattice data, we confirm the validity of the extrapolation function.
Using the new extrapolation function, we present updated lattice results 
for thermodynamics quantities in quenched QCD;
our results are consistent with the previous study [arXiv:1812.06444]
but we obtain smaller errors due to the reduction of systematic errors.
\end{abstract}

\subjectindex{B01, B31, B32, B38}
\maketitle

\section{Introduction}
The energy--momentum tensor (EMT) $T_{\mu \nu}(x)$ is a fundamental quantity in quantum field theory,
yet its lattice simulation is not straightforward due to explicit breaking of
the translation invariance in lattice regularization; see Ref.~\cite{Suzuki:2016ytc} and references therein.
In Refs.~\cite{Suzuki:2013gza,Makino:2014taa}, 
the so-called small flow time expansion (SF$t$X) method was devised
to solve this problem. In this method, one rewrites conserved currents in terms of 
the so-called flowed operators, defined from the gradient flow 
\cite{Narayanan:2006rf,Luscher:2009eq,Luscher:2010iy,Luscher:2011bx,Luscher:2013cpa}.
Due to the UV finiteness of flowed operators, the current represented with flowed operators
satisfies the correct Ward--Takahashi identity in any regularization. 
Then one can measure correctly normalized currents in lattice simulation.
For the EMT,  we can schematically write it in the form\fn{
In actual study, we mainly use the expression of Eqs.~\eqref{TL} and \eqref{S}.
We show Eq.~\eqref{eq:(1.1)} just for explanation.}
\be
T_{\mu \nu}(x)=\tilde{c}_1(t) \tilde{\mO}_{1, \mu \nu}(t,x)+\tilde{c}_2(t) \tilde{\mO}_{2, \mu \nu}(t,x)
+\tilde{c}_3(t)\tilde{\mO}_{3, \mu \nu}(t,x)+\tilde{c}_4(t) \tilde{\mO}_{4, \mu \nu}(t,x)+\mathcal{O}(t) . \label{eq:(1.1)} 
\ee
Here, $t$ is the flow time, whose mass dimension is $-2$; $\tilde{\mO}_{i, \mu \nu}(t,x)$'s are (dimension-four) flowed operators 
(whose explicit definitions are given below).
The coefficients $\tilde{c}_i(t)$'s can be perturbatively calculated  
via the small flow time expansion \cite{Luscher:2011bx} of the flowed operators
and the two-loop order results are known today \cite{Harlander:2018zpi};
see also Ref.~\cite{Artz:2019bpr}.
$\mathcal{O}(t)$ represents contributions from dimension-six operators.
In lattice simulation we measure the flowed operators $\tilde{\mO}_{i,\mu\nu}$ nonperturbatively and 
then multiply the perturbative coefficients $\tilde{c}_i(t)$ to obtain the EMT.
Many lattice simulations of the EMT with the SF$t$X method have been performed 
\cite{Asakawa:2013laa,Taniguchi:2016ofw, Kitazawa:2016dsl,Ejiri:2017wgd, Kitazawa:2017qab,Kanaya:2017cpp, Taniguchi:2017ibr, Yanagihara:2018qqg, Hirakida:2018uoy, Shirogane:2018zbp,Iritani:2018idk,Taniguchi:2019eid, Kitazawa:2019otp, Kanaya:2019okb, Taniguchi:2020mgg, Yanagihara:2020tvs, Shirogane:2020muc}
and its usefulness has been confirmed.

In lattice simulation using the SF$t$X method, one needs to take the small flow time limit ($t \to 0$).\fn{
In principle, the continuum limit ($a \to 0$ limit, where $a$ is lattice spacing) should be taken before the $t \to 0$ limit.
Hence, in our calculations we assume that the continuum limit is already taken.}
This is because the expression for the EMT in the SF$t$X method
becomes exact in the $t \to 0$ limit.
In this limit, we can get rid of higher orders in~$g(\mu(t))$, 
the flow time dependent running coupling, in the perturbative coefficients $\tilde{c}_i(t)$,
and $\mathcal{O}(t)$ contributions in Eq.~\eqref{eq:(1.1)}.
However, one cannot directly obtain lattice data at $t=0$ because 
lattice data  suffer from serious discretization effects when the flow time becomes too small compared to 
the lattice spacing, $t \lesssim a^2$.
Hence, one should take the $t \to 0$ limit by extrapolation of lattice data at $t \gg a^2$
assuming some function of $t$.
Thus, a $t \to 0$ extrapolation function can be critical to final results.

In earlier analyses with the SF$t$X method,
a linear function in $t$ was mainly used in the~$t\to 0$~extrapolation. 
This was motivated by the $\mathcal{O}(t)$ contribution in Eq.~\eqref{eq:(1.1)},
which is neglected in constructing the EMT in the SF$t$X method.
However, since we use fixed order perturbative results for $\tilde{c}_i(t)$,
a higher-order effect in $g(\mu(t))$ should also exist.
Parametrically, such an effect is dominant in small $t$ region.
This is because $g(\mu(t))^n \gg t\LMS^2 \sim e^{- 16 \pi^2/(\beta_0 g^2(\mu(t)))}$ 
($\beta_0$ is the first coefficient of the beta function),
where $t$ is an exponentially suppressed function for small $g(\mu(t))$.
It is therefore important to identify the higher-order effect in~$g(\mu(t))$
in order to accurately performe $t \to 0$ extrapolation analyses.

The purpose of this paper is to clarify the leading $t$ dependence of 
the fixed order formula for the EMT in
the SF$t$X method (which means that we use fixed order perturbative results for $\tilde{c}_i(t)$'s).
This study tells us which functional form should be used in the $t \to 0$ extrapolation,
and this knowledge will be essential in making the SF$t$X method a more precise framework.
Also, the recent two-loop order calculation of $\tilde{c}_i$'s \cite{Harlander:2018zpi}
can promote more accurate analyses, and it is timely 
to discuss this issue in the SF$t$X method.

In the rest of this this section, we briefly review the SF$t$X method 
and clarify the question studied in this paper more explicitly.
We also introduce quantities necessary for the subsequent studies.

The EMT in dimensional regularization is given by
\be
T_{\mu \nu}(x)=\frac{1}{g_0^2} \lt[\mathcal{O}_{1, \mu \nu}(x)-\frac{1}{4} \mathcal{O}_{2, \mu \nu}(x) \rt]+\frac{1}{4} \mathcal{O}_{3, \mu \nu}(x) , \label{EMT}
\ee
where $g_0$ is the bare gauge coupling; 
$\mathcal{O}_{i, \mu \nu}$'s are gauge invariant and symmetric dimension-four tensor operators defined as
\begin{align}
   \mathcal{O}_{1,\mu\nu}(x)&\equiv
   F_{\mu\rho}^a(x)F_{\nu\rho}^a(x),
\label{O1b}\\
   \mathcal{O}_{2,\mu\nu}(x)&\equiv
   \delta_{\mu\nu}F_{\rho\sigma}^a(x)F_{\rho\sigma}^a(x),
\label{O2b}\\
   \mathcal{O}_{3,\mu\nu}(x)&\equiv
   \sum_f\Bar{\psi}_f(x)
   \left(\gamma_\mu\overleftrightarrow{D}_\nu
   +\gamma_\nu\overleftrightarrow{D}_\mu\right)
   \psi_f(x),
\label{O3b}\\
   \mathcal{O}_{4,\mu\nu}(x)&\equiv
   \delta_{\mu\nu}
   \sum_f\Bar{\psi}_f(x)
   \overleftrightarrow{\Slash{D}}
   \psi_f(x),
\label{O4b}\\
   \mathcal{O}_{5,\mu\nu}(x)&\equiv
   \delta_{\mu\nu}
   \sum_fm_{f,0}\Bar{\psi}_f(x)
   \psi_f(x)  ,
\label{O5b}
\end{align}
where $m_{f,0}$ is the bare mass of the flavor~$f$.
These are bare composite operators and not finite in general.

In the SF$t$X method, one rewrites the EMT in terms of flowed operators.
The Yang--Mills gradient flow 
\cite{Narayanan:2006rf,Luscher:2009eq,Luscher:2010iy,Luscher:2011bx,Luscher:2013cpa}
 is defined by the following differential equations with respect to the flow time $t$.
For the gauge field, it is defined as
\be
\partial_t B_{\mu} (t,x)=D_{\nu} G_{\nu \mu}(t,x)+\alpha_0 D_{\mu} \partial_{\nu} B_{\nu} , \qquad B_{\mu}(t=0,x)=A_{\mu}(x) , \label{eq:(1.8)}
\ee
with a gauge parameter $\alpha_0$, where the covariant derivative on the gauge field is defined as
\be
D_{\mu}=\partial_{\mu}+[B_{\mu}, \cdot] .
\ee
$G_{\mu \nu}$ is the field strength of the flowed gauge field $B_{\mu}$.
For the fermion field, it is given by
\be
\partial_t \chi(t,x)=[\Delta-\alpha_0 \partial_{\mu} B_{\mu}(t,x)] \chi(t,x) , \qquad \chi(t=0,x)=\psi(x) ,
\ee
\be
\partial_t \bar{\chi}(t,x)=\bar{\chi}(t,x) [\overleftarrow{\Delta}+\alpha_0 \partial_{\mu} B_{\mu}(t,x)] , \qquad \bar{\chi}(t=0,x)=\bar{\psi}(x) ,
\ee
with 
\be
\Delta=D_{\mu} D_{\mu} , \qquad D_{\mu}=\partial_{\mu}+B_{\mu} ,
\ee
\be
\overleftarrow{\Delta}=\overleftarrow{D}_{\mu} \overleftarrow{D}_{\mu} ,  \qquad 
\overleftarrow{D}_{\mu}=\overleftarrow{\partial}_{\mu}-B_{\mu} .
\ee

We define analogous operators at positive flow time with tildes:
\begin{align}
   \Tilde{\mathcal{O}}_{1,\mu\nu}(t,x)&\equiv
   G_{\mu\rho}^a(t,x)G_{\nu\rho}^a(t,x),
\label{Of1}\\
   \Tilde{\mathcal{O}}_{2,\mu\nu}(t,x)&\equiv
   \delta_{\mu\nu}G_{\rho\sigma}^a(t,x)G_{\rho\sigma}^a(t,x),
\label{Of2}\\
   \Tilde{\mathcal{O}}_{3,\mu\nu}(t,x)&\equiv
   \sum_f\mathring{\Bar{\chi}}_f(t,x)
   \left(\gamma_\mu\overleftrightarrow{D}_\nu
   +\gamma_\nu\overleftrightarrow{D}_\mu\right)
   \mathring{\chi}_f(t,x),
\label{Of3}\\
   \Tilde{\mathcal{O}}_{4,\mu\nu}(t,x)&\equiv
   \delta_{\mu\nu}
   \sum_f\mathring{\Bar{\chi}}_f(t,x)
   \overleftrightarrow{\Slash{D}}
   \mathring{\chi}_f(t,x),
\label{Of4}\\
   \Tilde{\mathcal{O}}_{5,\mu\nu}(t,x)&\equiv
   \delta_{\mu\nu}
   \sum_f m_f\mathring{\Bar{\chi}}_f(t,x)
   \mathring{\chi}_f(t,x).
\label{Of5}
\end{align}
The above flowed operators are {\it finite} operators \cite{Luscher:2011bx}.\fn{
In Eq.~\eqref{Of5} $m_f$ denotes a renormalized mass parameter.
However, in this paper, we do not need to specify it 
because we do not use $\tilde{\mO}_{5, \mu \nu}$ in the following. 
}
We have accomplished the renormalization of flowed fermion fields using a ringed variable \cite{Makino:2014taa}:
\begin{align}
   \mathring{\chi}_f(t,x)
   &\equiv\sqrt{\frac{-2\dim(R)}
   {(4\pi)^2t^2
   \left\langle\Bar{\chi}_f(t,x)\overleftrightarrow{\Slash{D}}\chi_f(t,x)
   \right\rangle}}
   \,\chi_f(t,x),
\label{normal1}\\
   \mathring{\Bar{\chi}}_f(t,x)
   &\equiv\sqrt{\frac{-2\dim(R)}
   {(4\pi)^2t^2
   \left\langle\Bar{\chi}_f(t,x)\overleftrightarrow{\Slash{D}}\chi_f(t,x)
   \right\rangle}}
   \,\Bar{\chi}_f(t,x) \, .
\label{normal2}
\end{align}
Alternatively, one can also adopt the ${\rm{\overline{MS}}}$ scheme as in Ref.~\cite{Harlander:2018zpi}.
We denote such an operator set as $\Tilde{\mathcal{O}}^{\MSb}_{i, \mu \nu}(t,x)$.
The conversion from the ${\rm{\overline{MS}}}$ scheme to the ringed variable scheme 
is carried out by a matrix $R$ by
\be
\Tilde{\mathcal{O}}_{\mu \nu}(t,x)=R \, \Tilde{\mathcal{O}}^{\MSb}_{\mu \nu}(t,x) ,
\ee
with
\be
R=\left( \begin{array}{cccc}
1& 0 & 0 & 0 \\
0 & 1 & 0 & 0 \\
0 & 0 & r & 0 \\
0 & 0 & 0 & r
\end{array} \right) ,
\ee 
where $r$ is given in Eq.~(51) of Ref.~\cite{Harlander:2018zpi} as
\be
r=\mathring{Z_{\chi}}/Z_{\chi} .
\ee
When we suppress the index $i$ in the operators, it is understood as a four-component vector,
\be
\mathcal{O}_{\mu \nu}=\left( \begin{array}{c}
\mathcal{O}_{1, \mu \nu} \\
\mathcal{O}_{2, \mu \nu} \\
\mathcal{O}_{3, \mu \nu} \\
\mathcal{O}_{4, \mu \nu} 
\end{array} \right)  .
\ee
Note that we can eliminate $\mathcal{O}_{5, \mu \nu}$ by using the equation of motion (EOM),
$(1/2) \mO_{4, \mu \nu}+\mO_{5, \mu \nu}=0$, 
following Ref.~\cite{Harlander:2018zpi}.
Hence, it is sufficient to consider the four operators as a basis.

To rewrite the EMT in terms of the flowed operators, one needs to know the relation
between the flowed and unflowed operators.
This can be studied through the small flow time expansion \cite{Luscher:2011bx}:
\be
\Tilde{\mathcal{O}}_{\mu \nu}^{\MSb}(t,x)=\zeta (t) \mathcal{O}_{ \mu \nu}(x)+\mathcal{O}(t) , \label{SFT}
\ee
where $\zeta(t)$ is a $4 \times 4$ matrix, whose mass dimension is zero.
This matrix can be perturbatively calculated.
(This matrix has UV divergence, which is canceled against UV divergence of the bare operator
$\mathcal{O}_{\mu \nu}$, so that the left-hand side of Eq.~\eqref{SFT} becomes UV finite.)
Higher-dimensional operators are suppressed as $\mathcal{O}(t)$ for small $t$.
Here and hereafter, we implicitly assume that the vacuum expectation values
of the operators are subtracted. 
[This is the reason why the terms proportional to the identity operator are absent in Eq.~\eqref{SFT}.]

Once the relation between the flowed and unflowed operators is known, 
one can express the EMT, given by the unflowed operators,
in terms of the flowed operators using the inverse of the matrix $\zeta$.
It is conventionally expressed as\fn{
In the SF$t$X method, we neglect $\mathcal{O}(t)$ contributions in Eq.~\eqref{SFT},
and thus, strictly speaking, the $\mathcal{O}(t)$ error should be shown in the right-hand side
of Eq.~\eqref{GFEMT}.}
\begin{align}
T_{\mu \nu}(x)
&=c_1(t) \lt[\Tilde{\mathcal{O}}_{1, \mu \nu}(t,x)-\frac{1}{4} \Tilde{\mathcal{O}}_{2, \mu \nu}(t,x) \rt]+c_2(t) \Tilde{\mathcal{O}}_{2, \mu \nu}(t,x) \non
&\quad{}+c_3(t) \lt[ \Tilde{\mathcal{O}}_{3, \mu \nu}(t,x) -2 \Tilde{\mathcal{O}}_{4, \mu \nu}(t,x) \rt]+c_4(t) \Tilde{\mathcal{O}}_{4, \mu \nu}(t,x) . \label{GFEMT}
\end{align}
By decomposition of a (finite) symmetric tensor $S_{\mu \nu}$ into its traceless (TL) and scalar (S) parts as
 \be
S_{\mu \nu}^{\rm TL}=S_{\mu \nu} -\frac{\delta_{\mu \nu}}{4} \delta_{\rho \sigma} S_{\rho \sigma} ,
\ee
\be
S^{\rm S}=\delta_{\rho \sigma} S_{\rho \sigma} \, ,
\ee
the expression \eqref{GFEMT} is equivalent to
\be
T_{\mu \nu}(x)=T_{\mu \nu}^{\rm TL}(x)+\frac{\delta_{\mu \nu}}{4} T^{\rm S}(x) ,
\ee
with
\be
T_{\mu \nu}^{\rm TL}(x)=c_1(t) \Tilde{\mathcal{O}}^{\rm TL}_{1, \mu \nu}(t, x)+c_3(t) \Tilde{\mathcal{O}}^{\rm TL}_{3, \mu \nu}(t,x) \label{TL} ,
\ee
\be
T^S(x)=c_2(t) \Tilde{\mathcal{O}}^{\rm S}_{2}(t, x)+c'_4(t) \Tilde{\mathcal{O}}^{\rm S}_{4}(t,x)  , \label{S}
\ee
where
\be
c_4'(t)=c_4(t)- \frac{3}{2} c_3(t) .
\ee
The traceless flowed operators are given by 
$\Tilde{\mathcal{O}}^{\rm TL}_{1, \mu \nu}(t, x)
=\Tilde{\mathcal{O}}_{1, \mu \nu}(t, x)-(1/4) \Tilde{\mathcal{O}}_{2, \mu \nu}(t, x)$
and
$\Tilde{\mathcal{O}}^{\rm TL}_{3, \mu \nu}(t, x)
=\Tilde{\mathcal{O}}_{3, \mu \nu}(t, x)-(1/2) \Tilde{\mathcal{O}}_{4, \mu \nu}(t, x)$.

The coefficients $c^{(\prime)}_i(t)$ are given by perturbative series:
\be
c_1(t)=\frac{1}{g(\mu)^2} \sum_{n=0}^{\infty} k_1^{(n)}(L(\mu,t)) \lt[ \frac{g(\mu)^2}{(4 \pi)^2 } \rt]^n , \label{c1} 
\ee
\be
c_2(t)=\frac{1}{g(\mu)^2} \sum_{n=1}^{\infty} k_2^{(n)}(L(\mu,t)) \lt[ \frac{g(\mu)^2}{(4 \pi)^2 } \rt]^n , \label{c2}
\ee
\be
c_3(t)=\sum_{n=0}^{\infty} k_3^{(n)}(L(\mu,t)) \lt[ \frac{g(\mu)^2}{(4 \pi)^2 } \rt]^n ,  \label{c3}
\ee
\be
c'_4(t)=\sum_{n=0}^{\infty} k_4^{(n)}(L(\mu,t)) \lt[ \frac{g(\mu)^2}{(4 \pi)^2}  \rt]^n . \label{c4}
\ee
Here, $\mu$ is a renormalization scale and 
the perturbative coefficient $k_i^{(n)}$ is a polynomial of $L(\mu,t) \equiv \log{(2 \mu^2 e^{\gamma_E} t)}$.
Since the EMT and flowed operators are renormalization group (RG) invariant,
the coefficients $c_i(t)$'s are also RG invariant, i.e. independent of $\mu$.
In practical applications, we take the $t$-dependent renormalization scale, 
$\mu=\mu(t)=s/\sqrt{2 e^{\gamma_E} t}$, where $s$ is an $\mathcal{O}(1)$
numerical factor,
so that higher-order terms in $g(\mu)$ vanish in the $t \to 0$ limit.
The one-loop coefficients $k_i^{(1)}$'s were calculated
in Ref.~\cite{Suzuki:2013gza} for quenched QCD
and in Ref.~\cite{Makino:2014taa} for full QCD.
The two-loop coefficients $k_i^{(2)}$ were obtained in Ref.~\cite{Harlander:2018zpi}.
For quenched QCD, the three-loop coefficient $k_2^{(3)}$ (only for $i=2$) 
was obtained in  Ref.~\cite{Iritani:2018idk}.

The purpose of this paper is to investigate the (leading) $t$-dependence of 
the next-to-...-next-to-leading order (N$^k$LO) formula of the EMT in the SF$t$X method: 
\be
{T_{\mu \nu}^{\rm TL}}^{\text{(N$^k$LO)}}(x;t)=c^{\text{(N$^k$LO)}}_1(t) \tilde{\mO}_{1, \mu \nu}(t,x)+c^{\text{(N$^k$LO)}}_3(t) \tilde{\mO}_{3, \mu \nu}(t,x) , \label{TTLNkLO}
\ee
\be
{T^{\rm S}}^{\text{(N$^k$LO)}}(x;t)=c^{\text{(N$^k$LO)}}_2(t) \tilde{\mO}_{2}^{\rm S}(t,x)
+{c'}^{\text{(N$^k$LO)}}_4(t) \tilde{\mO}_{4}^{\rm S}(t,x) .  \label{TSNkLO}
\ee
Here, ${c^{(\prime)}_i}^{\text{(N$^k$LO)}}$ is given by the sum of the first $(k+1)$ terms of Eqs.~\eqref{c1}--\eqref{c4}:
\be
c_1^{\text{(N$^k$LO)}}(t)=\frac{1}{g(\mu)^2} \sum_{n=0}^{k} k_1^{(n)}(L(\mu,t)) \lt[ \frac{g(\mu)^2}{(4 \pi)^2 } \rt]^n , \label{c1NkLO}
\ee
\be
c_2^{\text{(N$^k$LO)}}(t)=\frac{1}{g(\mu)^2} \sum_{n=1}^{k+1} k_2^{(n)}(L(\mu,t)) \lt[ \frac{g(\mu)^2}{(4 \pi)^2 } \rt]^n , \label{c2NkLO}
\ee
\be
c_3^{\text{(N$^k$LO)}}(t)=\sum_{n=0}^{k} k_3^{(k)}(L(\mu,t)) \lt[ \frac{g(\mu)^2}{(4 \pi)^2 } \rt]^n , \label{c3NkLO}
\ee
\be
{c'_4}^{\text{(N$^k$LO)}}(t)=\sum_{n=0}^{k} k_4^{(k)}(L(\mu,t)) \lt[ \frac{g(\mu)^2}{(4 \pi)^2}  \rt]^n . \label{c4NkLO}
\ee 
Note that for $c_2(t)$ the upper edge of the sum is set to $k+1$.
Our main results of this paper are summarized in Sect.~\ref{sec:2}. 

In the following calculations we mainly use hatted operators, as in Ref.~\cite{Harlander:2018zpi},
to discuss renormalization of the bare operators,
instead of the set of Eqs.~\eqref{O1b}--\eqref{O4b}:
\be
\hat{\mathcal{O}}_{\mu \nu}(x)=H \mathcal{O}_{\mu \nu}(x) ,
\ee
where $\mathcal{O}_{\mu \nu}(x)$ is defined in Eqs.~\eqref{O1b}--\eqref{O4b} and $H$ denotes
\be
H=\left( \begin{array}{cccc}
1/g_0^2& 0 & 0 & 0 \\
0 & 1/g_0^2 & 0 & 0 \\
0 & 0 & 1 & 0 \\
0 & 0 & 0 & 1
\end{array} \right) .
\ee
$\hat{\mO}_{\mu \nu}$ are $D$-dimensional operators in dimensional regularization with $D=4 -2 \epsilon$.
We consider the renormalization of $\hat{\mO}_{\mu \nu}$ as
\be
\hat{\mO}_{\mu \nu}(x)=Z(g(\mu);\epsilon)  \hat{\mO}^R_{\mu \nu}(x; \mu) ,
\ee
where``$R$'' stands for renormalization of operators.
We use the $\overline{\rm MS}$ scheme to renormalize the unflowed operators.
Since the left-hand side of Eq.~\eqref{SFT},
\begin{align}
\Tilde{\mathcal{O}}^{\MSb}_{\mu \nu}(t, x)
=\zeta(t) \mathcal{O}_{\mu \nu}(x) +\mathcal{O}(t)
=\zeta(t)  H^{-1} Z(g(\mu)) \hat{\mathcal{O}}^R_{\mu \nu}(x; \mu) +\mathcal{O}(t) ,
\end{align}
is finite, one can determine $Z$ from the requirement that $\zeta(t)  H^{-1} Z$ is finite.
In this way, $Z$ has been calculated up to two-loop \cite{Harlander:2018zpi}.\fn{
In our convention, $Z$ is inverse of that of Ref.~\cite{Harlander:2018zpi}. \label{fn:4}}
Now, we can express the flowed operators with the renormalized (finite) quantities,
\be
\Tilde{\mathcal{O}}_{\mu \nu}(t, x)=\zeta^R(t; g(\mu), \mu) \hat{\mathcal{O}}_{\mu \nu}^R(x; \mu)+\mathcal{O}(t) \label{FtoUF} .
\ee
Here we define the renormalized matrix $\zeta^R$ as
\be
\zeta^R(t; g(\mu), \mu)=R \zeta(t) H^{-1} Z(g(\mu)) , \label{zetaRdef}
\ee
which is finite.

We also use the following relations:
\be
g_0=Z_g \lt(\frac{\mu e^{\gamma_E/2}}{\sqrt{4 \pi}} \rt)^{\epsilon} g(\mu) \, , \label{g0}
\ee
\be
\beta(g)=\mu \frac{d g}{d \mu}=-\sum_{i=-1}^{\infty} \beta_{i} g \lt[ \frac{g^2}{(4 \pi)^2} \rt]^{i+1} \, ,  \label{betafn}
\ee
\be
m_0=Z_m m(\mu) \, ,
\ee
\be
\gamma_m=-\mu \frac{d \log{m}}{d \mu}=\sum_{i=0}^{\infty}  \gamma_m^{(i)} \lt[ \frac{g^2}{(4 \pi)^2} \rt]^{i+1} \, ,
\ee
where $Z_g$ is given by
\be
Z_g=1-\frac{\beta_0}{2 \epsilon} \frac{g^2}{(4\pi)^2}+\lt(\frac{3 \beta_0^2}{8 \epsilon^2}-\frac{\beta_1}{4 \epsilon} \rt)  \lt[ \frac{g^2}{(4\pi)^2} \rt]^2+\dotsb \, . \label{Zg}
\ee
In the beta function, there is $\beta_{-1}=\epsilon$, which is non-zero in $D$-dimensional spacetime.
We also list the first few coefficients:
\be
\beta_0=\frac{11}{3} C_A-\frac{4}{3} T_F ,
\ee
\be
\beta_1=\frac{34}{3} C_A^2-\lt(4 C_F+\frac{20}{3} C_A \rt) T_F ,
\ee
\be
\gamma_m^{(0)}=6 C_F ,
\ee
\be
\gamma_m^{(1)}=\frac{97}{3} C_A C_F+3 C_F^2-\frac{20}{3} C_F T_F .
\ee
An explanation of the constants $C_A$, $T_F$, and $C_F$ is given in App.~\ref{app:A}.

The  rest of the paper is organized as follows.
In Sect.~\ref{sec:2}, we first explain our main results and show the functional form
to be used in the $t \to 0$ extrapolation,
for the convenience for those who are interested in practical analyses.
The derivations of these results are given in Sect.~\ref{sec:3},
the main part of this paper.
In this study, the necessary elements are minimal;
we only need the LO $\zeta^R$, which can be trivially obtained, 
and the one-loop anomalous dimension of the unflowed composite operators.
In Sect.~\ref{sec:4}, as an additional study, we give a general argument how we can 
study in detail the $t$ dependence caused by dimension-six operators
(which is denoted roughly by $\mO(t)$ above).
An explicit result is given for the traceless part of the EMT in quenched QCD.
In Sect.~\ref{sec:5}, we perform numerical analysis using lattice data. 
Here, we study the thermodynamic quantities, in particular the entropy density 
and trace anomaly, which are proportional to $T^{\rm TL}_{\mu \nu}(x)$ and $T^{\rm S}(x)$, respectively.
We confirm the validity of the $t \to 0$ extrapolation function by using actual lattice data.
In Sect.~\ref{sec:6}, we give our conclusions and discussion.
In App.~\ref{app:A}, our conventions are explained.
In App.~\ref{app:B}, the relation between the EMT and 
the renormalization of the dimension-four composite operators is reviewed.
In App.~\ref{app:C}, the anomalous dimension matrices for the dimension-four composite
operators are summarized.
In App.~\ref{app:D}, the $L(\mu,t)$ dependence of the perturbative series 
for the coefficients $c_i^{(\prime)}(t)$ is presented. 
Also the one-loop results are shown.
In App.~\ref{app:E}, we show the relations that 
the matrix $K$ (which is introduced in Sect.~\ref{sec:3}) should satisfy.
In App.~\ref{app:F}, we present the NLO $\zeta^R$, which becomes necessary if one wants to study 
higher-order behavior of the $t$ dependence remaining in the fixed-order formula for the EMT.
(The NLO $\zeta^R$ is used in App.~\ref{app:G}.)
In App.~\ref{app:G}, we give an argument for estimating higher-order effects 
which are neglected in Sect.~\ref{sec:3}.
In App.~\ref{app:H}, for reference, we present the results of 
the thermodynamics quantities obtained with linear-type extrapolation functions.

\section{$t \to 0$ extrapolation functions}
\label{sec:2}
We first explain our results in quenched QCD.
The lattice data obtained with the N$^k$LO formulae of the EMT, 
\be
{T_{\mu \nu}^{\rm TL}}^{\text{(N$^k$LO)}}(x;t)=c_1^{\text{(N$^k$LO)}}(t) \tilde{\mO}^{\rm TL}_{1, \mu \nu}(t,x) , \label{eq:(2.1)}
\ee
\be
{T^{\rm S}}^{\text{(N$^k$LO)}}(x;t)=c_2^{\text{(N$^k$LO)}}(t) \tilde{\mO}^{\rm S}_2(t,x),
\ee
where $c_1^{\text{(N$^k$LO)}}$ and $c_2^{\text{(N$^k$LO)}}$ are given by Eqs.~\eqref{c1NkLO} and \eqref{c2NkLO}, should be fitted with the functions of $t$,
\be
{T_{\mu \nu}^{\rm TL}}^{\text{(N$^k$LO)}}(x;t)=\lt(1-k_1^{(k+1)} \lt[\frac{g(\mu(t))^2}{(4 \pi)^2} \rt]^{k+1}\rt) T_{\mu \nu}^{\rm TL}(x) , \label{resquenchTL}
\ee
\be
{T^{\rm S}}^{\text{(N$^k$LO)}}(x;t)=\lt(1-\frac{8}{\beta_0} k_2^{(k+2)} \lt[\frac{g(\mu(t))^2}{(4 \pi)^2} \rt]^{k+1}\rt) T^{\rm S}(x) , \label{resquenchS}
\ee
in $t \to 0$ extrapolation analyses.
Here, $T_{\mu \nu}^{\rm TL}(x)$, $T^{\rm S}(x)$, $k_1^{(k+1)}$, and $k_2^{(k+2)}$  
are fit parameters. $g(\mu(t))$ is a running coupling\fn{One should consider
the $(k+1)$ or higher-loop beta function for the N$^k$LO calculation
in calculating the running. The same applies to the full QCD case.} and the renormalization scale $\mu(t)$
should be taken common to that of $c_{1,2}^{\text{(N$^k$LO)}}(t)$.
$T_{\mu \nu}^{\rm TL}(x)$ and $T^{\rm S}(x)$ are the EMT
we want to extract by the $t \to 0$ extrapolation analysis.

In full QCD, the lattice data obtained with the N$^k$LO formulae of the EMT,\fn{
Nowadays, $c_4$ is known to NNLO but $c_2$ is known to NLO.
If one uses the coefficients at the highest order available today,
it corresponds to the NLO formula of the EMT.}
given in Eqs.~\eqref{TTLNkLO} and \eqref{TSNkLO}, should be fitted with the functions,
\be
{T_{\mu \nu}^{\rm TL}}^{\text{(N$^k$LO)}}(x;t)=T_{\mu \nu}^{\rm TL}(x)+a(x) \lt[\frac{g(\mu(t))^2}{(4 \pi)^2} \rt]^{k+1} , \label{resfullTL}
\ee
\be
{T^{\rm S}}^{\text{(N$^k$LO)}}(x;t)=T^{\rm S}(x)+b (x) \lt[\frac{g(\mu(t))^2}{(4 \pi)^2} \rt]^{k+1} . \label{resfullS}
\ee
Here, $T_{\mu \nu}^{\rm TL}(x)$, $T^{\rm S}(x)$, $a(x)$, and $b(x)$ are fit parameters.
(Note that $a(x)$ and $b(x)$ are operators; see below.)
$T_{\mu \nu}^{\rm TL}(x)$ and $T^{\rm S}(x)$ correspond to the final results 
one is interested in.

In quenched QCD, the fit parameters $k_1^{(k+1)}$ and $k_2^{(k+2)}$ 
correspond to higher-order coefficients [cf. Eqs.~\eqref{c1} and \eqref{c2}].
In this sense, 
(i)  they do not depend on 
 $x$ or the typical scale $Q$ of a considered system
(for instance, when the expectation value of the one-point function of the EMT is considered 
at finite temperature $T$, the typical scale is $Q=T$),
and (ii) it can be predicted how they vary in response to the variation of the parameter $s$ in 
$\mu(t)=s/\sqrt{2 e^{\gamma_E} t}$ as long as the N$^k$LO coefficients are known; 
see Sect.~\ref{sec:5} and App.~\ref{app:D}.
Of course, these properties do not hold exactly due to systematic errors in fits.
Nevertheless, we can check the validity of the use of the above extrapolation function
by looking into these properties.
In our analyses in Sect.~\ref{sec:5}, we will take the fit parameters $k_1^{(k+1)}$ and $k_2^{(k+2)}$ 
common to all the simulated temperatures, taking the first property into account.
We check the validity of the use of the above extrapolation functions
by examining property~(ii),
the behavior of the fit parameters $k_1^{(k+1)}$ and $k_2^{(k+2)}$
upon the variation of the choice of renormalization scale.

On the other hand, in full QCD, 
the origin of the fit parameters $a(x)$ and $b(x)$ is not so simple. 
Hence we cannot expect parallel properties to the quenched QCD case.
For instance, they generally depend on $x$ and the typical scale $Q$ of a system.
Also, it is not apparent how they change with the choice of the renormalization scale.
We note that $a(x)$ and $b(x)$ are actually composite operators and 
cannot be regarded as $c$ numbers.
Therefore, for instance, the two-point function $\langle {T^{\rm S}}^{\text{(N$^k$LO)}}(x_1;t) {T^{\rm S}}^{\text{(N$^k$LO)}}(x_2;t) \rangle$
is approximated as
\begin{align}
&\langle {T^{\rm S}}^{\text{(N$^k$LO)}}(x_1;t) {T^{\rm S}}^{\text{(N$^k$LO)}}(x_2;t) \rangle \non
& \simeq \langle {T^{\rm S}}(x_1) {T^{\rm S}}(x_2) \rangle
+ \lt[\frac{g(\mu(t))^2}{(4 \pi)^2} \rt]^{k+1} [\langle b(x_1) {T^{\rm S}}(x_2) \rangle+\langle  {T^{\rm S}}(x_1) b(x_2) \rangle] .
\end{align}
Then, $\langle {T^{\rm S}}^{\text{(N$^k$LO)}}(x_1;t) {T^{\rm S}}^{\text{(N$^k$LO)}}(x_2;t) \rangle$ should be fitted 
with
\be
\langle {T^{\rm S}}^{\text{(N$^k$LO)}}(x_1;t) {T^{\rm S}}^{\text{(N$^k$LO)}}(x_2;t) \rangle
=\langle {T^{\rm S}}(x_1) {T^{\rm S}}(x_2) \rangle +b'(x_1-x_2)  \lt[\frac{g(\mu(t))^2}{(4 \pi)^2} \rt]^{k+1} ,
\ee
by treating $b'(x_1-x_2)$ as a fitting parameter. 
On the other hand, when a one-point function is studied, 
$b(x)$ can be treated as a $c$ number and $x$ dependence is eliminated.

\section{Derivation of $t$ dependence of N$^k$LO formula}
\label{sec:3}
We investigate the leading $t$ dependence of the N$^k$LO formula of the EMT in the SF$t$X method
and derive the results of Eqs.~\eqref{resquenchTL}, \eqref{resquenchS}, \eqref{resfullTL}, and \eqref{resfullS}.
Throughout this section, we neglect the $\mathcal{O}(t)$ effect, coming from dimension-six operators, 
which is a subleading effect for sufficiently small $t$. 

\subsection{Quenched QCD}
\label{sec:3.1}
In this section we consider quenched QCD.
We give two derivations. 
In Sect.~\ref{sec:3.1.1}, we give a simple derivation using a characteristic of quenched QCD.
In Sect.~\ref{sec:3.1.2}, we give an alternative derivation, which is relatively complicated
but can be generalized to full QCD straightforwardly.

\subsubsection{Derivation I}
\label{sec:3.1.1}
We can derive Eqs.~\eqref{resquenchTL} and \eqref{resquenchS} in a very simple manner.
The traceless part is given by
\be
T_{\mu \nu}^{\rm TL}(x)=c_1(t) \Tilde{\mathcal{O}}^{\rm TL}_{1, \mu \nu}(t,x)  .
\ee
By multiplying both sides by $c_1^{(\text{N$^k$LO})}(t) c_1(t)^{-1}$, we obtain, from Eq.~\eqref{eq:(2.1)},
\begin{align}
{T_{\mu \nu}^{\rm TL}}^{(\text{N$^k$LO})}(x;t)=c^{(\text{N$^k$LO})}_1(t) \Tilde{\mathcal{O}}_{1, \mu \nu}(t,x)
=c_1^{\text{(N$^k$LO)}}(t) c_1(t)^{-1} T_{\mu \nu}^{\rm TL}(x) \, . \label{multi}
\end{align}
Here, $c_1^{(\text{N$^k$LO})}(t) c_1(t)^{-1}$ deviates from 1 because of the lack of higher-order corrections.
If we write $c_1(t)^{-1}$ perturbatively as 
\be
c_1(t)^{-1}=g(\mu(t))^2 \sum_{i=0}^{\infty} p_1^{(i)}(L(\mu(t),t)) \lt[ \frac{g(\mu(t))^2}{(4 \pi)^2} \rt]^{i}  ,
\ee
$p_1^{(i)}$ satisfies
\begin{align}
&k_1^{(0)} p_1^{(0)}=1,  \non
& \sum_{i+j=n}  k_1^{(i)} p_1^{(j)}=0  \quad {\text{for}~n \geq 1}  . \label{const}
\end{align}
Then, noting that $c_1^{(\text{N$^k$LO})}=\frac{1}{g(\mu(t))^2}\sum_{n=0}^{k} k_1^{(n)}(L(\mu(t),t)) \lt[\frac{g(\mu(t))^2}{(4\pi)^2} \rt]^n$, we obtain
\begin{align}
c_1^{(\text{N$^k$LO})}(t) c_1(t)^{-1}
&\simeq 1+\sum_{\substack{i+j=k+1 \\ 0 \leq i \leq k}}  k_1^{(i)} p_1^{(j)} \lt[ \frac{g(\mu(t))^2}{(4 \pi)^2} \rt]^{k+1} \non
&=1-k_1^{(k+1)}(L(\mu(t),t)) \lt[ \frac{g(\mu(t))^2}{(4 \pi)^2} \rt]^{k+1} \, .
\end{align}
In the last line, we used Eq.~\eqref{const} and $p_1^{(0)}=(k_1^{(0)})^{-1}=1$.
Combining this and Eq.~\eqref{multi}, we obtain Eq.~\eqref{resquenchTL}.

The scalar part, 
\be
T^S(x)=c_2(t) \Tilde{\mathcal{O}}_{2}^{\rm S}(t,x) ,
\ee
can be investigated in a parallel manner.
In this case, one should note that there is no ``$k_2^{(0)}$."

\subsubsection{Derivation II}
\label{sec:3.1.2}
We have already clarified the $t$ dependence in the N$^k$LO formula in Derivation I.
However, Derivation I cannot be applied straightforwardly to full QCD
because the traceless and scalar parts of the EMT are given by linear combinations of 
flowed operators in full QCD. 
Hence, we consider another derivation which is applicable to full QCD. This is Derivation II.

We consider the difference between the EMT and that of the N$^k$LO formulae:
\be
T_{\mu \nu}^{\rm TL}(x)-{T_{\mu \nu}^{\rm TL}}^{(\text{N$^k$LO})}(x;t)
=[c_1 (t) -c_1^{(\text{N$^k$LO})}(t) ] \Tilde{\mathcal{O}}^{\rm TL}_{1, \mu \nu}(t,x)  , \label{diff1quench}
\ee 
\be
T^{\rm S}(x)-{T^{\rm S}}^{(\text{N$^k$LO})}(x;t)
=[c_2 (t) -c_2^{(\text{N$^k$LO})}(t) ] \Tilde{\mathcal{O}}^{\rm S}_{2}(t,x) . \label{diff2quench}
\ee
Because $T_{\mu \nu}^{\rm TL}(x)$ and $T^{\rm S}(x)$ are $t$ independent,
the $t$ dependence of the N$^k$LO formulae is exhibited by the right-hand sides of Eqs.~\eqref{diff1quench} and \eqref{diff2quench}.
Here, the $t$ dependence coming from the difference between the coefficients $c_i(t)$ is easily evaluated as
\be
c_1(t)-c_1^{(\text{N$^k$LO})}(t) \simeq \frac{1}{g(\mu(t))^2} k_1^{(k+1)} \lt[ \frac{g(\mu(t))^2}{(4 \pi)^2} \rt]^{k+1} , 
\ee
\be
c_2(t)-c_2^{(\text{N$^k$LO})}(t) \simeq \frac{1}{g(\mu(t))^2} k_2^{(k+2)} \lt[ \frac{g(\mu(t))^2}{(4 \pi)^2} \rt]^{k+2}  .
\ee
Thus, the remaining task is to investigate the $t$ dependence of the flowed operators 
$\Tilde{\mathcal{O}}^{\rm TL}_{1, \mu \nu}(t,x)$ and $\Tilde{\mathcal{O}}^{\rm S}_{2}(t,x)$.\fn{
In quenched QCD, $t$ dependence of these operators can actually be revealed easily. 
Since $T_{\mu \nu}^{\rm TL}(x)=c_1(t) \Tilde{\mathcal{O}}^{\rm TL}_{1, \mu \nu}(t,x)$ is $t$-independent,
the leading $t$-dependence is $\Tilde{\mathcal{O}}^{\rm TL}_{1, \mu \nu}(t,x)=c_1(t)^{-1}  T^{\rm TL}_{\mu \nu}(x) \simeq g(\mu(t))^2 T^{\rm TL}_{\mu \nu}(x)$. Similarly, $\Tilde{\mathcal{O}}^S_{2}(t,x) \simeq \frac{(4 \pi)^2}{k_2^{(1)}}  T^S(x)$ follows.
However, again for the reason that this argument cannot straightforwardly be applied to full QCD, 
we develop an argument not essentially relying on a characteristic of quenched QCD.}

From Eq.~\eqref{FtoUF}, we have
\be
\Tilde{\mathcal{O}}_{\mu \nu}(t,x)= \zeta^R(t; g(\mu), \mu) \hat{\mathcal{O}}_{\mu \nu}^R(x; \mu) \, . 
\ee
As explained in the introduction, we consider a $t$-dependent 
renormalization scale, $\mu=\mu(t)$. Accordingly, we set $\mu=\mu(t)$:
\be
\Tilde{\mathcal{O}}_{\mu \nu}(t,x)= \zeta^R(t; g(\mu(t)), \mu(t)) \hat{\mathcal{O}}^R_{\mu \nu}(x; \mu(t)) \, .
\ee
Note that $t$ dependence is exhibited not only by $\zeta^R(t; g(\mu(t)), \mu(t))$ but also by $\hat{\mathcal{O}}^R_{\mu \nu}(x; \mu(t))$. 
We can obtain the former matrix, $\zeta^R(t; g(\mu(t)), \mu(t))$, by renormalizing the bare results $\zeta(t)$ in Refs.~\cite{Suzuki:2013gza,Makino:2014taa,Harlander:2018zpi} according to Eq.~\eqref{zetaRdef}.
(As explained below, however, for the present purpose, 
we just need the tree-level result, which can be trivially obtained.)

Let us investigate the leading $t$ dependence of $\hat{\mathcal{O}}^R_{\mu \nu}(x; \mu(t))$.
Since the bare operator
\be
\hat{\mO}_{\mu \nu}(x)=Z(g(\mu)) \hat{\mO}^R_{\mu \nu}(x; \mu)
\ee
is independent of the renormalization scale, we have the RG equation
for the renormalized operators:
\be
\lt[ \mu \frac{d}{d \mu} +\gamma(g(\mu)) \rt] \hat{\mO}^R_{\mu \nu}(x; \mu)=0 \label{RGRenOp} ,
\ee
where $\gamma(g(\mu))$ is the anomalous dimension matrix,
\be
\gamma(g)=Z^{-1} \mu \frac{d}{d \mu} Z=Z^{-1} \beta(g) \frac{d}{d g} Z . \label{gammaDef}
\ee
The solution to Eq.~\eqref{RGRenOp} is given by
\be
\hat{\mathcal{O}}_{\mu \nu}^R(x; \mu)=K(\mu; \mu_0) \hat{\mathcal{O}}_{\mu \nu}^R(x; \mu_0) ,
\ee
with
\be
K(\mu; \mu_0)=P \exp\lt[-\int_{g(\mu_0)}^{g(\mu)} d x \frac{\gamma(x)}{\beta(x)} \rt] . \label{Kfirstappear}
\ee
Here, $P$ denotes the path-ordered product.\fn{
More specifically, the path-order product orders a product as follows:
the operator whose variable is closest to $g(\mu)$ is brought to the most left side, 
and the second closest one is to the second location from left, and so on.}

To summarize, the flowed operator can be written as 
\be
\Tilde{\mathcal{O}}_{\mu \nu}(t,x)= \zeta^R(t; g(\mu(t)), \mu(t)) K(\mu(t);\mu_0) \hat{\mathcal{O}}_{\mu \nu}^R(x; \mu_0)  . \label{OTrewrite} 
\ee
The $t$ dependence of the flowed operaor is exhibited by $\zeta^R(t; g(\mu(t)), \mu(t))$ and $K(\mu(t);\mu_0)$.
Then, we calculate these two matrices.
Since we are interested in the leading behavior, we consider $\zeta^R$ and $K$ at LO.

At LO, $\zeta^R$ [whose definition is given in Eq.~\eqref{zetaRdef}] is trivially given by
\be
\zeta^R(t; g(\mu), \mu)=
\begin{pmatrix}
g(\mu)^2 & 0 \\
0 & g(\mu)^2
\end{pmatrix} ,
\ee 
and $K(\mu; \mu_0)$ is given by 
\be
K(\mu; \mu_0)=
\begin{pmatrix}
1 & \frac{1}{4} \lt[ \lt(\frac{g(\mu_0)}{g(\mu)} \rt)^2-1 \rt]  \\
0 &  \lt(\frac{g(\mu_0)}{g(\mu)} \rt)^2
\end{pmatrix} \, ,
\ee
from the anomalous dimension matrix of Eq.~\eqref{gamma0quench}.

From these results, we obtain the $t$ dependence of the flowed operators at LO as
\be
\Tilde{\mO}_{1, \mu \nu}^{\rm TL}(t,x)=g(\mu(t))^2 \lt[\hat{\mO}^R_{1, \mu \nu}(x; \mu_0)-\frac{1}{4} \hat{\mO}^R_{2, \mu \nu}(x; \mu_0)+\mathcal{O}(g(\mu_0)^2) \rt]+\mathcal{O}(g(\mu(t))^4) , \label{TLquenchDerivation2}
\ee
\be
\Tilde{\mO}_2^{\rm S}(t,x)=[g(\mu_0)^2 + \mathcal{O}(g(\mu_0)^4) ]\hat{\mO}^{R, {\rm S}}_{2}(x; \mu_0)+\mathcal{O}(g(\mu(t))^2) .  \label{SquenchDerivation2}
\ee
Thus, from Eqs.~\eqref{diff1quench} and \eqref{diff2quench}, we conclude that
\be
{T_{\mu \nu}^{\rm TL}}^{(\text{N$^k$LO})}(x;t)=T_{\mu \nu}^{\rm TL}(x)+\mathcal{O}(g(\mu(t))^{2(k+1)}) \, ,
\ee
\be
{T^{\rm S}}^{(\text{N$^k$LO})}(x;t)={T^{\rm S}}(x)+\mathcal{O}(g(\mu(t))^{2 (k+1)}) \, .
\ee
We have revealed the leading dependence on $g(\mu(t))$,
which agrees with Derivation I.

However, these are not as precise as Eqs.~\eqref{resquenchTL} and \eqref{resquenchS}.
In fact, we can reproduce Eqs.~\eqref{resquenchTL} and \eqref{resquenchS} as follows.
First, let us consider the traceless part. 
Actually, the quantity inside the square brackets in Eq.~\eqref{TLquenchDerivation2}
can be written as
$\hat{\mO}^R_{1, \mu \nu}(x; \mu_0)-\frac{1}{4} (1+\mathcal{O}(g(\mu_0)^2))\hat{\mO}^R_{2, \mu \nu}(x; \mu_0)$;
the higher-order effects in $g(\mu_0)$ appear only in the coefficient of $\hat{\mO}^R_{2, \mu \nu}$,
and the coefficient of $\hat{\mO}^R_{1, \mu \nu}$ is exactly one.
This follows from $K(\mu; \mu_0)_{11}=1$ at any order of perturbation theory,
because $Z_{11}=1$ and $Z_{21}=0$ to all orders [see App.~\ref{app:B}, in particular Eq.~\eqref{quenchZ}].
[Note that $g(\mu_0)$ dependence comes only from $K(\mu,\mu_0)$ and not from $\zeta^R(t;g(\mu(t)),\mu(t))$.]
Now we note that $\Tilde{\mathcal{O}}_{1, \mu \nu}^{\rm TL}(t,x)$
is traceless. Then $\hat{\mO}^R_{1, \mu \nu}(x; \mu_0)-\frac{1}{4} (1+\mathcal{O}(g(\mu_0)^2))\hat{\mO}^R_{2, \mu \nu}(x; \mu_0)$ should also be traceless.
Therefore, $\frac{1}{4} (1+\mathcal{O}(g(\mu_0)^2))\hat{\mO}^R_{2, \mu \nu}(x; \mu_0)$
should be $(\delta_{\mu \nu}/4) \hat{\mO}^{R, {\rm S}}_{1}(x;\mu_0)$.
Using Eq.~\eqref{TLrenquench}, we then conclude that
\begin{align}
\Tilde{\mO}_{1, \mu \nu}^{\rm TL}(t,x)
&=g(\mu(t))^2 \lt[\hat{\mO}^R_{1, \mu \nu}(x; \mu_0)-\frac{\delta_{\mu \nu}}{4} \hat{\mO}_1^{R, {\rm S}}(x; \mu_0) \rt]+\mathcal{O}(g(\mu(t))^4)  \non
&=g(\mu(t))^2 T_{\mu \nu}^{\rm TL}(x)+\mathcal{O}(g(\mu(t))^4) \, .
\end{align}
This precisely gives Eq.~\eqref{resquenchTL}.

Secondly, let us consider $\Tilde{\mO}^{\rm S}_2(t,x)$. 
The $g(\mu(t))^0$-term of $\Tilde{\mO}^{\rm S}_2(t,x)$,
whose LO result is shown in Eq.~\eqref{SquenchDerivation2},
is given by $f(g(\mu_0)) \mathcal{O}_2^{R, {\rm S}}(x; \mu_0)$ beyond LO with some function $f$.
This should be $\mu_0$ independent.
Hence, $f(g(\mu_0))$ should be $f(g(\mu_0))=const. \times (-\frac{\beta(g(\mu_0))}{8 g(\mu_0)})$.
Here we have used the fact that the trace of the EMT, which is  given by Eq.~\eqref{Srenquench},
is $\mu$ independent.
From the LO result \eqref{SquenchDerivation2},
the constant is determined as $const.=\frac{8 (4 \pi)^2}{\beta_0}$.
Thus, we can rewrite Eq.~\eqref{SquenchDerivation2} as
\begin{align}
\Tilde{\mO}_2^{\rm S}(t,x)
&=\frac{8 (4 \pi)^2}{ \beta_0} \lt(-\frac{\beta(g(\mu_0))}{8 g(\mu_0)} \rt) \hat{\mO}^{R, {\rm S}}_{2}(x; \mu_0)+\mathcal{O}(g(\mu(t))^2) \non
&=\frac{8 (4 \pi)^2}{ \beta_0}  T_{\rho \rho}(x) +\mathcal{O}(g(\mu(t))^2) .
\end{align}
This and Eq.~\eqref{diff2quench} give Eq.~\eqref{resquenchS}; note that $k_2^{(1)}=\beta_0/8$ in quenched QCD.

\subsection{Full QCD}
We investigate the leading $t$ dependence of the N$^k$LO formulae of the EMT in full QCD
in a manner parallel to Derivation II.
The N$^k$LO formulae differ from the exact ones by
\be
T_{\mu \nu}^{\rm TL}(x)-{T_{\mu \nu}^{\rm TL}}^{(\text{N$^k$LO})}(x;t)
=[c_1 (t) -c_1^{(\text{N$^k$LO})}(t) ] \Tilde{\mathcal{O}}^{\rm TL}_{1, \mu \nu}(t,x) 
+[c_3 (t) -c_3^{(\text{N$^k$LO})}(t) ] \Tilde{\mathcal{O}}^{\rm TL}_{3, \mu \nu}(t,x)  ,
\ee
\be
T^{\rm S}(x)-{T^{\rm S}}^{(\text{N$^k$LO})}(x;t)
=[c_2 (t) -c_2^{(\text{N$^k$LO})}(t) ] \Tilde{\mathcal{O}}^{\rm S}_{2}(t,x)
+[c'_4(t) -{c_4^{\prime}}^{(\text{N$^k$LO})}(t) ] \Tilde{\mathcal{O}}^{\rm S}_{4}(t,x)  ,
\ee
where
\be
c_1(t)-c_1^{(\text{N$^k$LO})}(t) \simeq \frac{1}{g(\mu(t))^2} k_1^{(k+1)} \lt[ \frac{g(\mu(t))^2}{(4 \pi)^2} \rt]^{k+1}  ,
\ee
\be
c_2(t)-c_2^{(\text{N$^k$LO})}(t) \simeq \frac{1}{g(\mu(t))^2} k_2^{(k+2)} \lt[ \frac{g(\mu(t))^2}{(4 \pi)^2} \rt]^{k+2}   ,
\ee
\be
c_3(t)-c_3^{(\text{N$^k$LO})}(t) \simeq k_3^{(k+1)} \lt[ \frac{g(\mu(t))^2}{(4 \pi)^2} \rt]^{k+1}  ,
\ee
\be
c'_4(t)-{{c_4}^{\prime}}^{(\text{N$^k$LO})}(t) \simeq k_4^{(k+1)} \lt[ \frac{g(\mu(t))^2}{(4 \pi)^2} \rt]^{k+1}   .
\ee
Then, we investigate the leading $t$ dependence of $\Tilde{\mathcal{O}}^{\rm TL}_{1, \mu \nu}(t,x)$,
$\Tilde{\mathcal{O}}^{\rm TL}_{3, \mu \nu}(t,x)$, $\Tilde{\mathcal{O}}^{\rm S}_{2}(t,x)$, and~$\Tilde{\mathcal{O}}^{\rm S}_{4}(t,x)$.

The $t$ dependence of the flowed operators can be investigated from
\be
\Tilde{\mO}_{\mu \nu}(t,x)=\zeta^R(t; g(\mu(t)), \mu(t)) K(\mu(t); \mu_0) \hat{\mO}^R_{\mu \nu}(x; \mu_0) , \label{tildeOrewrite}
\ee
where $K(\mu;\mu_0)$ denotes
\be
K(\mu;\mu_0)=P \exp\lt[-\int_{g(\mu_0)}^{g(\mu)} d x \frac{\gamma(x)}{\beta(x)} \rt] . \label{K}
\ee
The anomalous dimension matrix is defined in a parallel manner to Eq.~\eqref{RGRenOp}.

At LO, $\zeta^R(t; g(\mu), \mu)$  is given by
\begin{align}
\zeta^R(t; g(\mu), \mu)=
\begin{pmatrix}
g(\mu)^2  & 0 & 0 & 0 \\
0  & g(\mu)^2 & 0 &  0\\
0 & 0 & 1 & 0 \\
0 & 0 & 0 & 1\\
\end{pmatrix} .
\end{align} 
Now we calculate $K(\mu; \mu_0)$ at LO, where we set  $\gamma=\gamma_0 g^2/(4 \pi)^2$ and $\beta=-\beta_0 g [g^2/(4 \pi)^2]$ in Eq.~\eqref{K}. 
The $g$ dependence of $K$ is determined by eigenvalues of  $\gamma_0$.
We give $\gamma_0$ in Eq.~\eqref{gamma0full}.
The eigenvalues are given by $\lambda=-2 \beta_0, 0,  \frac{8}{3} (2 C_F+ T_F)$. 
(The eigenvalue $0$ is degenerate.)
Then we obtain
\begin{align}
&K_{11}=\frac{2 C_F}{2 C_F+ T_F} +\frac{T_F}{2 C_F+T_F} \lt( \frac{g(\mu)}{g(\mu_0)} \rt)^{\frac{8 (2C_F+T_F)}{3 \beta_0}} , \non
& K_{12}=\frac{1}{4} \lt(\frac{g(\mu_0)}{g(\mu)} \rt)^2-\frac{C_F}{2 (2C_F+T_F)}-\frac{T_F}{4(2C_F+T_F)} \lt( \frac{g(\mu)}{g(\mu_0)} \rt)^{\frac{8 (2C_F+T_F)}{3 \beta_0}} , \non
& K_{13}=\frac{C_F}{2 (2C_F+T_F)}-\frac{C_F}{2 (2C_F+T_F)} \lt( \frac{g(\mu)}{g(\mu_0)} \rt)^{\frac{8 (2C_F+T_F)}{3 \beta_0}} , \non
& K_{14}=\frac{3 C_F}{2 \beta_0} \lt(\frac{g(\mu_0)}{g(\mu)} \rt)^2
-\lt[\frac{C_F}{4(2 C_F+T_F)}+\frac{3 C_F}{2 \beta_0} \rt]
+\frac{C_F}{4(2 C_F+T_F)} \lt( \frac{g(\mu)}{g(\mu_0)} \rt)^{\frac{8 (2C_F+T_F)}{3 \beta_0}} , \non 
& K_{21}=0  , \non
& K_{22}=\lt(\frac{g(\mu_0)}{g(\mu)} \rt)^2  , \non
& K_{23}=0 , \non
& K_{24}=\frac{6 C_F}{\beta_0} \lt(\frac{g(\mu_0)}{g(\mu)} \rt)^2-\frac{6 C_F}{\beta_0} , \non
& K_{31}=\frac{4 T_F}{2 C_F+T_F} -\frac{4 T_F}{2 C_F+T_F} \lt( \frac{g(\mu)}{g(\mu_0)} \rt)^{\frac{8 (2C_F+T_F)}{3 \beta_0}} , \non
& K_{32}=-\frac{T_F}{2 C_F+T_F} +\frac{T_F}{2 C_F+T_F} \lt( \frac{g(\mu)}{g(\mu_0)} \rt)^{\frac{8 (2C_F+ T_F)}{3 \beta_0}} , \non
& K_{33}=\frac{T_F}{2 C_F+T_F}+\frac{2 C_F}{2 C_F+T_F} \lt( \frac{g(\mu)}{g(\mu_0)} \rt)^{\frac{8 (2C_F+T_F)}{3 \beta_0}} , \non
& K_{34}=\frac{C_F}{2 C_F+T_F}-\frac{C_F}{2 C_F+T_F} \lt( \frac{g(\mu)}{g(\mu_0)} \rt)^{\frac{8 (2C_F+T_F)}{3 \beta_0}} , \non
& K_{41}=K_{42}=K_{43}=0 , \non
& K_{44}=1 . \label{Kfull}
\end{align}
One can confirm that this result satisfies the required relations for $K$, given in App.~\ref{app:E}.

From these matrices, the $t$ dependence of the flowed operators of interest is given by
\begin{align}
&\Tilde{\mO}_{1, \mu \nu}^{\rm TL}(t,x) \non
&=g(\mu(t))^2 \bigg[ \frac{2 C_F}{2 C_F+T_F} \lt(\hat{\mO}_{1, \mu \nu}^R(x;\mu_0)-\frac{1}{4}\hat{\mO}_{2, \mu \nu}^R(x;\mu_0)+\frac{1}{4} \hat{\mO}_{3, \mu \nu}^R(x;\mu_0)-\frac{1}{8} \hat{\mO}_{4, \mu \nu}^R(x;\mu_0)  \rt)+\mathcal{O}(g(\mu_0)^2)\bigg] , \non
&\quad{}+\mathcal{O}( g(\mu(t))^2 (g(\mu(t))^2/g(\mu_0)^2)^{\frac{4 (2C_F+ T_F)}{3 \beta_0}})  \label{O1TL}  \\ 
&\Tilde{\mO}_{3, \mu \nu}^{\rm TL}(t,x) \non
&=\bigg[ \frac{4 T_F}{2 C_F+T_F} \lt(\hat{\mO}_{1, \mu \nu}^R(x;\mu_0)-\frac{1}{4}\hat{\mO}_{2, \mu \nu}^R(x;\mu_0)+\frac{1}{4} \hat{\mO}_{3, \mu \nu}^R(x;\mu_0)-\frac{1}{8} \hat{\mO}_{4, \mu \nu}^R(x;\mu_0) \rt)+\mathcal{O}(g(\mu_0)^2) \bigg] \non
&\quad{}+\mathcal{O}((g(\mu(t))^2/g(\mu_0)^2)^{\frac{4 (2C_F+ T_F)}{3 \beta_0}}) ,  \label{O3TL}
\end{align}
and
\begin{align}
\Tilde{\mO}_{2}^{\rm S}(t,x)
&=g(\mu_0)^2 \lt(\hat{\mO}_2^{R, {\rm S}}(x;\mu_0)+\frac{6 C_F}{\beta_0} \hat{\mO}_4^{R, {\rm S}}(x;\mu_0)+\mathcal{O}(g(\mu_0)^2) \rt)+\mathcal{O}(g(\mu(t))^2)  , \label{O2S} \\
\Tilde{\mO}_{4}^{\rm S}(t,x)
&=\hat{\mO}_4^{R, {\rm S}}(x;\mu_0)+\mathcal{O}(g(\mu_0)^2)+\mathcal{O}(g(\mu(t))^2) . \label{O4S}
\end{align}
Hence, we obtain
\be
{T_{\mu \nu}^{\rm TL}}^{(\text{N$^k$LO})}(x;t)=T_{\mu \nu}^{\rm TL}(x)+\mathcal{O}([g(\mu(t))^2]^{k+1}) ,
\ee
\be
{T^{\rm S}}^{(\text{N$^k$LO})}(x;t)=T^{\rm S}(x)+\mathcal{O}([g(\mu(t))^2]^{k+1}) .
\ee
We have obtained the main results, Eqs.~\eqref{resfullTL} and \eqref{resfullS}.

For systematic calculation, for instance to study higher-order effects, 
it would be convenient to decompose dimension-four operators into traceless parts and scalar operators.
Then it is enough to treat two-by-two matrices, instead of four-by-four matrices, for each vector space.
We discuss higher-order effects in this treatment in Appendix~\ref{app:G}.
In addition, in Appendix~\ref{app:G} we explain a systematic way to estimate the errors of the LO results,
which are shown by the symbol $\mathcal{O}(\dots)$ in Eqs.~\eqref{O1TL}--\eqref{O4S}.

One might wonder if the $t$-dependent term of the N$^k$LO formulae for the EMT 
is proportional to the EMT as in the quenched QCD case.
However, this is not necessarily expected in full QCD.
The reason why we obtained such a result in quenched QCD can be explained as follows. 
In quenched QCD,  $T_{\mu \nu}^{\rm TL}(x)$ ($T^S(x)$) 
is the only traceless (scalar) operator which is RG invariant.
Then, $\tilde{\mathcal{O}}_{1, \mu \nu}^{\rm TL}(t,x)$ ($\tilde{\mathcal{O}}_{2}^{\rm S}(t,x)$), which is traceless (scalar) 
and RG invariant (or more precisely $\mu_0$ independent), 
should be proportional to $T_{\mu \nu}^{\rm TL}$ ($T^S$).
Hence, the $t$ dependent term of the N$^k$LO EMT expression is proportional to 
the exact EMT. This is the conclusion of Sect.~\ref{sec:3.1.2}.
On the other hand, in the full QCD case, considering scalar operators, we have two RG invariant operators: 
$T^{\rm S}$ and $\hat{\mathcal{O}}^R_4$.
Then it is not necessary that 
the RG invariant flowed operators $\tilde{\mathcal{O}}_2^{\rm S}$ and $\tilde{\mathcal{O}}_4^{\rm S}$
are proportional to $T^{\rm S}$.
This expectation can be denied more explicitly by considering the difference between the LO and NLO 
formulae for $T^{\rm S}$. It is not proportional to
$[1/(4 \pi)^2] \lt(\frac{11}{24} C_A-\frac{3}{8} T_F \rt)\tilde{\mathcal{O}}_2^S+\frac{1}{8} \tilde{\mathcal{O}}_4^S$,
which is $T^{\rm S}$ as $t \to 0$ due to Eq.~\eqref{S} with the LO coefficients $c_2(t)$ and $c'_4(t)$.

\section{$\mathcal{O}(t)$ correction: Contribution from dimension-six operators}
\label{sec:4}

So far, we have neglected the contribution from dimension-six operators
in the small flow time expansion, which is suppressed for small $t$ roughly as $\mathcal{O}(t)$.
This is parameterically smaller than the $\mO(g(\mu(t))^n)$ terms, which we have obtained in Sect.~\ref{sec:3},
in a sufficiently small $t$ region.
Nevertheless, it might be possible that the $\mO(t)$ effect dominates over 
the $\mO(g(\mu(t))^n)$ effect in the region of $t$ where lattice simulation is practically carried out.
Hence, as an additional study, we discuss how we can detect the detailed $t$ dependence coming
from dimension-six operators, and show an explicit result of the detailed $t$ dependence for a specific example.

We first give a general argument for how to investigate the detailed $t$ dependence 
coming from dimension-six operators.
We go back to the small flow time expansion of the flowed operators $\Tilde{\mathcal{O}}_{i, \mu \nu}(t,x)$:
\be
\Tilde{\mathcal{O}}_{i, \mu \nu}(t,x)
=\zeta^R_{i j}(t; g(\mu), \mu) \hat{\mathcal{O}}^R_{j, \mu \nu} (x; \mu)+t \eta_{i j}(t) \mathcal{O}^{(6)}_{j, \mu \nu}(x)
+\mathcal{O}(t^2) \, , \label{SFTE2}
\ee
where the second term, which was neglected in the previous section, is now our focus;
$\mathcal{O}^{(6)}_{i, \mu \nu}$ is a bare dimension-six operator and
$\eta_{i j}$ is a coefficient matrix.
We denote the $\mathcal{O}(t)$ contribution by
\be
\delta \Tilde{\mathcal{O}}_{i, \mu \nu}(t,x) \equiv t \eta_{i j}(t) \mathcal{O}^{(6)}_{j, \mu \nu}(x) .
\ee
Hereafter, the symbol $\delta$ means an $\mathcal{O}(t)$ contribution.
The gradient flow representation of the EMT [see Eq.~\eqref{eq:(1.1)}], 
\be
T^{\rm GF}_{\mu \nu}(x;t)
=\sum_i \tilde{c}_i (t; g(\mu)) \Tilde{O}_{i, \mu \nu}(t,x)  ,
\ee
differs from the actual EMT by the dimension-six operators $\delta  \Tilde{\mathcal{O}}_{i, \mu \nu}(t,x)$,
which is roughly~$\mathcal{O}(t)$.
(We are now assuming that $\tilde{c}_i$'s are exactly known and this part does not induce any error.)
This is the reason why we show the superscript ``GF'' and $t$ dependence. 
The difference is given by
\be
\delta T^{\rm GF}_{\mu \nu}(x;t)=T^{\rm GF}_{\mu \nu}(x;t)-T_{\mu \nu}(x)
=\sum_i  \tilde{c}_i (t; g(\mu))  \delta  \Tilde{\mathcal{O}}_{i, \mu \nu}(t,x) . \label{diffOt}
\ee
We investigate the leading $t$ dependence of the right-hand side.
For this purpose, as before, we study the $t$ dependence of $\delta  \Tilde{\mathcal{O}}_{i, \mu \nu}(t,x)$ 
by rewriting it as
\begin{align}
 \delta  \Tilde{\mathcal{O}}_{i, \mu \nu}(t,x)
 &=t \eta_{i j}^{R}(t; g(\mu), \mu) \mathcal{O}^{(6) R}_{j, \mu \nu} (x; \mu) \non
 &=t  \eta_{i j}^{R}(t; g(\mu(t)), \mu(t)) K^{(6)}_{j k}  (\mu(t), \mu_0) \mathcal{O}^{(6) R}_{k, \mu \nu} (x; \mu_0) . \label{dim6tdep}
\end{align}
In the first equality, we have used the fact that the flowed operator is finite and thus
can be rewritten by renormalized  (finite) quantities.
The product of the two renormalized  quantities is independent of the renormalization scale $\mu$.
In the second equality, we have considered the RG evolution of the dimension-six renormalized operators,
\be
\lt[ \mu \frac{d}{d \mu}+\gamma^{(6)}(g(\mu)) \rt] \mathcal{O}_{\mu \nu}^{(6) R}=0 \, ,
\ee
and $K^{(6)}$ is given by
\be
K^{(6)}(\mu; \mu_0)=P \exp \lt[- \int^{g(\mu)}_{g(\mu_0)} dx \frac{\gamma^{(6)}(x) }{\beta(x)} \rt] .
\ee
To reveal the $t$ dependence, we need to know $\eta_{i j}^{R}(t; g(\mu(t)), \mu(t))$ and $K^{(6)}_{j k}  (\mu(t), \mu_0)$.

To obtain the leading $t$ dependence concretely, we need to know the matrix $\eta_{i j}$ at leading order
and the anomalous dimension matrix of dimension-six operators at one-loop.\fn{ 
In this analysis, it is sufficient to know $\tilde{c}_i (t; g(\mu))$ at LO.
}
The former can be calculated without loop calculations
and by using the flow equation alone. (We will show this explicitly in quenched QCD below.)
However, it is rather complicated to know the anomalous dimension matrix $\gamma^{(6)}$,
whose results are not completely known even now.
Here, we limit ourselves to the traceless part of the EMT in quenched QCD.
We demonstrate how we can calculate the leading $t$ dependence
coming from the dimension-six operators for this case.

We consider $T^{\rm GF, TL}_{\mu \nu}(x;t)$ in quenched QCD 
[$T^{\rm GF, TL}_{\mu \nu}(x;t)=c_1(t; g(\mu(t))) \tilde{\mO}_{1,\mu\nu}^{\rm TL}(t,x)$] and
investigate the detailed $t$ dependence caused by dimension-six operators.
We calculate $\eta_{ij}$ by studying the $\mathcal{O}(t)$ contribution of ${\rm tr} \, (G_{\mu \rho} G_{\nu \rho})$.
We will represent the $\mathcal{O}(t)$ contribution of the flowed operator ${\rm tr} \, (G_{\mu \rho} G_{\nu \rho})$ using the basis 
\begin{align}
& \mathcal{O}^{(6)}_{1, \mu \nu}=\frac{1}{g_0^2} {\rm tr} (D_{\nu} F_{\rho \sigma} D_{\mu} F_{\rho \sigma})+(\mu \leftrightarrow \nu)  , \non
& \mathcal{O}^{(6)}_{2, \mu \nu}=\frac{1}{g_0^2} {\rm tr} (D_{\rho} F_{\mu \rho} D_{\sigma} F_{\nu \sigma})+(\mu \leftrightarrow \nu) , \non
& \mathcal{O}^{(6)}_{3, \mu \nu}=\frac{1}{g_0^2} {\rm tr} (D_{\nu} F_{\mu \rho} D_{\sigma} F_{\rho \sigma})+(\mu \leftrightarrow \nu)  ,
\end{align}
taking into account that a similar basis is adopted in Ref.~\cite{Kim:2015ywa}.
We use 
\begin{align}
\delta G_{\mu \nu}
&=\delta (\partial_{\mu} B_{\nu}-\partial_{\nu} B_{\mu}+[B_{\mu}, B_{\nu}]) \non
&=\partial_{\mu} \delta B_{\nu}+[A_{\mu}, \delta B_{\nu}]-\partial_{\nu} \delta B_{\mu}-[A_{\nu}, \delta B_{\mu}] \non
&=D_{\mu} \delta B_{\nu}-D_{\nu} \delta B_{\mu} \non
&=t (D_{\mu} D_{\rho} F_{\rho \nu}-D_{\nu} D_{\rho} F_{\rho \mu}) ,
\end{align}
which follows from the flow equation $\partial_t B_{\mu}^a(t,x)=D_{\nu} G_{\nu \mu}^a(t,x)$ in Eq.~\eqref{eq:(1.8)}\fn{
In right-hand side of Eq.~\eqref{eq:(1.8)}, we have $\alpha_0 \times (\text{gauge non-invariant operator})$.
However, this part does not contribute to the flow time evolution of gauge invariant operators.}.
Then we have
\begin{align}
\delta {\rm tr} (G_{\mu \sigma} G_{\nu \sigma})
&={\rm tr} (\delta  G_{\mu \sigma} \cdot F_{\nu \sigma})+(\mu \leftrightarrow \nu) \non
&=t \, {\rm tr} (D_{\rho} F_{ \sigma \rho} D_{\mu} F_{\nu \sigma}) - t \, {\rm tr} (D_{\rho} F_{\mu \rho} D_{\sigma} F_{\nu \sigma })  \non
&\quad{}-t \, \partial_{\mu} {\rm tr} (D_{\rho} F_{\sigma \rho } \cdot F_{\nu \sigma}) +t \,\partial_{\sigma} {\rm tr} (D_{\rho}F_{\mu \rho} \cdot F_{\nu \sigma} )  \non
&\quad{}+(\mu \leftrightarrow \nu) 
\end{align}
Hereafter, we neglect the derivative terms. This is valid when we 
consider the matrix element with the states which are translational invariant.
Then, we obtain
\be
\delta {\rm tr} (G_{\mu \sigma} G_{\nu \sigma})
=-t g_0^2\mathcal{O}^{(6)}_{2, \mu \nu} + t  g_0^2 \mathcal{O}^{(6)}_{3, \mu \nu} .
\ee
We consider its traceless part and also renormalization:
\be
\delta {\rm tr} ([G_{\mu \sigma} G_{\nu \sigma}]^{\rm TL})
=-t  g(\mu(t))^2 \mathcal{O}^{(6)\, R,{\rm TL}}_{2, \mu \nu}(x;\mu(t))+ t g(\mu(t))^2 \mathcal{O}^{(6)\, R,{\rm TL}}_{3, \mu \nu}(x;\mu(t)) , \label{eta}
\ee
where we take the renormalization scale $\mu=\mu(t)$.

Now we calculate the $K^{(6)}$ matrix.
The renormalization factor for $\{\mathcal{O}^{(6)\, R,{\rm TL}}_{1, \mu \nu}, \mathcal{O}^{(6)\, R,{\rm TL}}_{2, \mu \nu},
\mathcal{O}^{(6)\, R,{\rm TL}}_{3, \mu \nu} \}$ was calculated in Ref.~\cite{Kim:2015ywa}:
\be
\mathcal{O}^{(6) \, {\rm TL}}_{i, \mu \nu}=Z_{i j}  \mathcal{O}^{(6) \, R, {\rm TL}}_{j, \mu \nu} ,
\ee
with
\be
Z=\left( \begin{array}{ccc}
1-\frac{3 C_A}{\epsilon} \frac{g^2}{(4 \pi)^2}& \frac{C_A}{3 \epsilon} \frac{g^2}{(4\pi)^2} & -\frac{8 C_A}{3 \epsilon} \frac{g^2}{(4 \pi)^2}  \\
0 & 1-\frac{4 C_A}{3 \epsilon} \frac{g^2}{(4 \pi)^2} & -\frac{ C_A}{6 \epsilon} \frac{g^2}{(4 \pi)^2} \\
0 & -\frac{2 C_A}{3 \epsilon} \frac{g^2}{(4 \pi)^2}  & 1-\frac{7 C_A}{6 \epsilon} \frac{g^2}{(4 \pi)^2} 
\end{array} \right) .
\ee
From this, we obtain the anomalous dimension:
\be 
\lt(\mu \frac{d}{d \mu} +\gamma^{(6)} \rt) \mathcal{O}_{\mu \nu}^{(6) \, R,{\rm TL}}=0 ,
\ee
with
\begin{align}
\gamma^{(6)}
&=\left( \begin{array}{ccc}
6 & -\frac{2}{3 }  & \frac{16}{3}  \\
0 & \frac{8}{3}  & \frac{1}{3}  \\
0 & \frac{4}{3}  & \frac{7}{3} 
\end{array} \right)  C_A  \frac{g^2}{(4 \pi)^2}+\mathcal{O}(g^4) .
\end{align}
By calculating $K^{(6)}=P \exp\lt[-\int_{g(\mu_0)}^{g(\mu)} dx \frac{\gamma^{(6)}(x)}{\beta(x)} \rt] $ at LO,
we obtain the RG evolution of the renormalized operators: 
\begin{align}
\mathcal{O}^{(6) R,{\rm TL}}_{2, \mu \nu}(x;\mu(t))
&=\frac{1}{\sqrt{17}} \lt(\frac{1+\sqrt{17}}{2} \lt(\frac{g(\mu(t))}{g(\mu_0)} \rt)^{\lambda_2}
+\frac{-1+\sqrt{17}}{2} \lt(\frac{g(\mu(t))}{g(\mu_0)} \rt)^{\lambda_3} \rt) \mathcal{O}^{(6) R,{\rm TL}}_{2, \mu \nu}(x;\mu_0) \non
&\quad{}+\frac{1}{\sqrt{17}} \lt(\lt(\frac{g(\mu(t))}{g(\mu_0)} \rt)^{\lambda_2}-\lt(\frac{g(\mu(t))}{g(\mu_0)} \rt)^{\lambda_3} \rt) \mathcal{O}^{(6) R,{\rm TL}}_{3, \mu \nu}(x;\mu_0)  , \label{K61}
\end{align}
\begin{align}
\mathcal{O}^{(6) R,{\rm TL}}_{3, \mu \nu}(x;\mu(t))
&=\frac{4}{\sqrt{17}} \lt(\lt(\frac{g(\mu(t))}{g(\mu_0)} \rt)^{\lambda_2}-\lt(\frac{g(\mu(t))}{g(\mu_0)} \rt)^{\lambda_3} \rt) \mathcal{O}^{(6) R,{\rm TL}}_{2, \mu \nu}(x;\mu_0) \non
&\quad{}+\frac{1}{2 \sqrt{17}} \lt((-1+\sqrt{17})\lt(\frac{g(\mu(t))}{g(\mu_0)} \rt)^{\lambda_2}
+(1+\sqrt{17}) \lt(\frac{g(\mu(t))}{g(\mu_0)} \rt)^{\lambda_3} \rt)  \non
&\qquad{}\times\mathcal{O}^{(6) R,{\rm TL}}_{3, \mu \nu}(x;\mu_0)  , \label{K62}
\end{align}
where $\lambda_{1,2,3}$ are eigenvalues of the matrix $\frac{3}{11 C_A} \gamma_0^{(6)}$,
$\lambda_1=\frac{18}{11}, \lambda_2=\frac{1}{22} (15+\sqrt{17}), \lambda_3=\frac{1}{22} (15-\sqrt{17})$,
whose numerical values are $\lambda_1 \approx1.64$, $\lambda_2 \approx 0.87$, $\lambda_3 \approx 0.49$.
($\lambda_1$ does not appear in the result.) 

From Eqs.~\eqref{eta}, \eqref{K61} and \eqref{K62}, we obtain the leading behavior as
\be
\delta {\rm tr} ([G_{\mu \rho} G_{\nu \rho}]^{\rm TL}) \simeq  C  g(\mu(t))^{2+\lambda_3} t ,
\ee
with a (dimension-six) constant $C$, because $\lambda_3$ is the smallest eigenvalue.
Finally from Eq.~\eqref{diffOt} we obtain
\be
\delta T_{\mu \nu}^{\rm GF, TL}=c_1(t) \delta \tilde{\mathcal{O}}^{\rm TL}_{1, \mu \nu}
\simeq  C'  g(\mu(t))^{\lambda_3} t , \label{TLdim6}
\ee
with a (dimension-six) constant $C'$.

\section{Numerical analysis of thermodynamics quantities in quenched QCD}
\label{sec:5}
In this section, we carry out lattice simulation of the EMT with the SF$t$X method in quenched QCD.
We study thermodynamic quantities \cite{Boyd:1996bx,Okamoto:1999hi, Borsanyi:2012ve, Borsanyi:2013bia,Bazavov:2014pvz,Shirogane:2016zbf, Giusti:2014ila, Giusti:2016iqr, Caselle:2018kap}.
We use the $t\to 0$ extrapolation function studied in Sect.~\ref{sec:3} (or given in Sect.~\ref{sec:2})
and also examine its validity.
Our analysis is similar to that of Ref.~\cite{Iritani:2018idk}.
The main difference is that here we use Eqs.~\eqref{resquenchTL} and \eqref{resquenchS}
in the $t\to 0$ extrapolation, but in Ref.~\cite{Iritani:2018idk} a linear function in $t$ was mainly used.

We use the lattice data obtained in Ref.~\cite{Kitazawa:2016dsl} for flowed operators.\fn{
The authors are grateful to Takumi Iritani  and Masakiyo Kitazawa for letting us use the data.}
See this reference for the details of the lattice setup.
We study the finite temperature effect, i.e. the difference between 
finite temperature and zero temperature, of the (dimensionless) 
entropy density $s/T^3$ and trace anomaly $\Delta/T^4$, which are given by
\be
s T=\epsilon+p=-\frac{4}{3} T^{\rm TL}_{44} , \label{entropy}
\ee
\be
\Delta=\epsilon-3p=-T^S . \label{TA}
\ee
We measure these quantities at temperature 
$T/T_c=0.93$, $1.02$, $1.12$, $1.40$, $1.68$, $2.10$, $2.31$, and $2.69$,
where $T_c$ denotes the critical temperature.\fn{
The trace anomaly is studied only at $T/T_c=0.93$, $1.02$, $1.12$, $1.40$, and $1.68$.
This is due to lack of zero temperature simulations, which require more numerical costs.
On the other hand, we do not need zero temperature simulations for the entropy density
because it is exactly zero at zero temperature.}

Our lattice analysis consists of three steps.
First, we carry out $a \to 0$ extrapolation with lattice data at three lattice spacings  
to obtain continuum limit results for the flowed operator $\tilde{\mathcal{O}}^{\rm TL}_{1, 44}$
and $\tilde{\mathcal{O}}^{\rm S}_{2}$. We obtain these results as
a function of $t T^2$.
In order to keep discretization effects under good control
we need $2 e^{\gamma_E} t  /a^2 \gg 1$. 
(Here we assume that the typical scale of the flowed operators is $1/\sqrt{2 e^{\gamma_E} t}$.)
Noting that $T=1/(N_{\tau} a)$ and $N_{\tau} \geq 12$ 
(where $N_{\tau}$ is the number of sites in the Euclidean time direction)
in our lattice setup, we see that the above condition corresponds to $t T^2 \gg 0.002$.
Then we basically use the continuum limit results at $t T^2 \geq 0.010$.
Secondly, after obtaining the continuum limit results for the flow operators, 
we multiply the flowed operators by coefficients $c_i(t)$ [and trivial factors in Eqs.~\eqref{entropy} and~\eqref{TA}]
such that they correspond to the thermodynamic quantities.
We use NLO or NNLO coefficients, namely we use the NLO or NNLO formula of the EMT.
Finally, we extrapolate the data to the zero flow time limit using Eqs.~\eqref{resquenchTL}
and \eqref{resquenchS} with $k=1$ or $2$.
Then we obtain the final results.

\begin{figure}[tbhp]
\begin{center}
\includegraphics[width=10cm]{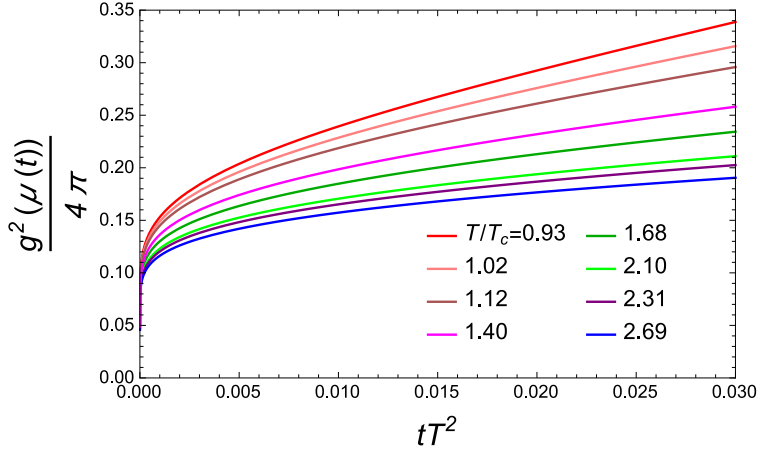}
\end{center}
\caption{The size of the running coupling constant $g^2(\mu(t))/(4 \pi)$ as a function of $t T^2$.
It is shown for various $T/T_c$.}
\label{fig:coupling}
\end{figure}

In the second and third steps, we rely on perturbation theory.
This is valid when the flow time satisfies $t \ll \Lambda_{\overline{\rm MS}}^{-2}$.
In Fig.~\ref{fig:coupling}, we show the size of the running coupling as a function of $t T^2$ 
in the relevant region, taking the renormalization scale of the running coupling as 
$\mu(t)=1/(2 e^{\gamma_E} t)^{1/2}$.
 Our analysis is mainly performed for $t T^2 \leq 0.015$.
To obtain the running coupling as a function of $t T^2$, which is 
originally a function of $\mu(t)/\Lambda_{\overline{\rm MS}}$, we use \cite{Kitazawa:2016dsl}
\be
w_0 T_c=0.2524 , \quad{} w_0 \Lambda_{\overline{\rm MS}}=0.2154 , \label{scaleinputs}
\ee
where $w_0$ is a reference scale. (Here we only need the ratio $T_c/\Lambda_{\overline{\rm MS}}$.)
Here and hereafter we use the three-loop beta function.

We summarize the setup of our central analysis:
\begin{align}
&{\text{range of the used lattice data:}}~ 0.010 \leq t T^2 \leq 0.015 , \non
&{\text{scale setting parameters:  Eq.~\eqref{scaleinputs}}}, \non
&{\text{renormalization scale:}}~ \mu_s(t)=s /(2 e^{\gamma_E} t)^{1/2}~{\text{with}~} s=1.
\end{align}
We estimate systematic errors by varying the above conditions.
In the systematic error analysis of the range, we extend the
range to smaller $t$ region and use the data at $0.005 \leq t T^2 \leq 0.015$.
For the scale setting parameters, we neglect the small error of $w_0 T_c$ estimated in Ref.~\cite{Kitazawa:2016dsl}, 
but we consider the error of $\Lambda_{\overline{\rm MS}}$ of $2.7 \%$ \cite{Aoki:2019cca}.
We also vary the renormalization scale within $s=1/\sqrt{2}$ to $2$.

\begin{figure}[tbhp]
\begin{minipage}{0.5\hsize}
\begin{center}
\vspace{1mm}
\includegraphics[width=7.25cm]{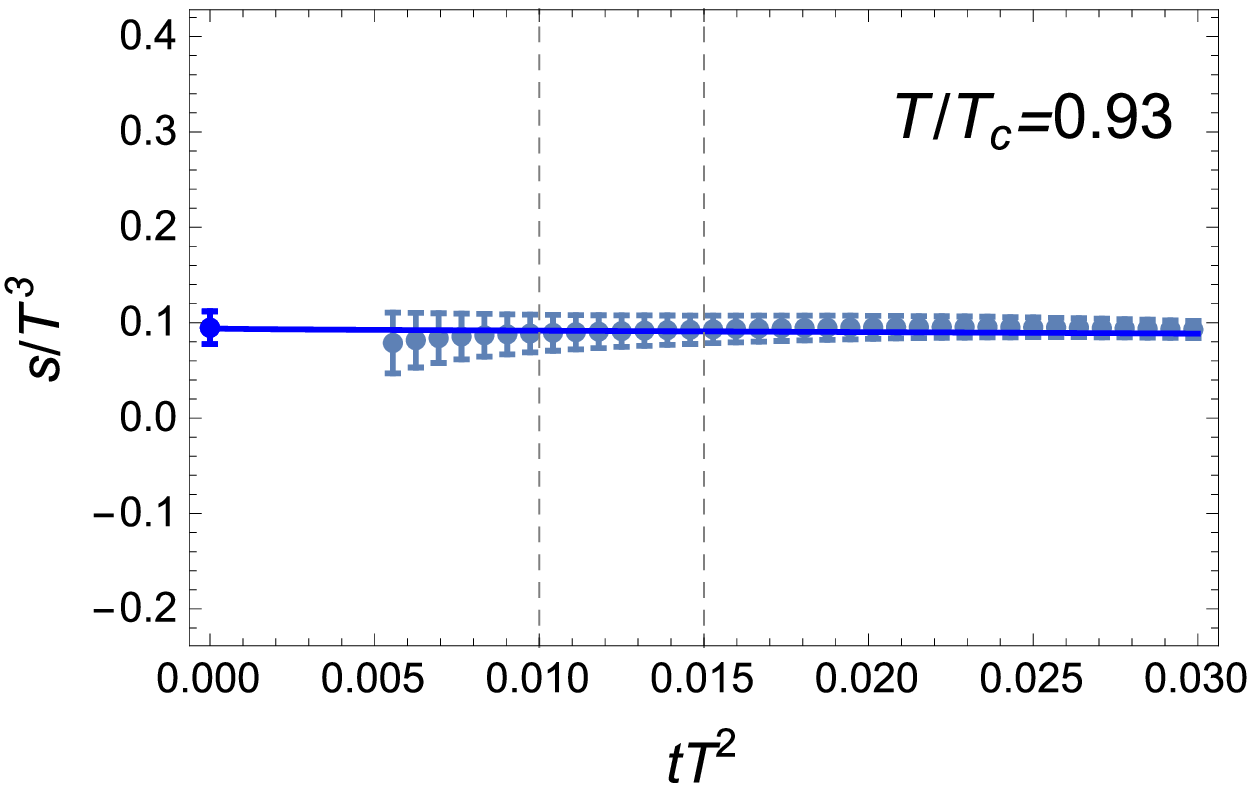}
\end{center}
\end{minipage}
\begin{minipage}{0.5\hsize}
\begin{center}
\includegraphics[width=7cm]{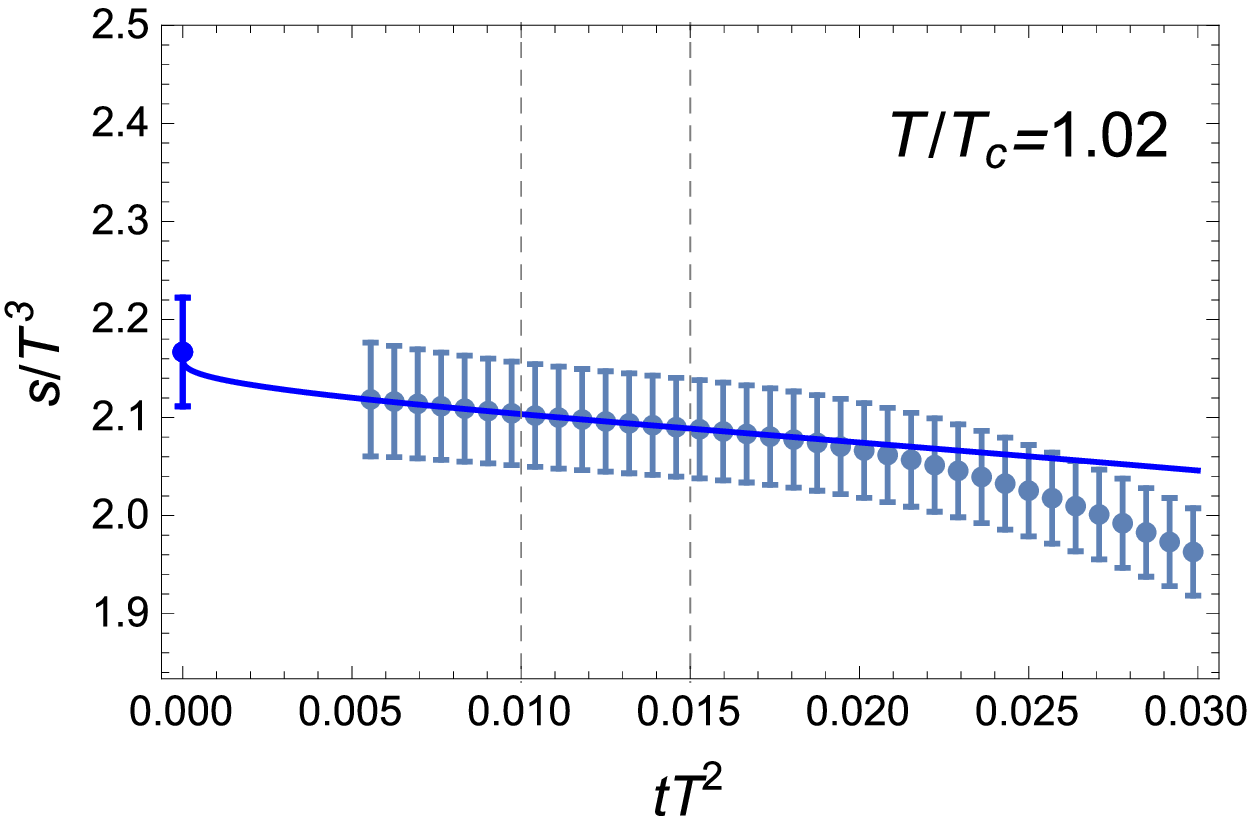}
\end{center}
\end{minipage}
\begin{minipage}{0.5\hsize}
\begin{center}
\vspace{1mm}
\includegraphics[width=7.05cm]{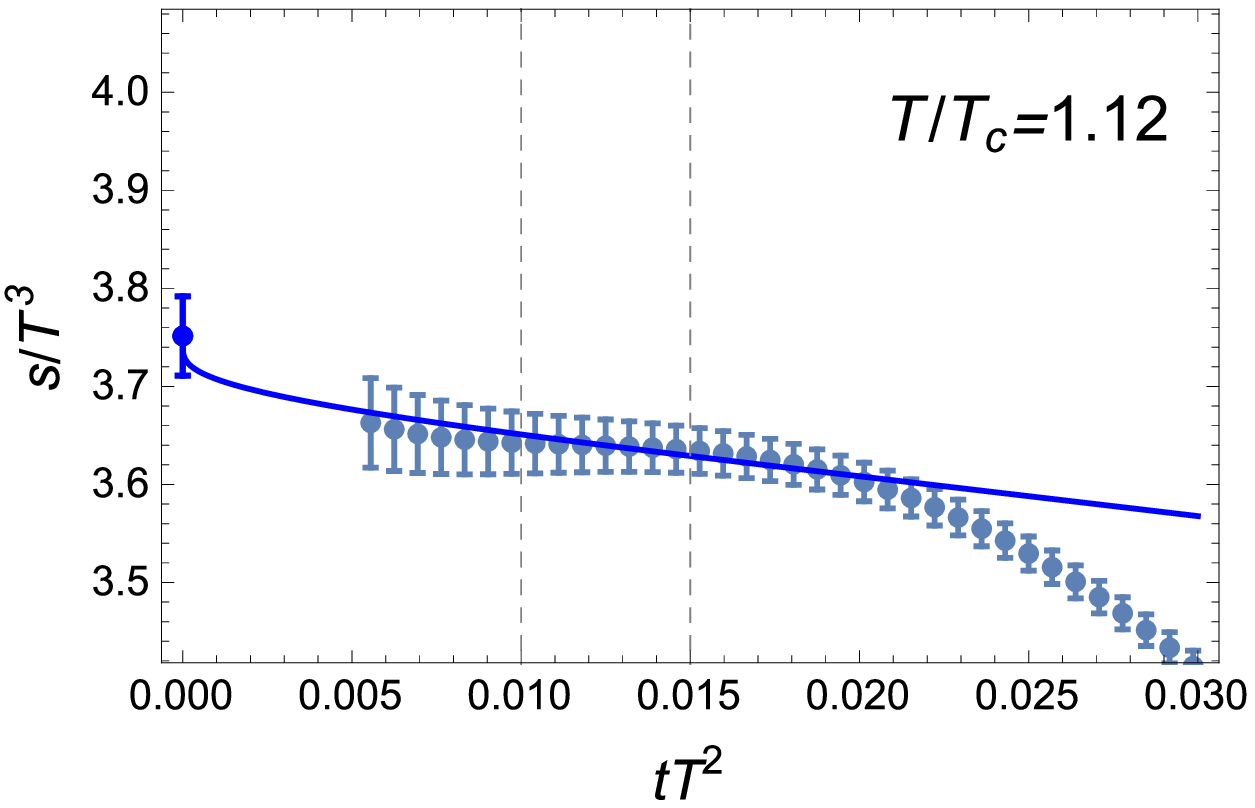}
\end{center}
\end{minipage}
\begin{minipage}{0.5\hsize}
\begin{center}
\includegraphics[width=7cm]{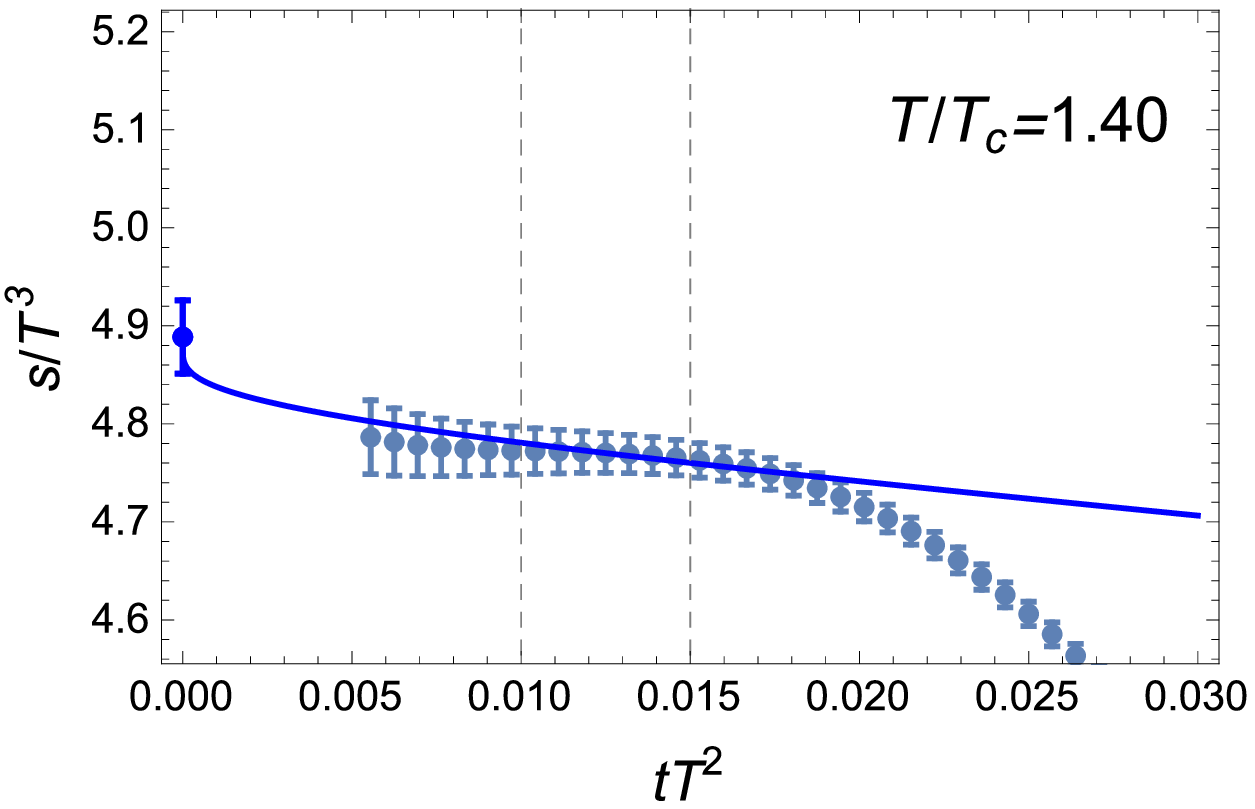}
\end{center}
\end{minipage}
\begin{minipage}{0.5\hsize}
\begin{center}
\vspace{1mm}
\includegraphics[width=7.05cm]{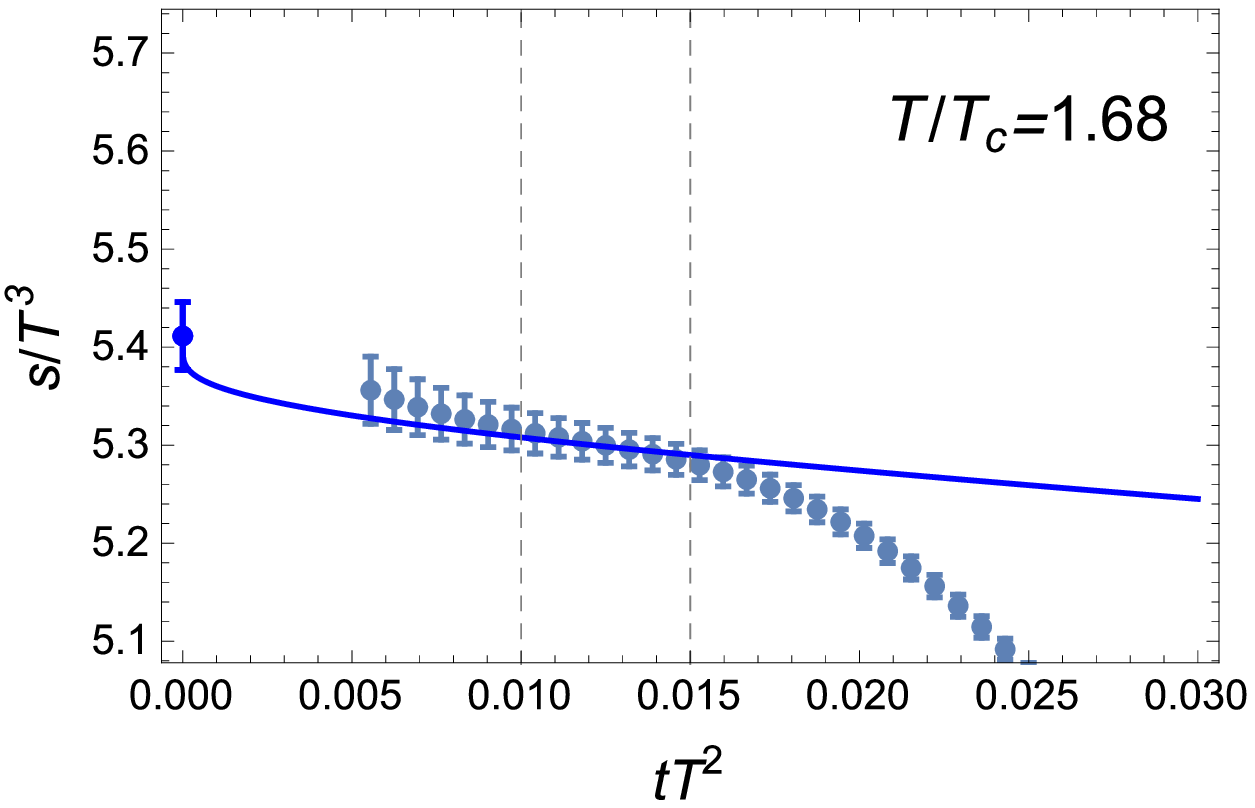}
\end{center}
\end{minipage}
\begin{minipage}{0.5\hsize}
\begin{center}
\includegraphics[width=7cm]{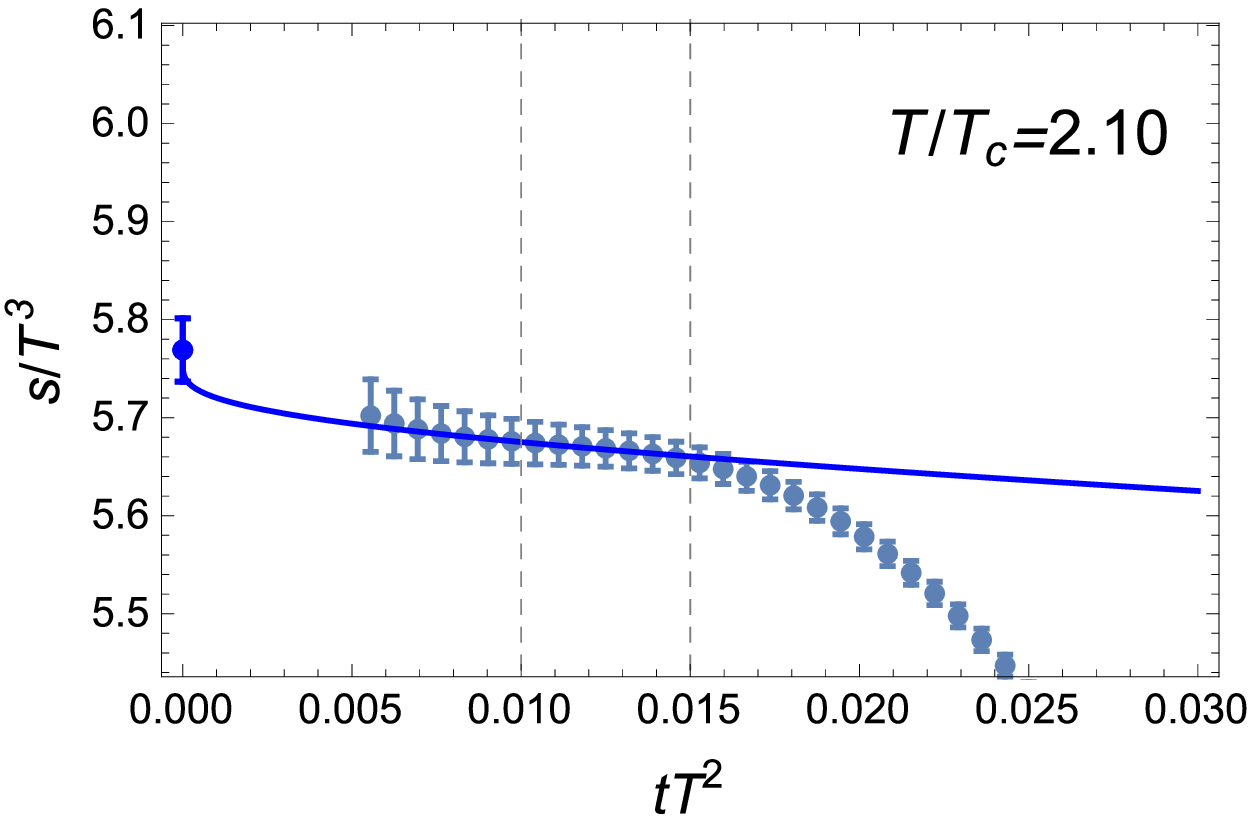}
\end{center}
\end{minipage}
\begin{minipage}{0.5\hsize}
\begin{center}
\vspace{1mm}
\includegraphics[width=7.05cm]{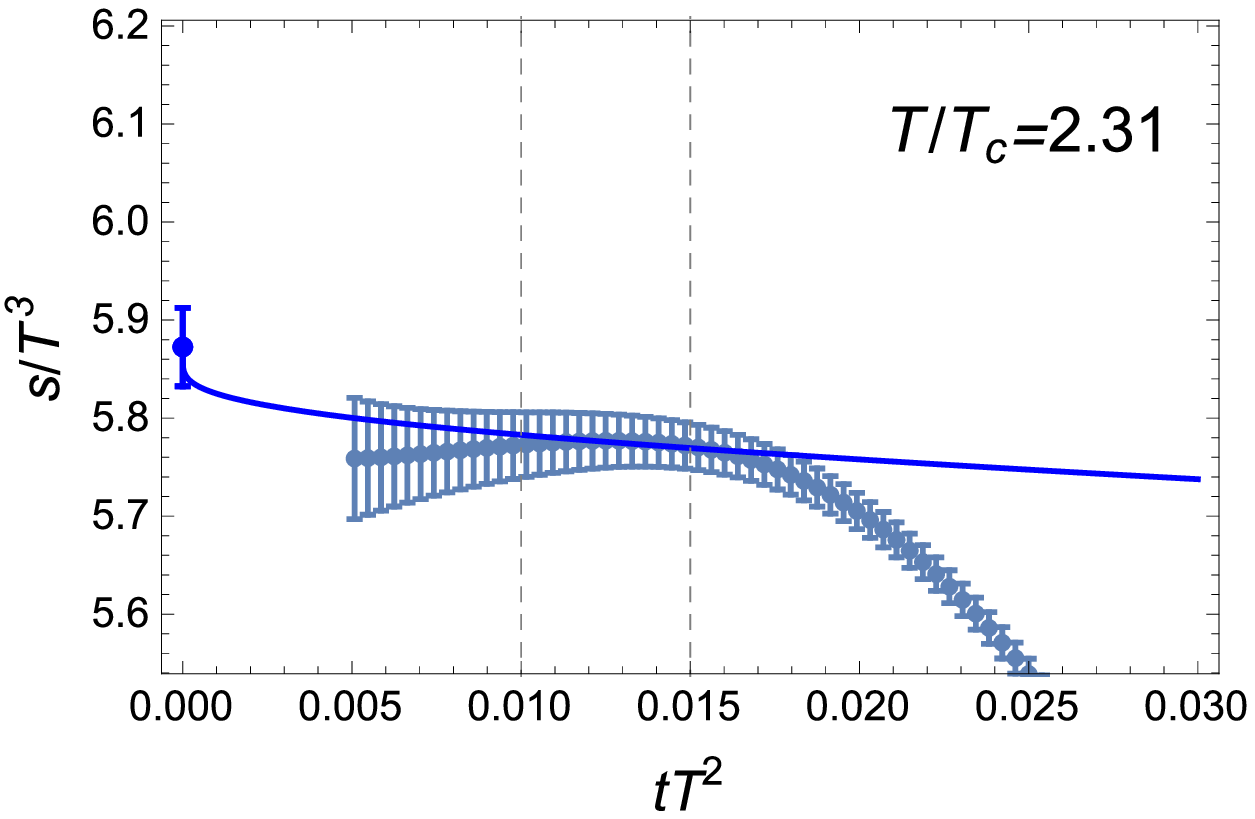}
\end{center}
\end{minipage}
\begin{minipage}{0.5\hsize}
\begin{center}
\includegraphics[width=7cm]{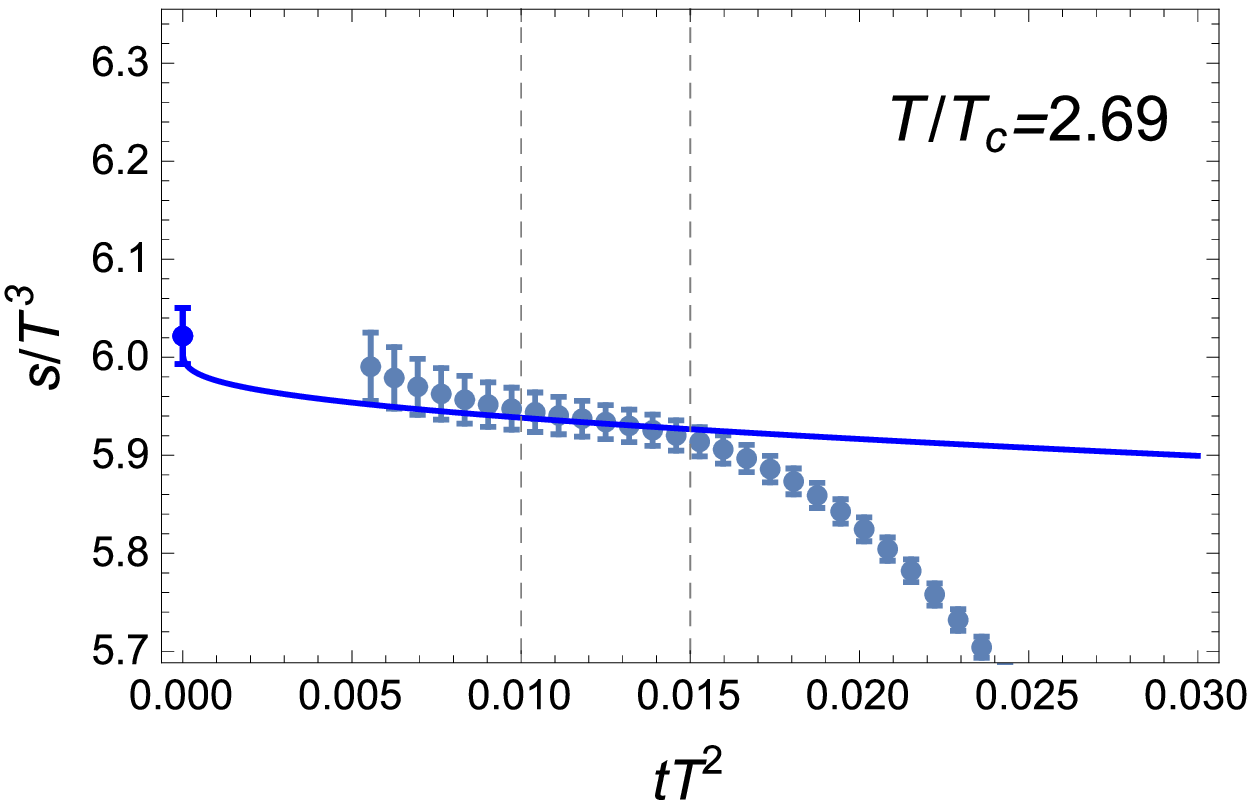}
\end{center}
\end{minipage}
\caption{NLO analysis of the entropy density. 
The blue solid lines are the $t \to 0$ extrapolation function of Eq.~\eqref{resquenchTL} with $k=1$.
The error bars show statistic errors only.
 (The same applies to Figs.~\ref{fig:entropyhighercoeff}-\ref{fig:higherordercoeffTA}.)
The gray dashed lines show the fit range used in the $t \to 0$ extrapolation.
In each figure, the simulation temperature is shown.}
\label{fig:NLOentropy}
\end{figure}

Now we present our numerical results.
In Fig.~\ref{fig:NLOentropy}, we show the $t \to 0$ extrapolation 
of the lattice data. This is the NLO analysis of the entropy density. 
Then we use the extrapolation function of Eq.~\eqref{resquenchTL} with $k=1$.
We regard the fit parameter $k_1^{(2)}$ as independent of temperature
and it is taken to be common to all the simulation temperatures.
A characteristic of the extrapolation function \eqref{resquenchTL}
is that it rises sharply around $t \sim 0$.
This stems from the singularity of $g^2(\mu(t)) \sim \frac{1}{\log{(1/(t \LMS^2))}}$ at $t=0$.
The extrapolation function is consistent with the (non-trivial) behavior of the lattice data in the small $t$ region, 
$t T^2=0.005$--$0.015$, in Fig.~\ref{fig:NLOentropy}, although we use the data at $0.010$--$0.015$ in the fit. 
This indicates the validity of the extrapolation function.

We can also confirm the validity of the extrapolation function by
comparing the result of the fit parameter $k_1^{(2)}$,
which corresponds to the NNLO coefficient for $c_1(t)$, with 
its already known result.
From the fit, we obtain
\be
k_1^{(2)}=88 (18) , \label{k12fitres}
\ee
where the value inside the parentheses denotes the statistical error,
and the exact NNLO result is $k_1^{(2)}=87.23...$ \cite{Harlander:2018zpi}.
Our result agrees well with the exact result.

Even in the case where the higher order perturbative coefficient $k_1^{(k+1)}$ is not known,
the following analysis is possible to check the validity of the use of the extrapolation function \eqref{resquenchTL}.
We focus on the property that the $L(\mu,t)=\log(2 \mu^2 e^{\gamma_E} t)$ dependence of
the N$^{k+1}$LO perturbative coefficient is totally determined by the perturbative 
coefficients up to N$^k$LO and the beta function; see App.~\ref{app:D}.
Then we can compare the $L(\mu,t)$ dependence
of the fit parameter $k_1^{(k+1)}$ with the predicted dependence on $L(\mu,t)$.

\begin{figure}
\begin{minipage}{0.5\hsize}
\begin{center}
\includegraphics[width=7.5cm]{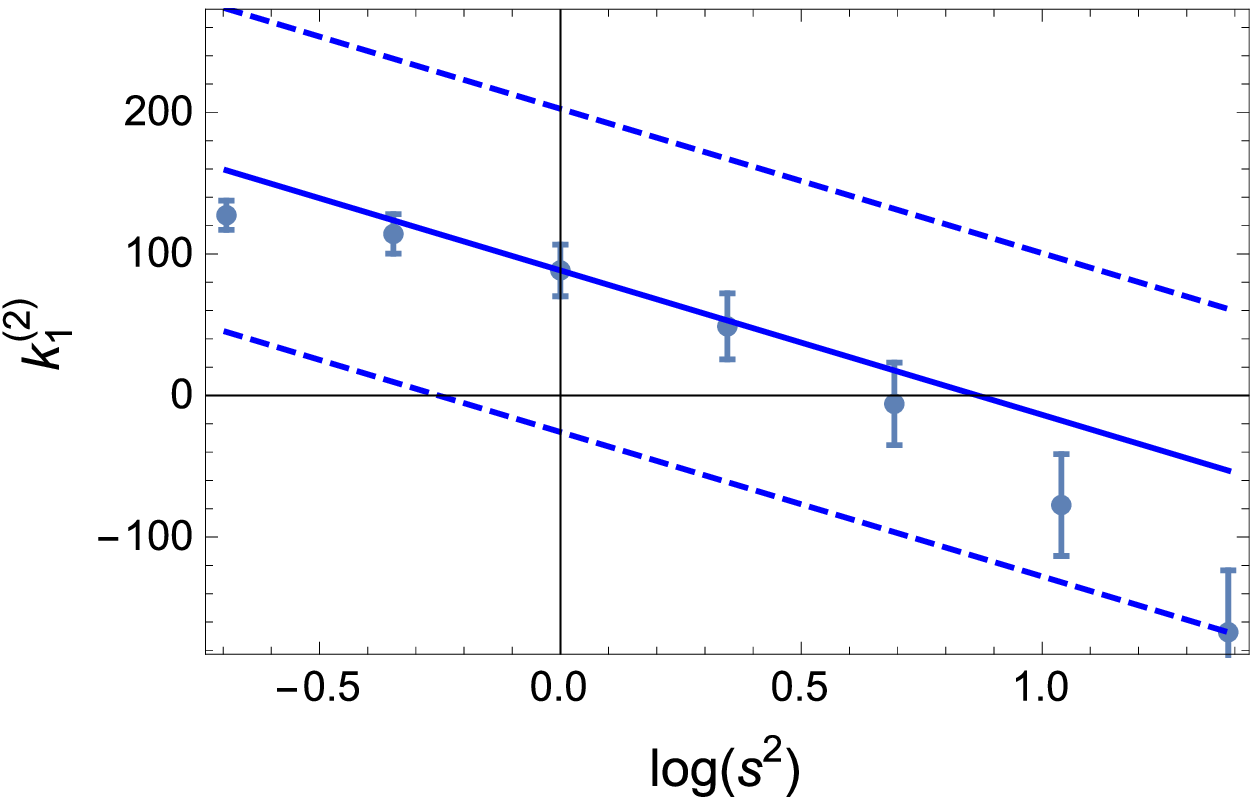}
\end{center}
\end{minipage}
\begin{minipage}{0.5\hsize}
\begin{center}
\includegraphics[width=7.5cm]{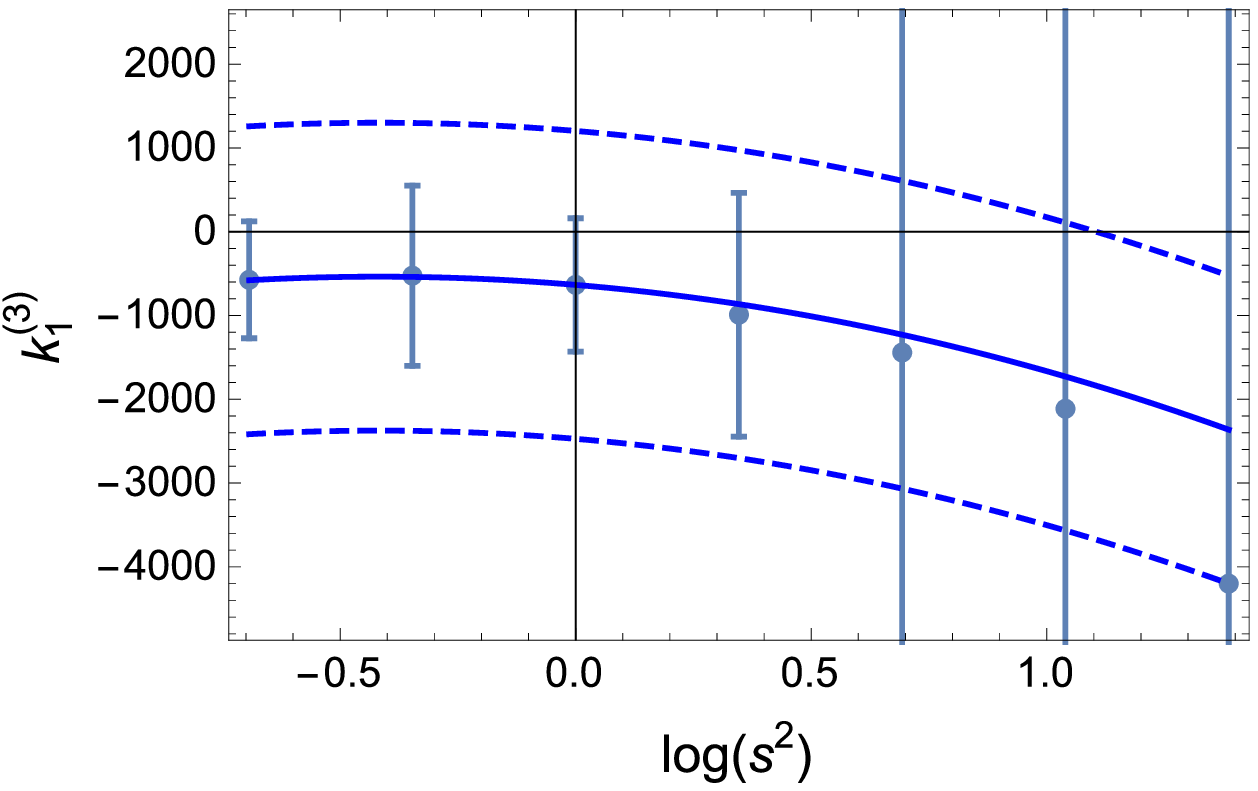}
\end{center}
\end{minipage}
\caption{The fit parameter $k_1^{(2)}$ or $k_1^{(3)}$ as a function of $\log{s^2}$. 
The data points with error bars show the results of the fit parameter
$k_1^{(2)}$ or $k_1^{(3)}$ obtained in the $t \to 0$ extrapolation analyses for various $s$. 
The blue lines show $\log{s^2}$ dependence of the perturbative coefficients $k_1^{(2)}$ and $k_1^{(3)}$
dictated from the RG equation; see Eq.~\eqref{eq:(D2)}.
We assume different $\log{s^2}$ independent constants for three blue lines;
see the main text.}
\label{fig:entropyhighercoeff}
\end{figure}

In the left panel of Fig.~\ref{fig:entropyhighercoeff}, we show the $\log{s^2}(=L(\mu_s(t),t))$ dependence of the fit parameter $k_1^{(2)}$.
It indeed exhibits similar dependence to the blue lines,
which show the exact $\log{s^2}$ dependence of $k_1^{(2)}$, predicted from the RG equation;
see the third equation of Eq.~\eqref{eq:(D2)}.
We assume different log independent constants $k_1^{(2)}(L=0)$ for the three blue lines.
For the solid line we set $k_1^{(2)}(L=0)$ in Eq.~\eqref{eq:(D2)} to $k_1^{(2), \rm{fit}}(L=0)$,
which is the fit parameter obtained in our central analysis with $s=1$ (or $\log{(s^2)}=0$)
[i.e. the central value of Eq.~\eqref{k12fitres}]. 
For the dashed line below, we set $k_1^{(2)}(L=0)$ in Eq.~\eqref{eq:(D2)}
such that $k_1^{(2)}(L=\log{(2^2)})$ coincides with $k_1^{(2), \rm{fit}}(L=\log{(2^2)})$.
Here $k_1^{(2), \rm{fit}}(L=\log{(2^2)})$ is the fit parameter obtained when we set $s=2$.\fn{
The largest difference in the estimate for $k_1^{(2)}(L=0)$ is caused by the $s=1$ and $s=2$ cases.
This is why we focused on $s=2$ as the error estimate.} 
The difference in $k_1^{(2)}(L=0)$ between the two analyses 
can be regraded as an error of the estimate of $k_1^{(2)}(L=0)$.
For the dashed line above, the same size variation of $k_1^{(2)}(L=0)$
is assumed with the reversed sign.

We now estimate systematic errors. 
We vary one of the following conditions from the central analysis:
the fit range, $\LMS$, and the renormalization scale.
We have already explained how we change these conditions above.
We examine the variations from the central values caused by changing a condition,
and the (largest) difference is given as a systematic error. 
Our final result for the NLO analysis of the entropy density is summarized in Table~\ref{tab:1}.

\begin{figure}[tbhp]
\begin{minipage}{0.5\hsize}
\begin{center}
\vspace{1mm}
\includegraphics[width=7.25cm]{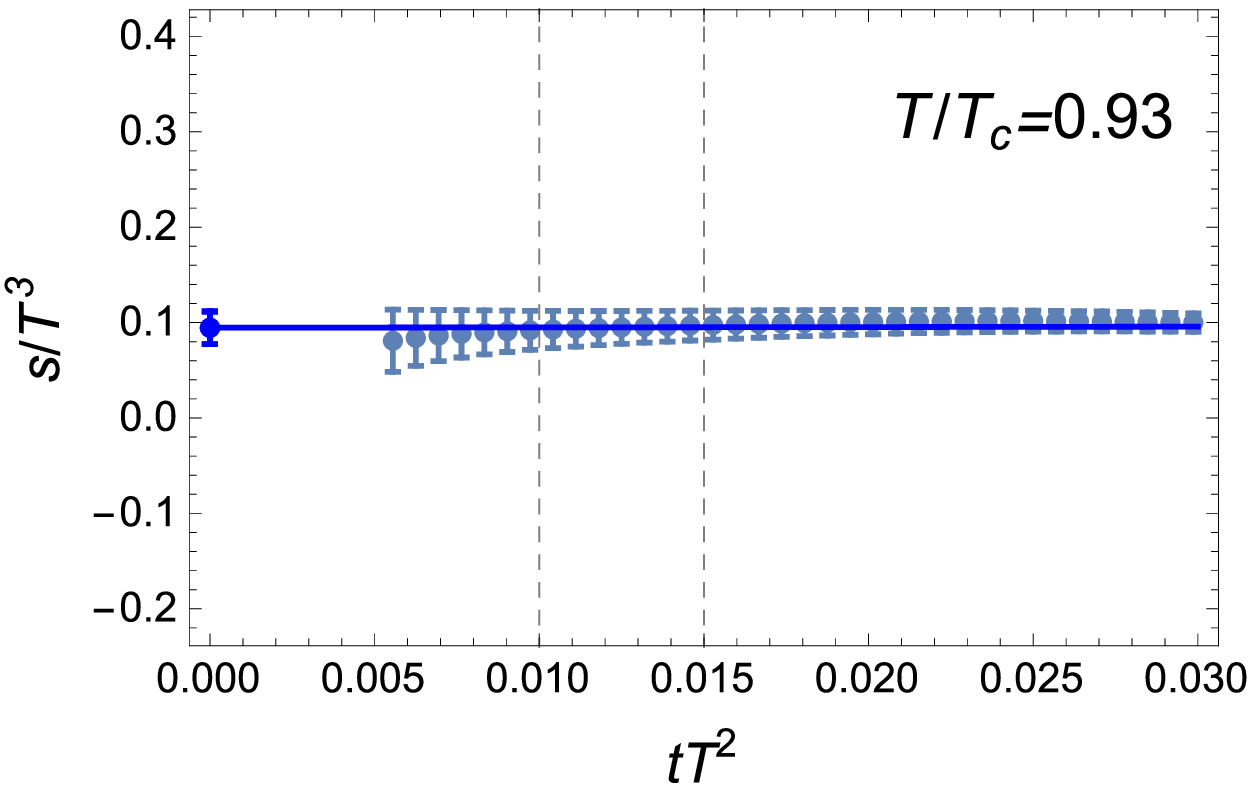}
\end{center}
\end{minipage}
\begin{minipage}{0.5\hsize}
\begin{center}
\includegraphics[width=7cm]{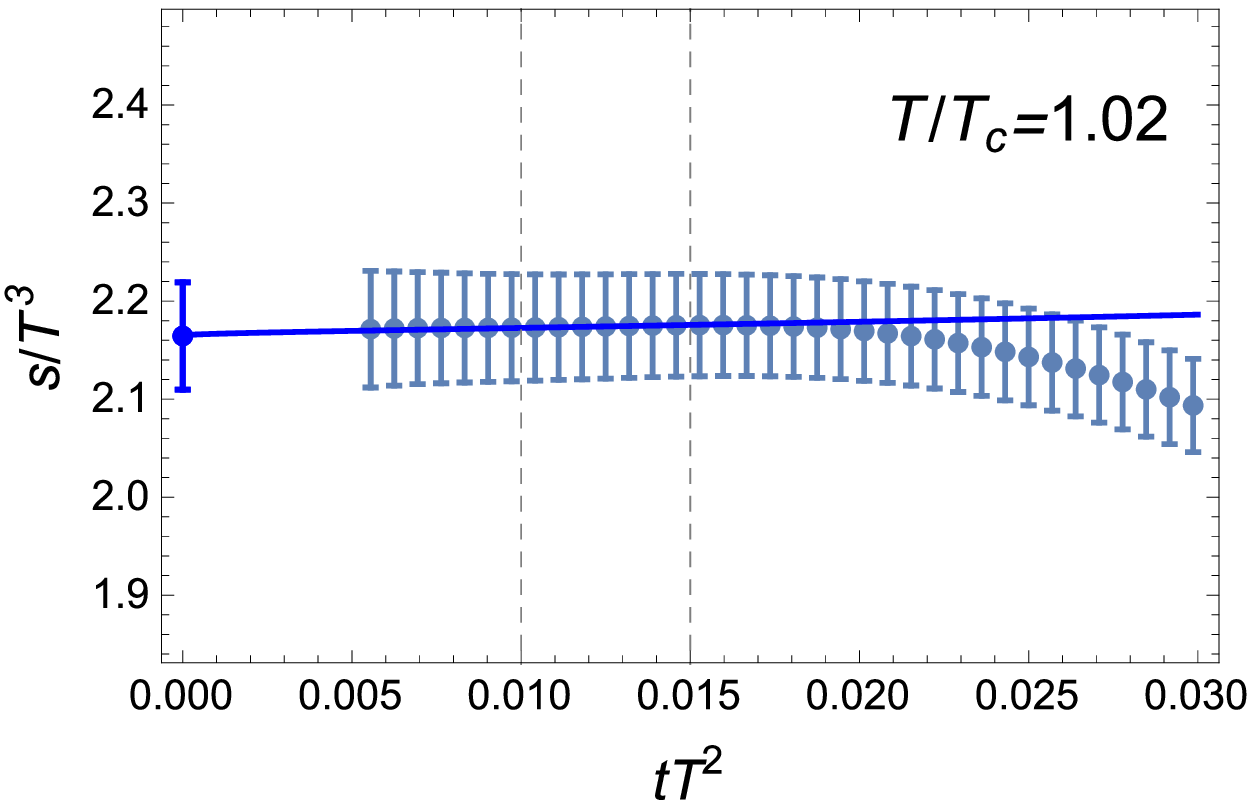}
\end{center}
\end{minipage}
\begin{minipage}{0.5\hsize}
\begin{center}
\vspace{1mm}
\includegraphics[width=7.05cm]{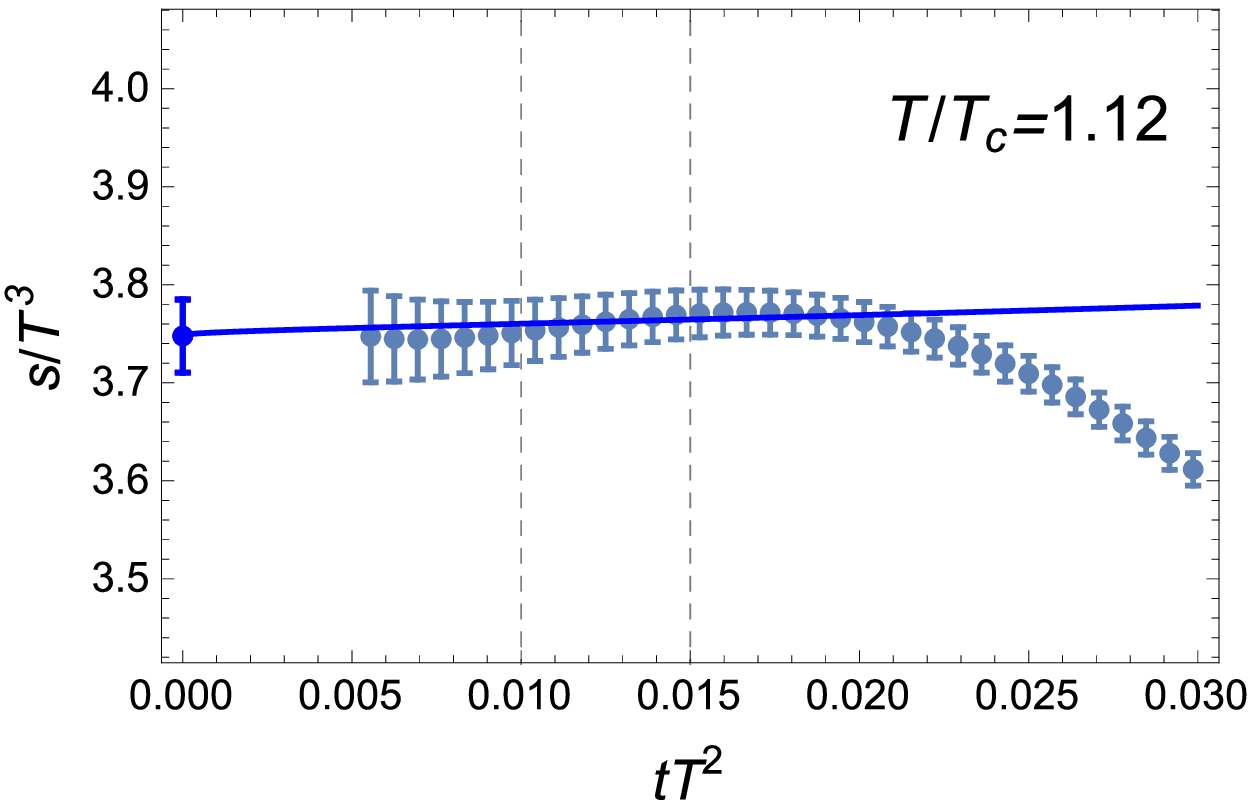}
\end{center}
\end{minipage}
\begin{minipage}{0.5\hsize}
\begin{center}
\includegraphics[width=7cm]{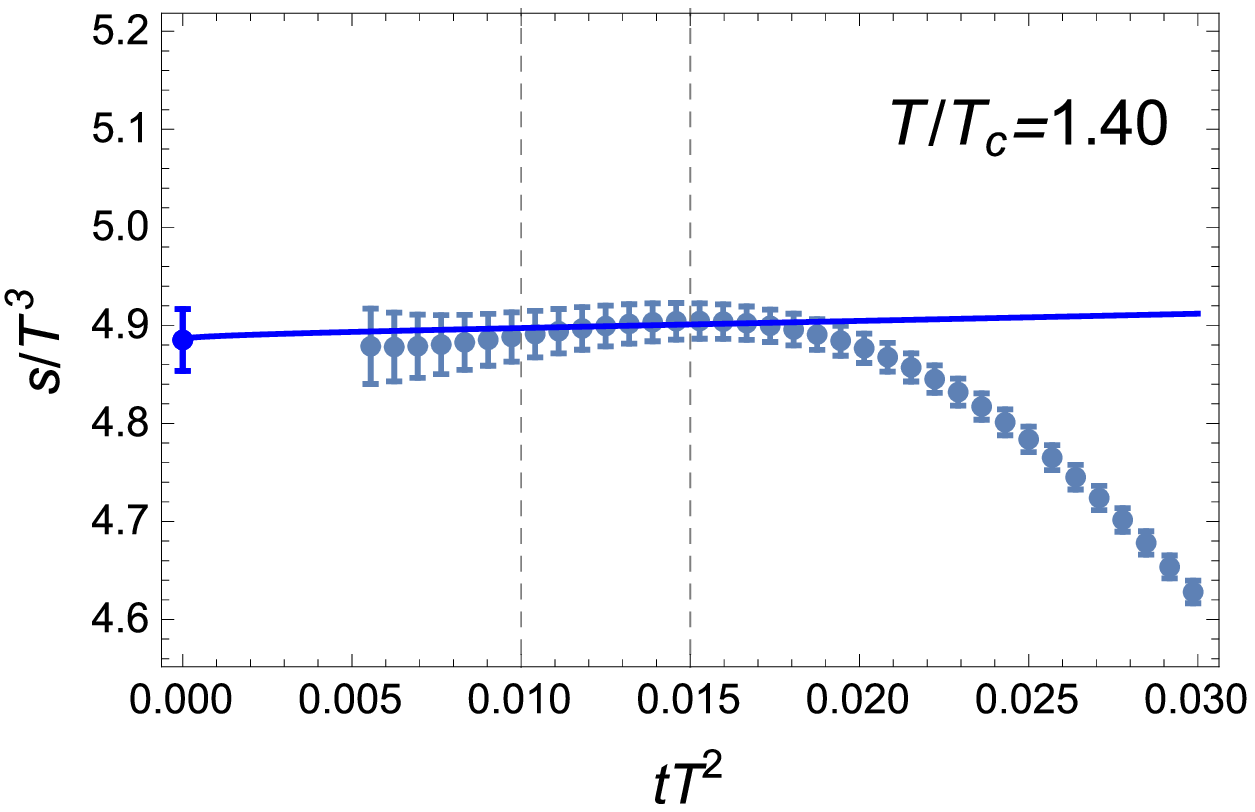}
\end{center}
\end{minipage}
\begin{minipage}{0.5\hsize}
\begin{center}
\vspace{1mm}
\includegraphics[width=7.05cm]{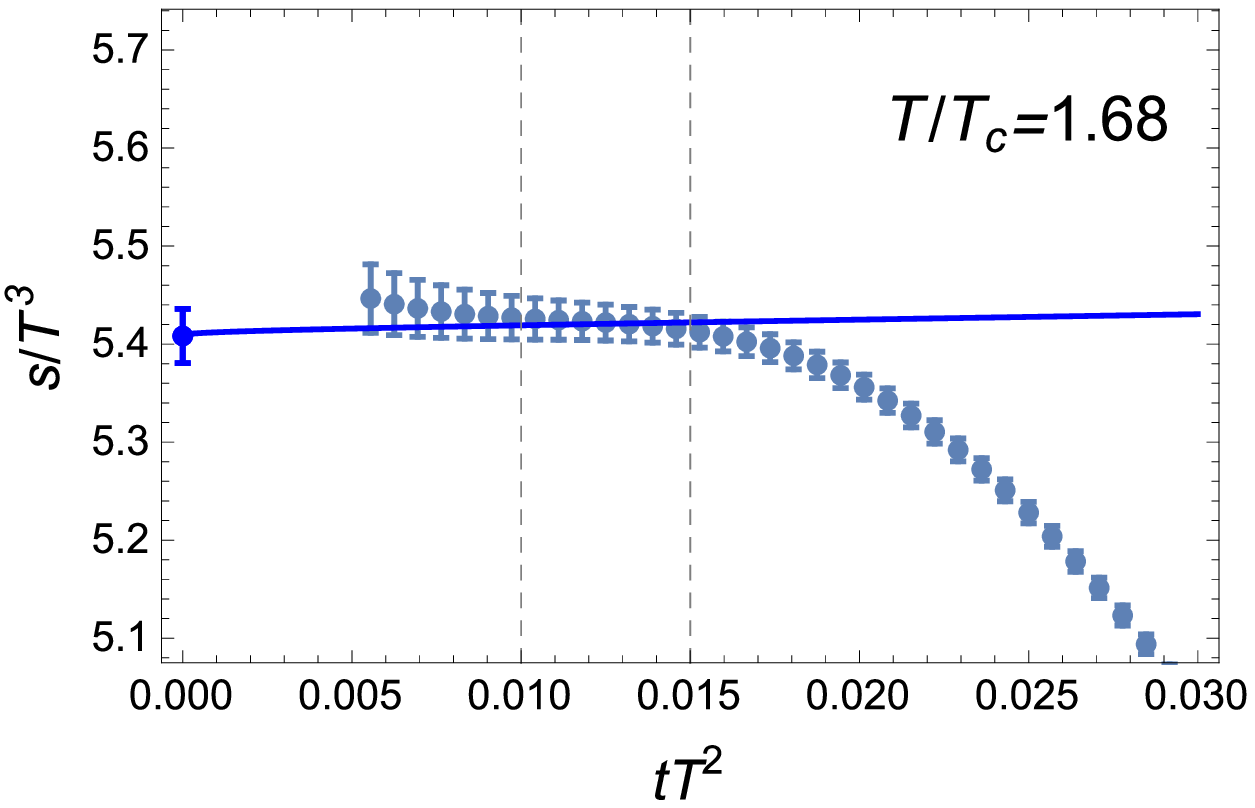}
\end{center}
\end{minipage}
\begin{minipage}{0.5\hsize}
\begin{center}
\includegraphics[width=7cm]{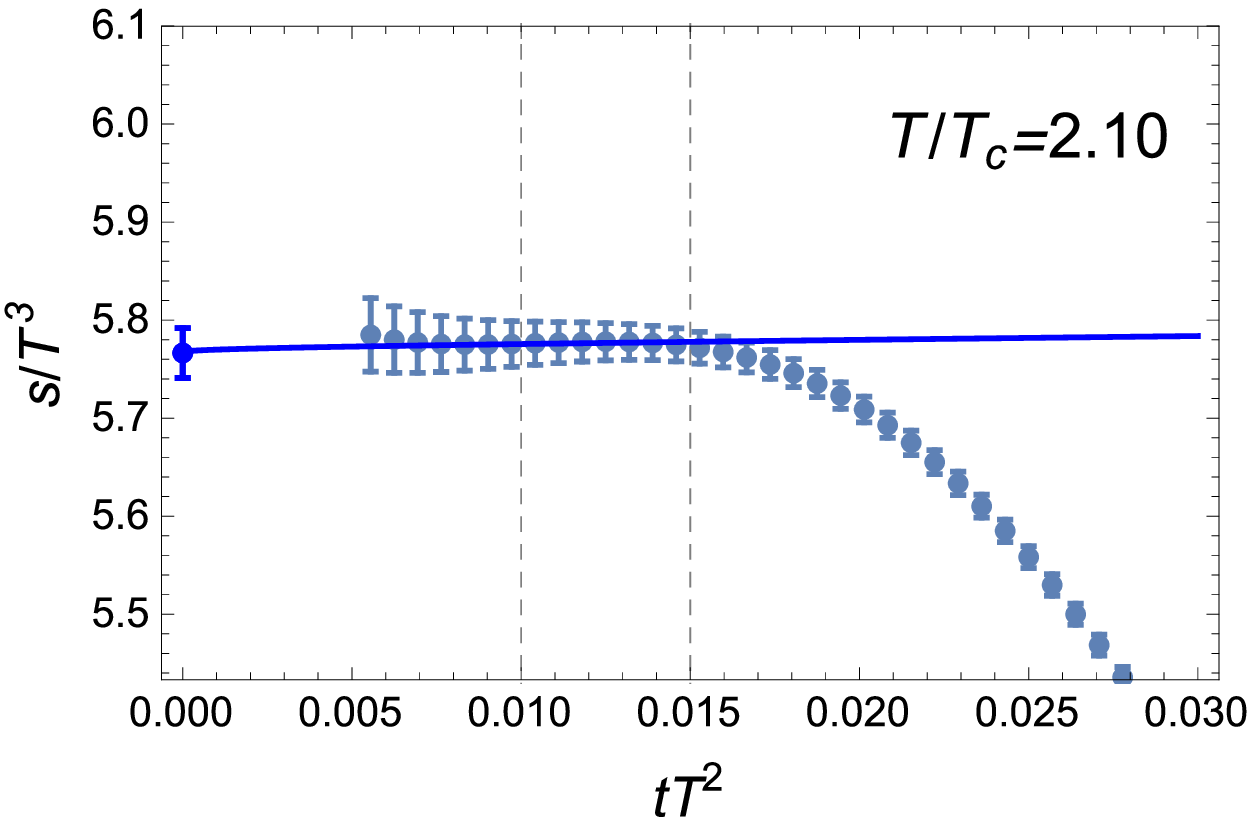}
\end{center}
\end{minipage}
\begin{minipage}{0.5\hsize}
\begin{center}
\vspace{1mm}
\includegraphics[width=7.05cm]{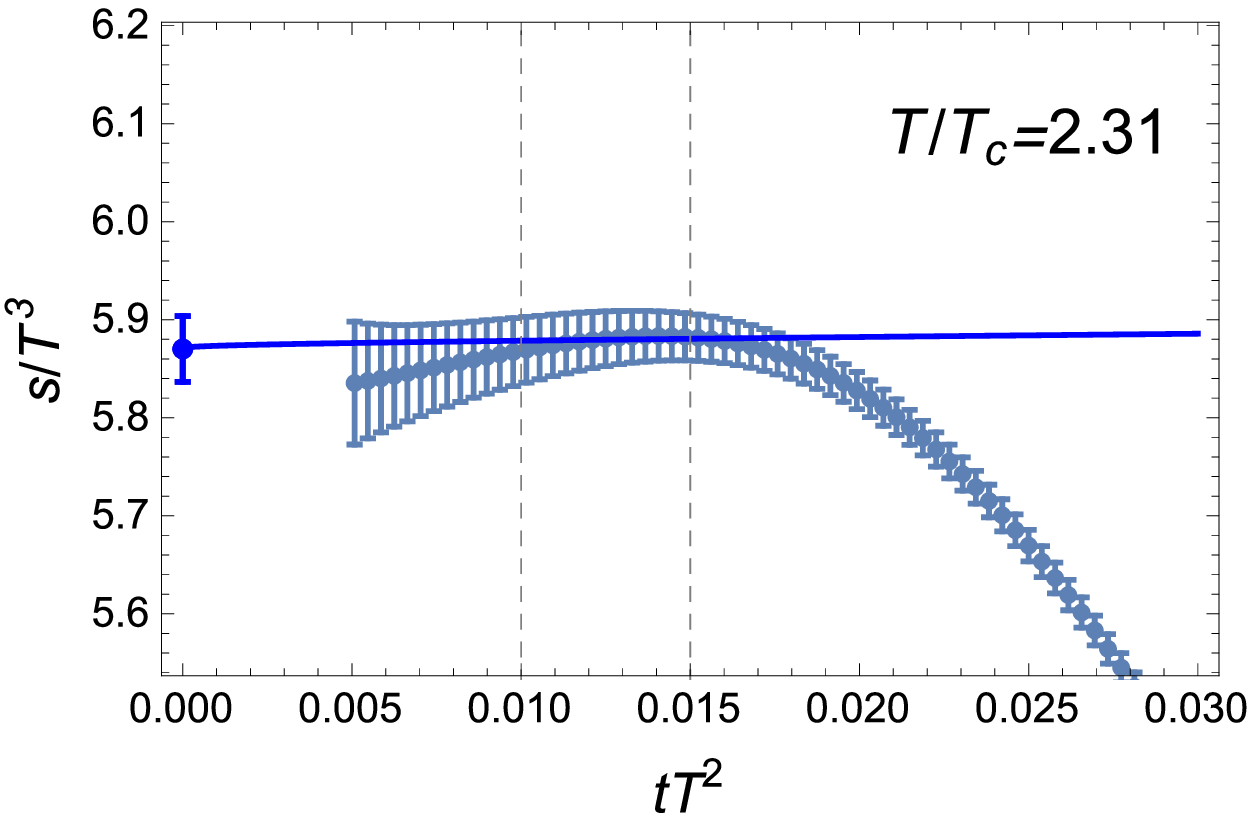}
\end{center}
\end{minipage}
\begin{minipage}{0.5\hsize}
\begin{center}
\includegraphics[width=7cm]{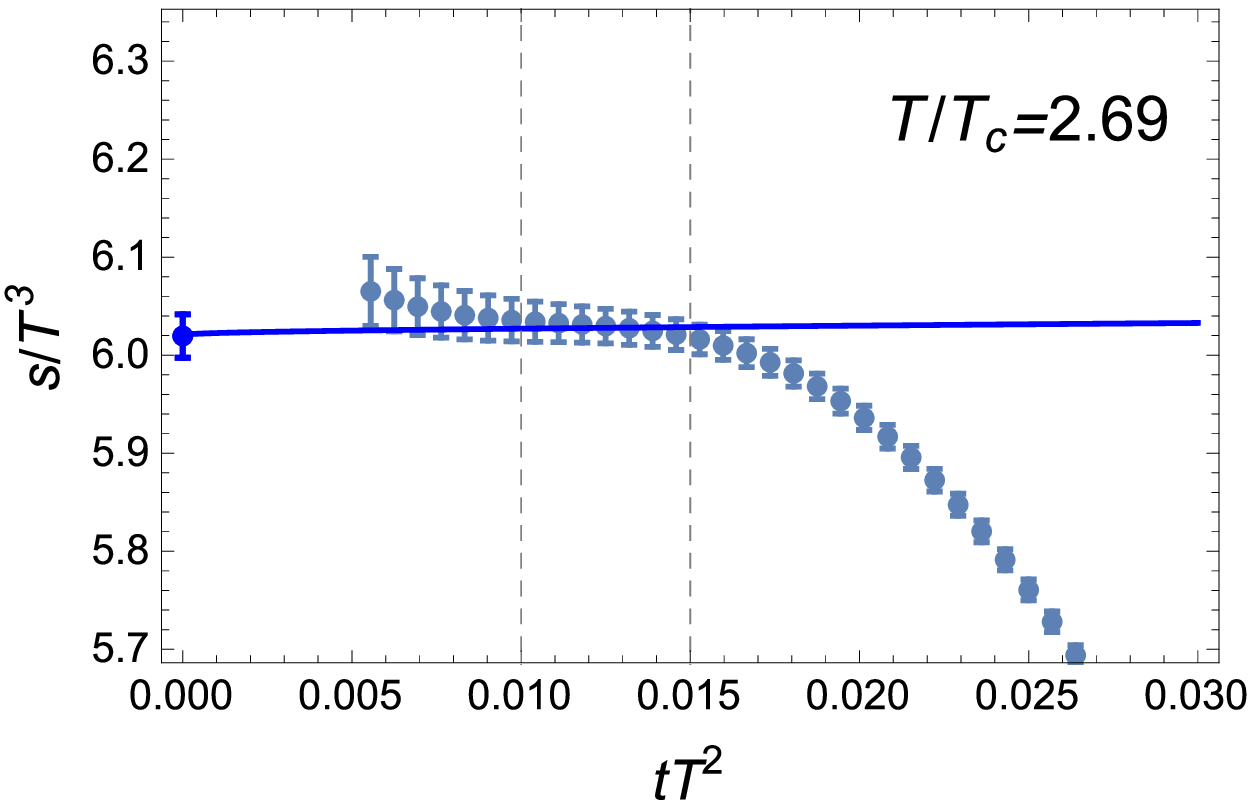}
\end{center}
\end{minipage}
\caption{NNLO analysis of the entropy density. 
The blue solid lines are the $t \to 0$ extrapolation function of Eq.~\eqref{resquenchTL} with $k=2$.
The gray dashed lines show the fit range used in the $t \to 0$ extrapolation.}
\label{fig:NNLOentropy}
\end{figure}

We move on to the NNLO analysis of the entropy density.
In Fig.~\ref{fig:NNLOentropy}, one can see that the $t$ dependence of the data becomes milder at NNLO \cite{Iritani:2018idk}.
This can be naturally understood from our theoretical study in Sect.~\ref{sec:3};
the remaining $t$ dependence of the NNLO formula is given by $\mathcal{O}(g(\mu(t))^6)$ at NNLO, 
but $\mathcal{O}(g(\mu(t))^4)$ at NLO.
Actually, the data can be regarded as flat within the statistical errors.
Then final results are insensitive to extrapolation functions.

We again check the validity of the extrapolation function of Eq.~\eqref{resquenchTL} (with $k=2$) 
in the right panel of Fig.~\ref{fig:entropyhighercoeff}.
As explained above, we examine the $\log{s^2}$ dependence of the fit parameter $k_1^{(3)}$.
Although the statistical errors are large, the behavior of the central values looks highly 
consistent with the expected behavior. (In drawing the blue lines in Fig.~\ref{fig:entropyhighercoeff},
we used the exact value of $k_1^{(2)}$.)
Hence, we consider the extrapolation function to also be valid in the NNLO analysis.

We summarize the NNLO result of the entropy density in Table~\ref{tab:1} with systematic errors.

We make some comments.
First, the NLO and NNLO analyses are mutually consistent.
It is worth noting that the variations of the central values are quite small.
Here, the extrapolation function \eqref{resquenchTL} plays an important role.
For a comparison, see App.~\ref{app:H}.
Secondly, at NNLO, the error associated with the renormalization scale, which 
is one of the dominant uncertainties at NLO, is considerably reduced. 
This shows that the SF$t$X method enables us to perform
systematic and accurate analyses of the EMT.
At NNLO, the dominant uncertainties come from the statistical error and the error in~$\LMS$.

\begin{figure}[tbhp]
\begin{minipage}{0.5\hsize}
\begin{center}
\vspace{1mm}
\includegraphics[width=7.25cm]{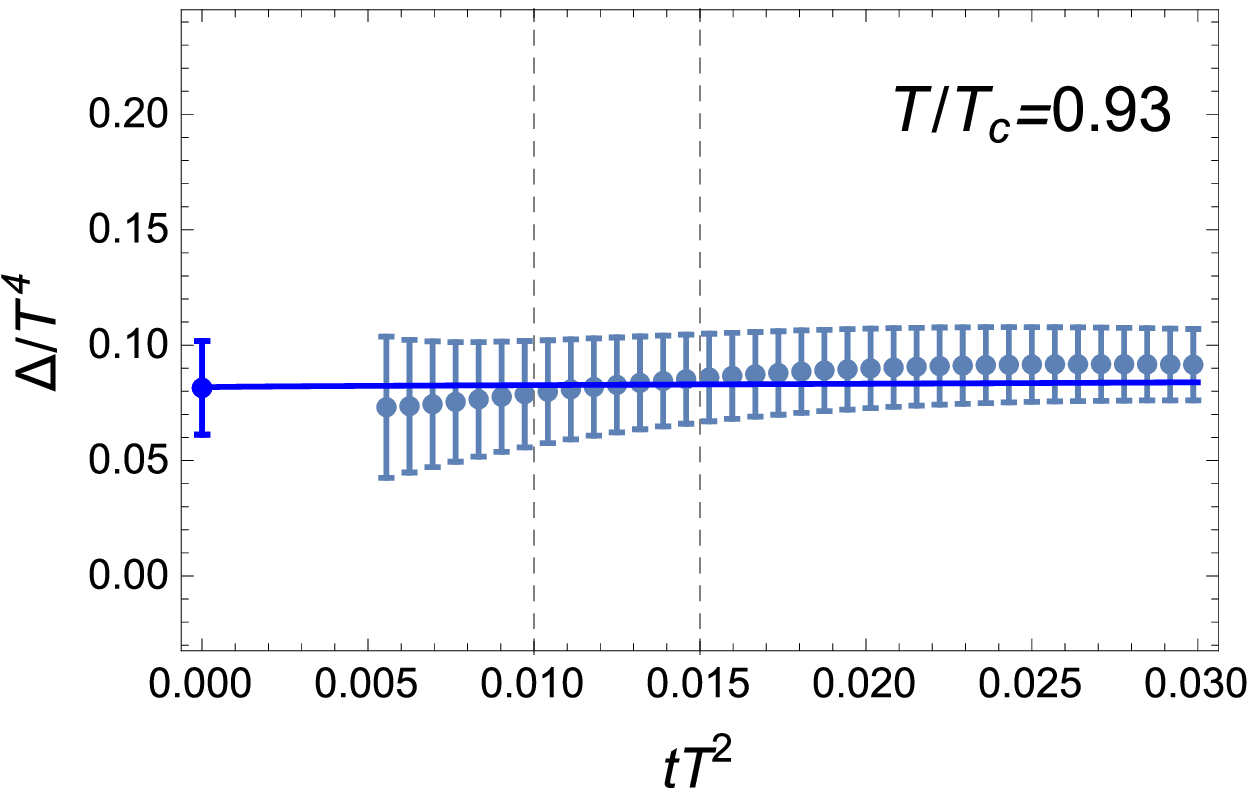}
\end{center}
\end{minipage}
\begin{minipage}{0.5\hsize}
\begin{center}
\includegraphics[width=7cm]{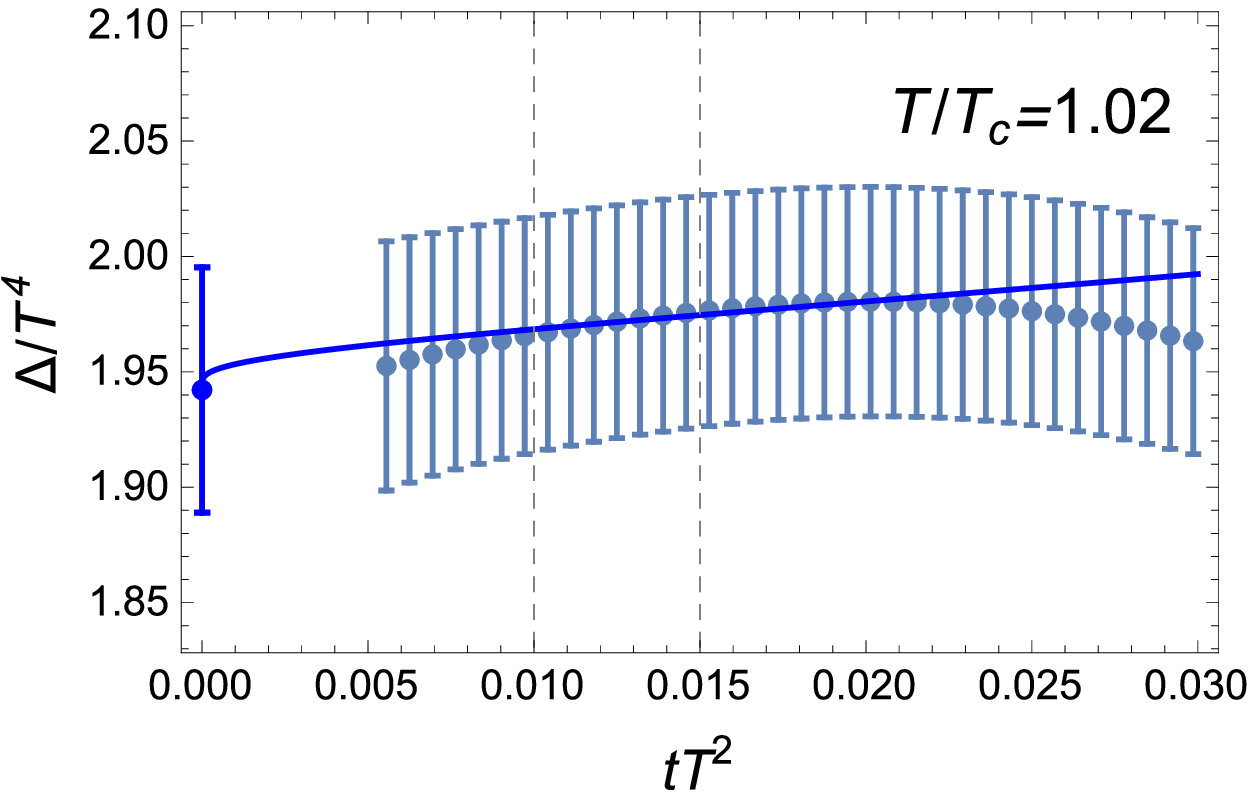}
\end{center}
\end{minipage}
\begin{minipage}{0.5\hsize}
\begin{center}
\vspace{1mm}
\includegraphics[width=7.05cm]{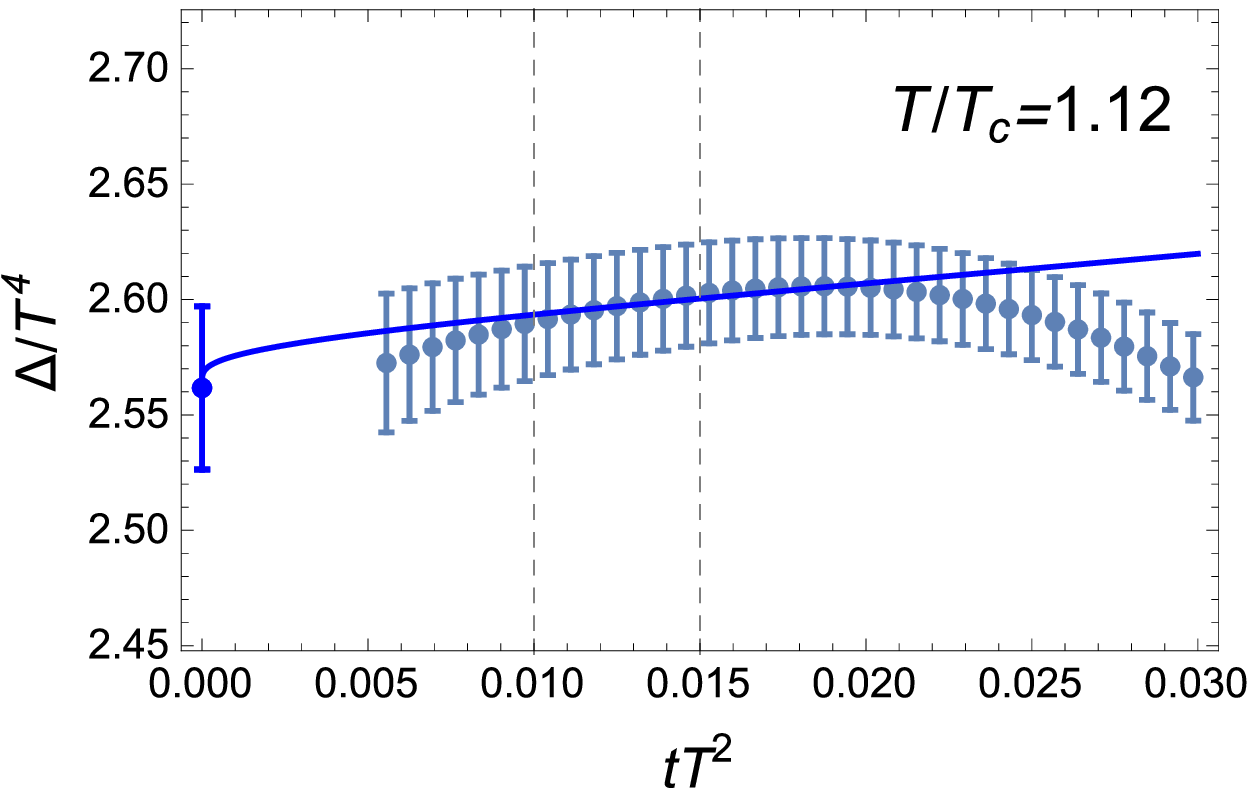}
\end{center}
\end{minipage}
\begin{minipage}{0.5\hsize}
\begin{center}
\includegraphics[width=7cm]{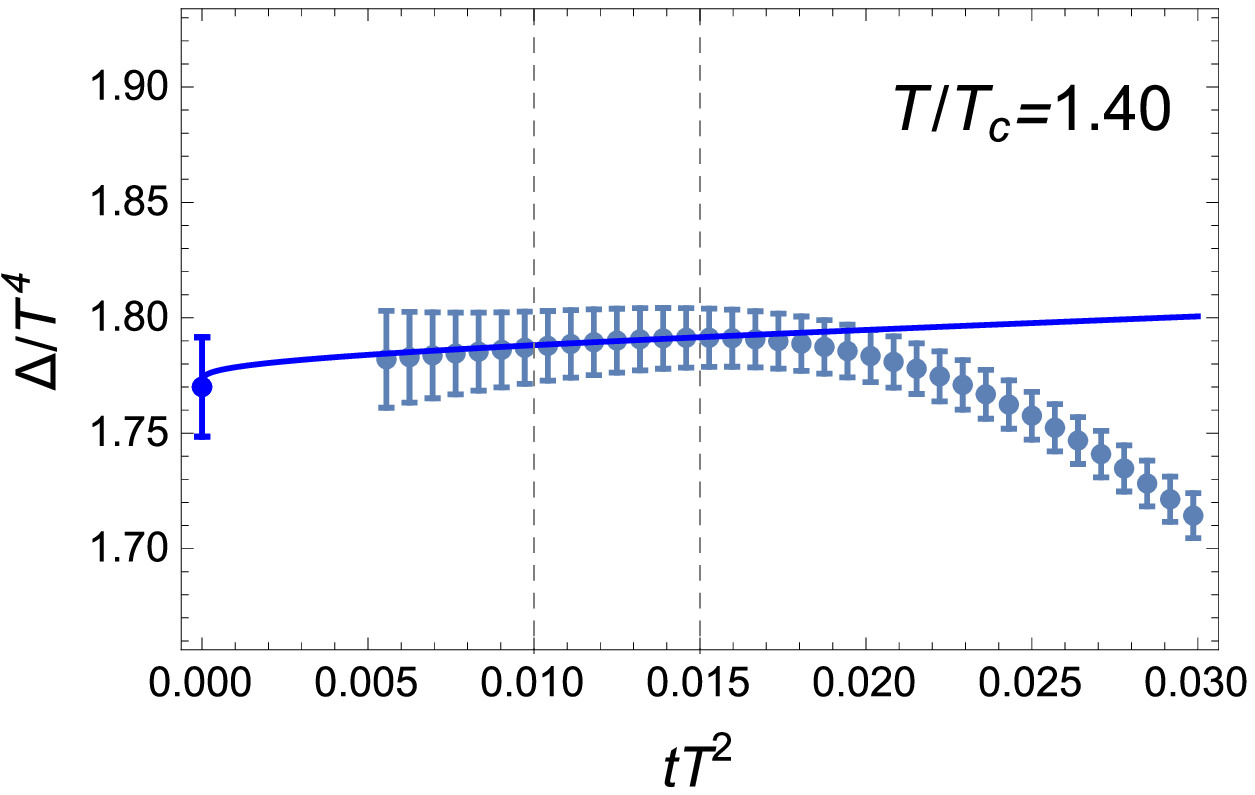}
\end{center}
\end{minipage}
\begin{minipage}{0.5\hsize}
\begin{center}
\vspace{1mm}
\includegraphics[width=7.05cm]{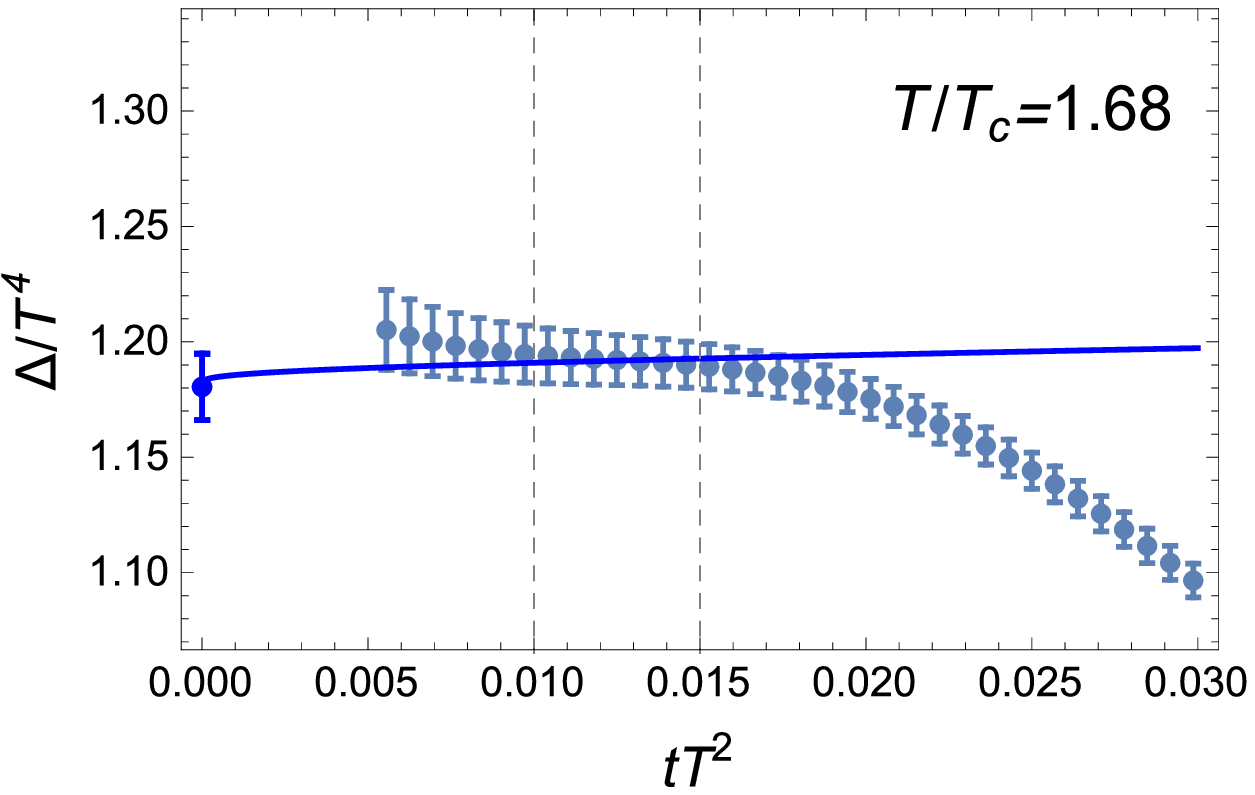}
\end{center}
\end{minipage}
\caption{NLO analysis of the trace anomaly. 
The blue solid lines are the $t \to 0$ extrapolation function of Eq.~\eqref{resquenchS} with $k=1$.
The gray dashed lines show the fit range used in the $t \to 0$ extrapolation.}
\label{fig:NLOTA}
\end{figure}

\begin{figure}[tbhp]
\begin{minipage}{0.5\hsize}
\begin{center}
\vspace{1mm}
\includegraphics[width=7.25cm]{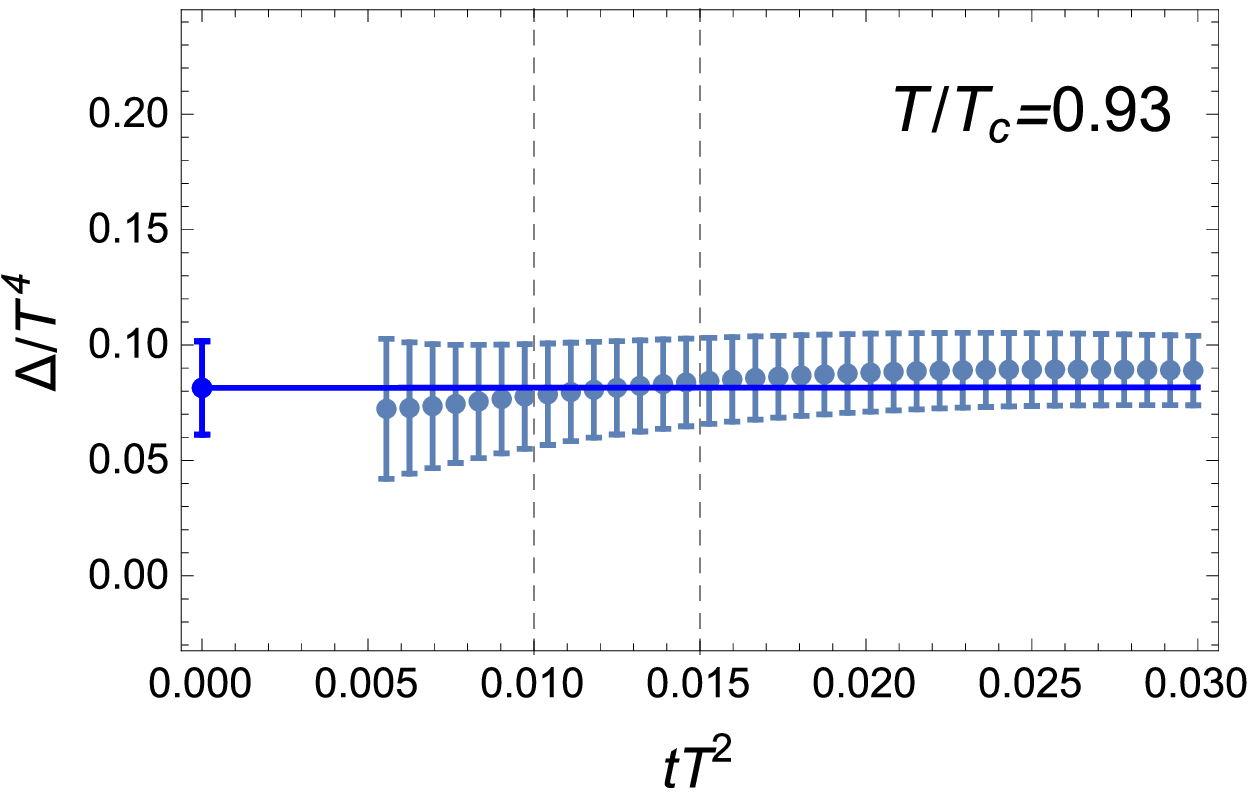}
\end{center}
\end{minipage}
\begin{minipage}{0.5\hsize}
\begin{center}
\includegraphics[width=7cm]{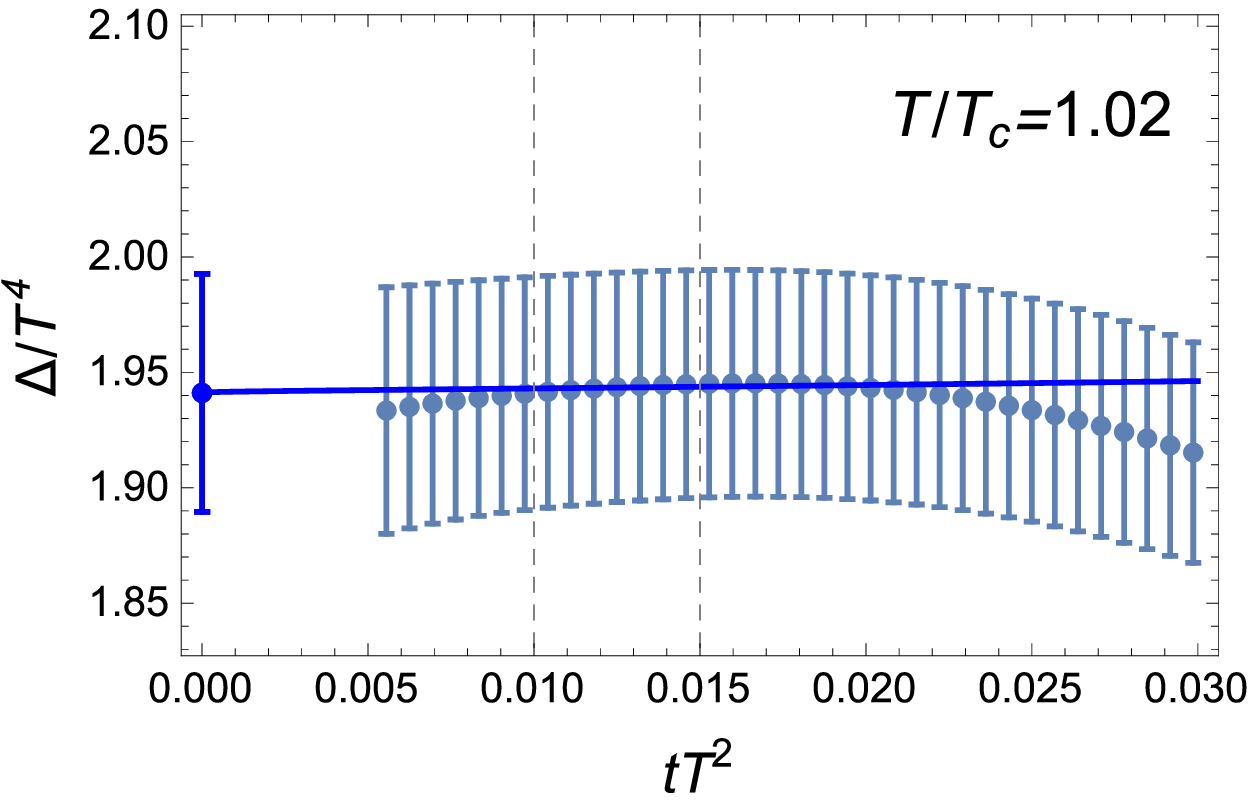}
\end{center}
\end{minipage}
\begin{minipage}{0.5\hsize}
\begin{center}
\vspace{1mm}
\includegraphics[width=7.05cm]{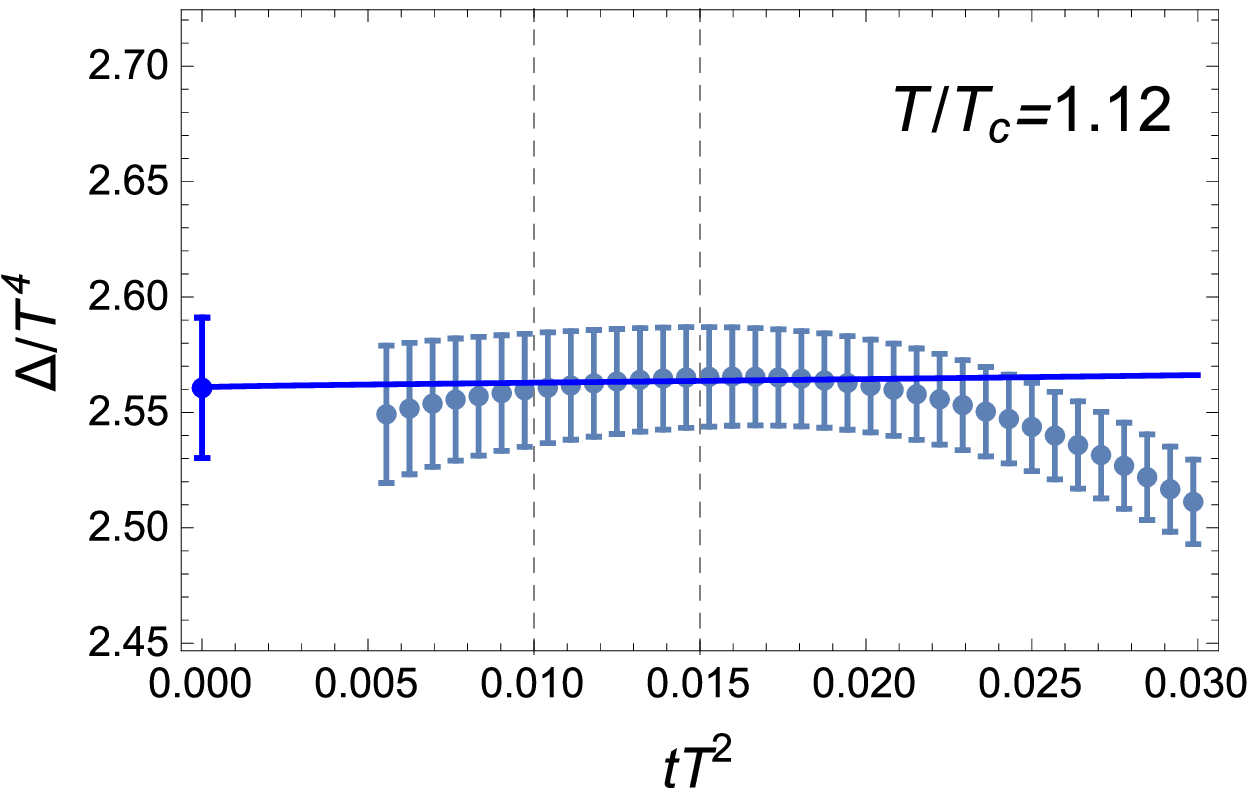}
\end{center}
\end{minipage}
\begin{minipage}{0.5\hsize}
\begin{center}
\includegraphics[width=7cm]{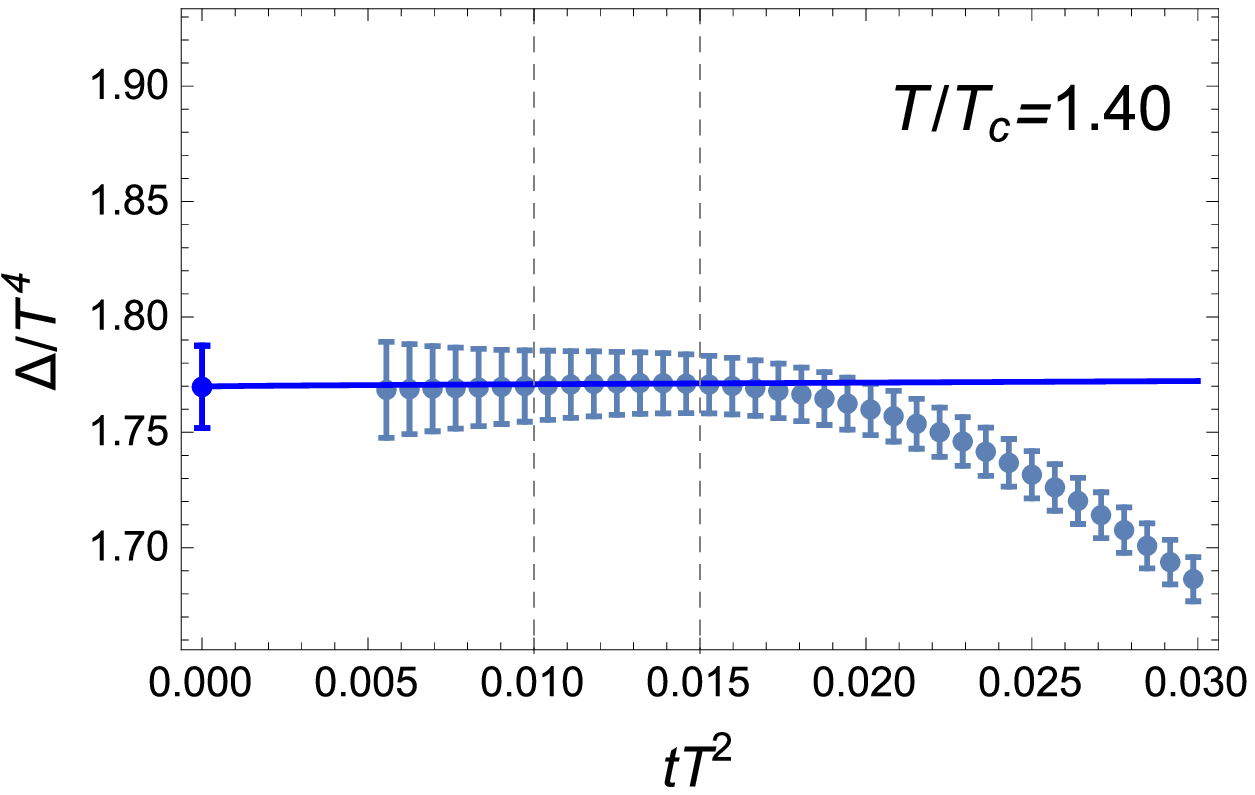}
\end{center}
\end{minipage}
\begin{minipage}{0.5\hsize}
\begin{center}
\vspace{1mm}
\includegraphics[width=7.05cm]{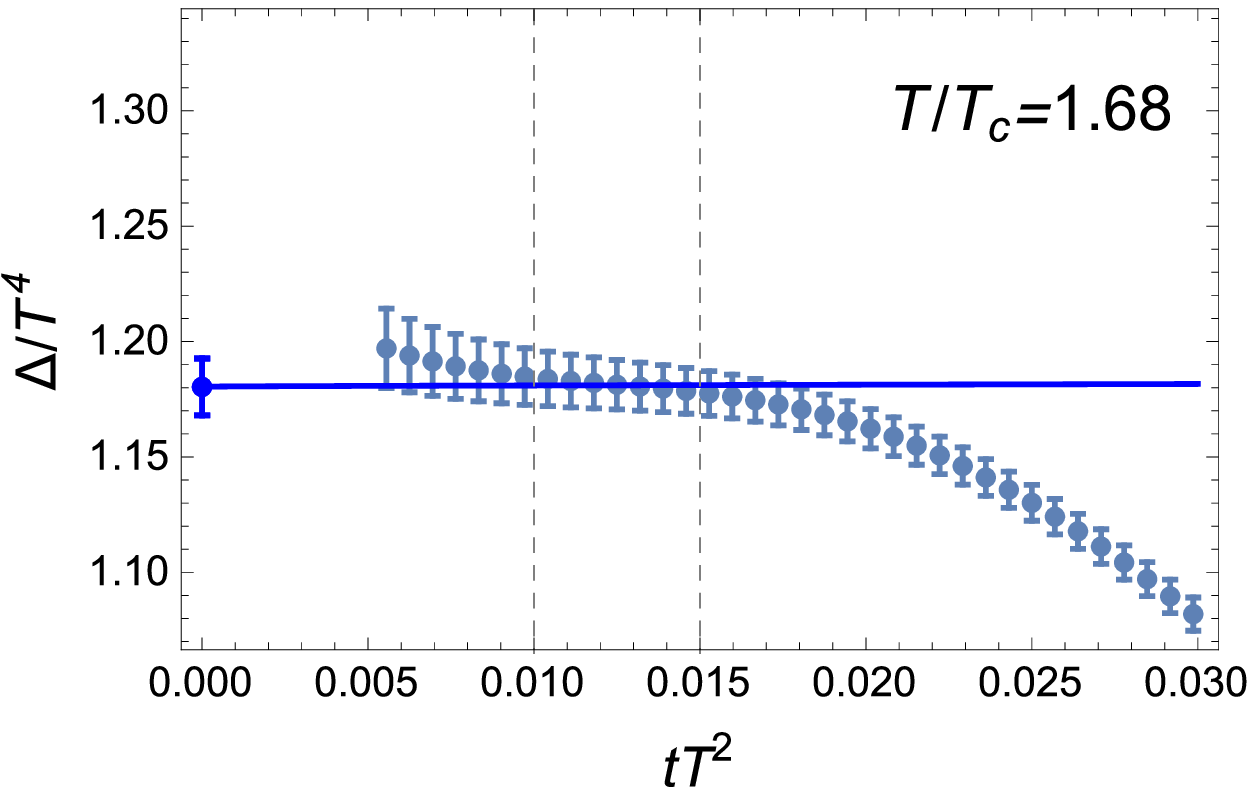}
\end{center}
\end{minipage}
\caption{NNLO analysis of the trace anomaly. 
The blue solid lines are the $t \to 0$ extrapolation function of Eq.~\eqref{resquenchS} with $k=2$.
The gray dashed lines show the fit range used in the $t \to 0$ extrapolation.}
\label{fig:NNLOTA}
\end{figure}

\begin{figure}
\begin{minipage}{0.5\hsize}
\begin{center}
\includegraphics[width=7.5cm]{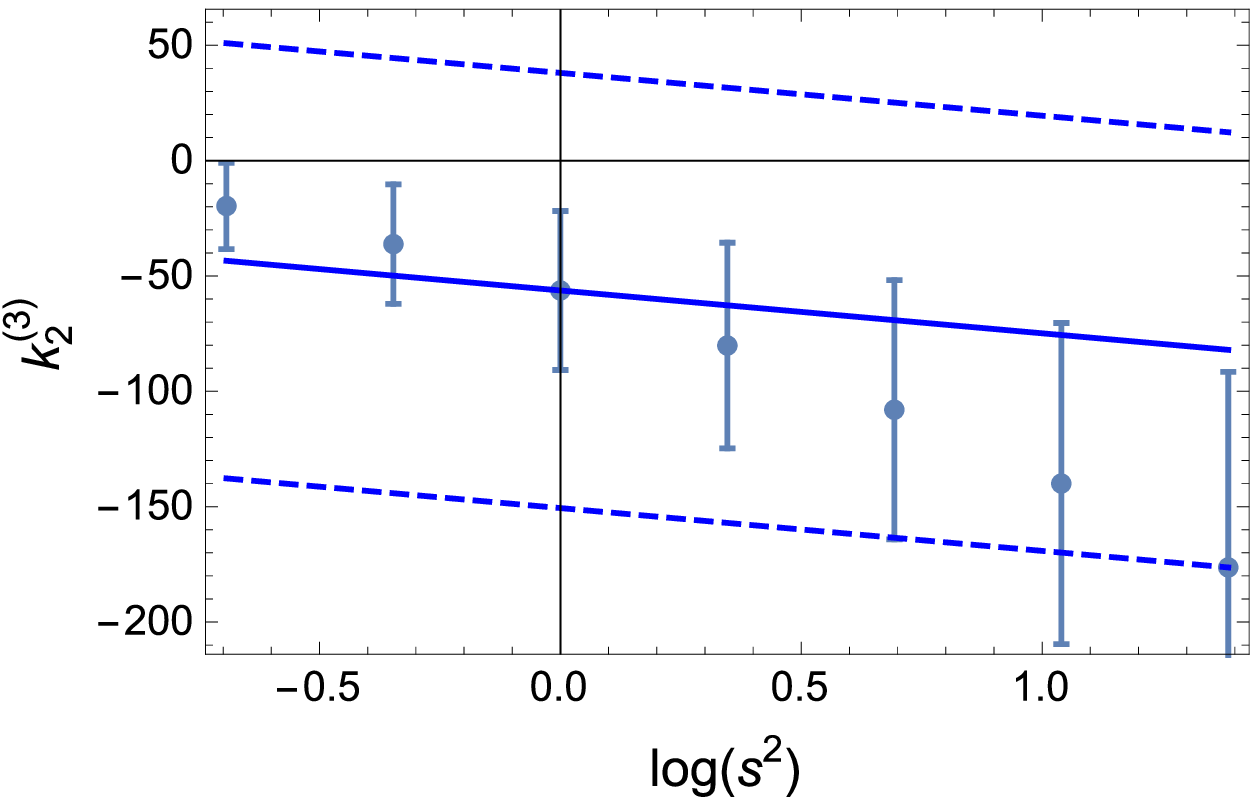}
\end{center}
\end{minipage}
\begin{minipage}{0.5\hsize}
\begin{center}
\includegraphics[width=7.5cm]{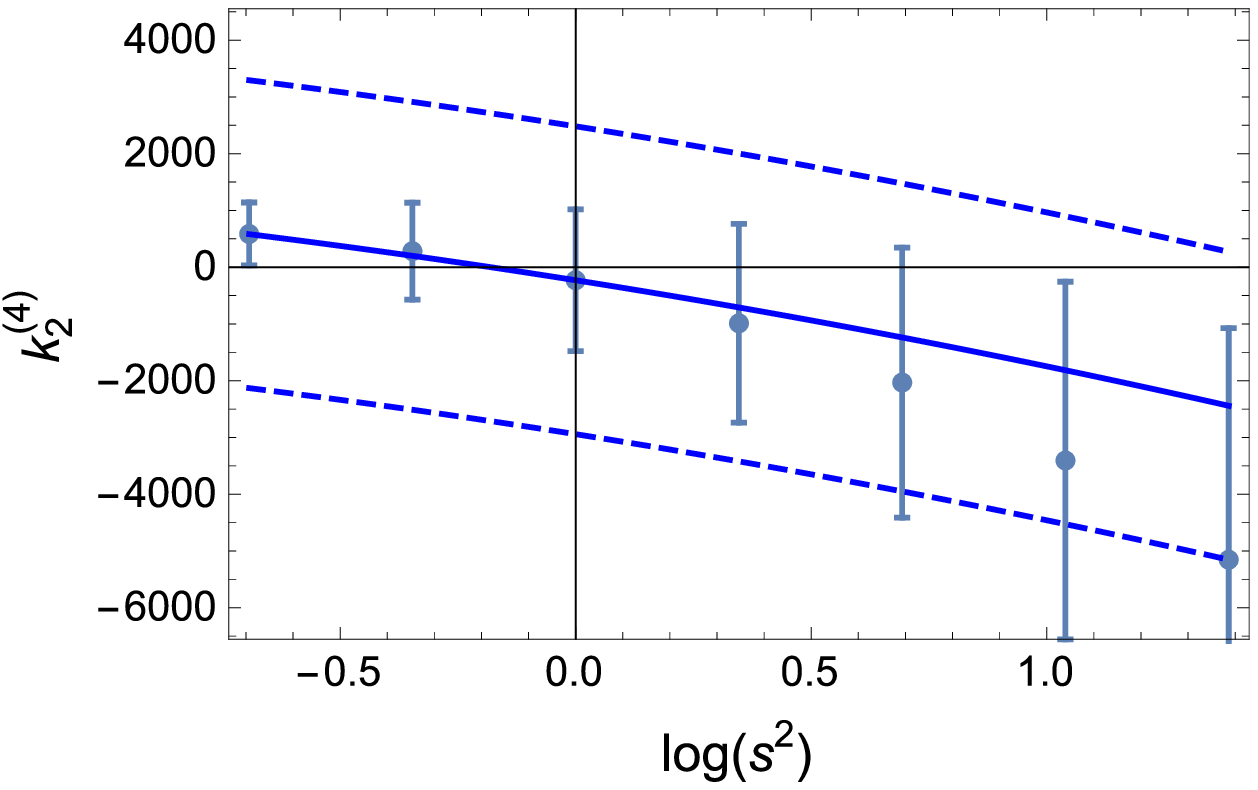}
\end{center}
\end{minipage}
\caption{The fit parameter $k_2^{(3, 4)}$ as a function of $\log{s^2}$. 
See the caption in Fig.~\ref{fig:entropyhighercoeff}.}
\label{fig:higherordercoeffTA}
\end{figure}

\begin{table}[t!]
\begin{center}
\begin{tabular}{l|l|l} 
\hline\hline
   $s/T^3$ \\
\hline
   $T/T_c$ & NLO & $\text{N}^2\text{LO}$  \\
\hline
0.93 & $0.095(17)(02)(00)(00)\,[17]$ & $0.095(17)(02)(01)(00)\,[17]$  \\
1.02 & $2.167(55)(02)(10)(10)\,[57]$ & $2.164(55)(01)(14)(03)\,[57]$  \\
1.12 & $3.751(41)(07)(17)(20)\,[49]$ & $3.748(37)(06)(25)(04)\,[46]$  \\
1.40 & $4.889(37)(07)(23)(31)\,[54]$ & $4.885(32)(06)(35)(02)\,[47]$ \\
1.68 & $5.411(35)(02)(26)(36)\,[56]$ & $5.408(28)(03)(38)(02)\,[47]$ \\
2.10 & $5.769(32)(02)(27)(38)\,[57]$ & $5.766(25)(01)(40)(02)\,[48]$ \\
2.31 & $5.873(40)(10)(28)(39)\,[63]$ & $5.870(34)(08)(40)(03)\,[53]$  \\
2.69 & $6.022(29)(03)(28)(39)\,[56]$ & $6.020(22)(05)(41)(03)\,[47]$  \\
\hline\hline
   $\Delta/T^4$ \\
\hline
   $T/T_c$ & $\text{N}\text{LO}$ & $\text{N}^2\text{LO}$  \\
\hline
0.93 & $0.081(20)(03)(00)(00)\,[20]$ & $0.081(20)(03)(00)(00)\,[20]$ \\
1.02 & $1.942(53)(07)(00)(09)\,[54]$ & $1.941(52)(06)(00)(01)\,[52]$ \\
1.12 & $2.562(35)(08)(00)(12)\,[38]$ & $2.561(30)(07)(00)(02)\,[31]$ \\
1.40 & $1.770(22)(03)(00)(09)\,[23]$ & $1.770(18)(03)(00)(02)\,[18]$ \\
1.68 & $1.180(14)(01)(00)(06)\,[16]$ & $1.180(12)(02)(00)(01)\,[13]$ \\
\hline\hline
\end{tabular}
\end{center}
\caption{NLO and NNLO results for the entropy density and trace anomaly.
The values inside parentheses show errors associated with  (statistic)(range)($\Lambda_{\overline{\rm MS}}$)(renormalization scale) in this order.
The values inside square brackets show total errors, which are given by combining all errors in quadrature.}
\label{tab:1}
\end{table}

As a by-product, we give our estimates of the perturbative coefficients $k_1^{(2)}$ and $k_1^{(3)}$:
\be
k_1^{(2)}(L=0)=88 (18) (2) (29) (114) ,
\ee
\be
k_1^{(3)}(L=0)=-635 (796) (78) (1319) (1837) .
\ee
We note that we already know $k_1^{(2)}(L=0)$ exactly and the above estimate is consistent with it (as already mentioned above).
The values in parentheses are statistical and systematic errors, which are shown in the same order
as in Table~\ref{tab:1}.

We also perform the NLO and NNLO analyses for the trace anomaly;
we show them in Figs.~\ref{fig:NLOTA} and \ref{fig:NNLOTA}.
The results are summarized in Table~\ref{tab:1}.
We check the validity of the perturbative extrapolation function in Fig.~\ref{fig:higherordercoeffTA}.
Since the situation is almost parallel to the case of the entropy density, we do not repeat the explanation. 
However, we note that the error in $\LMS$ does not induce a dominant error in the trace anomaly.

We give our estimate of the perturbative coefficients $k_2^{(3)}$ and $k_2^{(4)}$:
\be
k_2^{(3)}(L=0)=-56 (34) (9) (4) (94) , 
\ee
\be
k_2^{(4)}(L=0)=-230 (1247) (394) (144) (2712) .
\ee
The exact result of the NLO coefficient is known to be $k_2^{(3)}(L=0)=-51.84...$ \cite{Iritani:2018idk},
and we find a good agreement.

Finally, we compare our results with recent precise studies.
Our results for the entropy density are consistent 
with Refs.~\cite{Giusti:2016iqr} and \cite{Caselle:2018kap}.
For the trace anomaly, our results are consistent with Ref.~\cite{Caselle:2018kap}
but not consistent with Ref.~\cite{Borsanyi:2012ve} within our and their final errors.

We note that our results for both quantities 
are consistent with Ref.~\cite{Iritani:2018idk}\fn{
Although in Ref.~\cite{Iritani:2018idk} the ``NNLO" coefficient $c_2(t)$ 
meant the result up to two-loop order,
we regard $c_2(t)$ up to two-loop order as the NLO coefficient in this paper.
Then the ``NNLO" analysis of the trace anomaly in Ref.~\cite{Iritani:2018idk}
corresponds to the NLO analysis in this paper and
the ``NNNLO" analysis of the trace anomaly in Ref.~\cite{Iritani:2018idk} to the NNLO analysis in this paper.
Since there is no tree-level perturbative coefficient in $c_2(t)$, 
we notice that it is natural to call the one-loop order coefficient LO for the trace part
as in the present paper.
A merit of adopting this order counting is that
the $t$ dependence of the N$^k$LO formulae is $\mO(g(\mu(t))^{2(k+1)})$ 
both for the traceless and scalar (or trace) parts.
This convention is common to that in a recent paper~\cite{Shirogane:2020muc}.
} and that we obtained smaller errors in this paper due to reduction
of the systematic errors. One of the key elements in the smaller errors is that 
in the present analysis we do not need to consider the uncertainty 
associated with the functional form of a $t \to 0$ extrapolation function;
we have confirmed that using the extrapolation functions of Eqs.~\eqref{resquenchTL} and \eqref{resquenchS} is valid.
Also the systematic error associated with the renormalization scale gets smaller 
compared to Ref.~\cite{Iritani:2018idk}, in particular for the NLO analysis of the entropy density; 
see Table 2 of Ref.~\cite{Iritani:2018idk}.\fn{
Note that the way to estimate this systematic error is different.
In Ref.~\cite{Iritani:2018idk}, 
$\mu(t) \in [\frac{0.354}{\sqrt{t}}, \frac{0.530}{\sqrt{t}}]$ is used and
in the present paper $\mu(t) \in [\frac{0.375}{\sqrt{t}}, \frac{1.06}{\sqrt{t}}]$ is used.
}
We also note that we did not perform a fit with the range $0.010 \leq t T^2 \leq 0.020$,
which is used in Ref.~\cite{Iritani:2018idk} in a systematic error analysis.
This is because our extrapolation function obtained based on
perturbation theory does not look reasonable in this larger $t$ region.

\newpage

\section{Conclusions and discussion}
\label{sec:6}
The SF$t$X method is a powerful method for simulating the EMT on the lattice,
and in this paper we revealed the functional form to be used in $t \to 0$ extrapolation,
which is required to obtain final results of the EMT.
We explained our main results in Sect.~\ref{sec:2}.
This understanding allows us to perform more precise analyses
of the EMT using the SF$t$X method.
As an additional theoretical study, we also revealed 
the detailed $t$ dependence caused by dimension-six operators for 
the traceless part of the EMT in quenched QCD.

We carried out numerical analyses of the thermodynamics quantities in quenched QCD
with our new extrapolation functions (Sect.~\ref{sec:5}).  
Our extrapolation function is shown to be reasonable from the lattice data
and it serves to reduce systematic errors, compared with the conventionally used
linear function in $t$.
As a by-product, we also gave estimates of the NNNLO coefficients 
$k_1^{(3)}$ and $k_2^{(4)}$ in quenched QCD.

Our lattice analysis is carried out in a sufficiently small $t$ region,
where the coupling constant is given by $(\alpha_s=)g^2/(4 \pi) \lesssim 0.25$.
In full QCD, it might be difficult to obtain lattice data for such a small $t$ region
because lattice spacing tends to be larger. 
(We noted in Sect.~\ref{sec:5} that  the extrapolation functions
do not look sufficiently reasonable for larger $t$.)
In the case where the extrapolation functions \eqref{resfullTL} and \eqref{resfullS}
do not look consistent with available lattice data,
it would be difficult to carry out $t \to 0$ extrapolation in a reasonable way.
For instance, if one includes subleading $t$ dependence 
(e.g. higher-order effects in $g(\mu(t))$ or a linear function in $t$) in a fitting function,
the fit would be destabilized due to the difficulty in distinguishing different functions. 
One conservative attitude would be giving the systematic error associated with
extrapolation function by trying some functions.
However, such a difficulty may be systematically overcome by going to a finer lattice
and/or going to a higher order of perturbation theory.

Finally we mention that it would be possible to reveal proper $t \to 0$ extrapolation functions
in other SF$t$X methods (for other conserved currents) in a similar manner.

\section*{Acknowledgments}
The authors thank Takumi Iritani and Masakiyo Kitazawa 
for fruitful discussions and for letting us use lattice data.
This work was supported by JSPS Grant-in-Aid for Scientific Research Grant
Numbers, JP16H03982 and~JP20H01903 (H.S.) and JP19K14711 (H.T.).

\appendix

\section{Convention}
\label{app:A}
We set the normalization of anti-Hermitian generators~$T^a$ of the
representation~$R$ of the gauge group~$G$ as~${\rm tr}_R(T^aT^b)=-T_R \delta^{ab}$
and~$T^a T^a=-C_R \bold{1}$.
We denote ${\rm tr}_R(1)=\dim(R)$. From the
structure constants defined by~$[T^a,T^b]=f^{abc}T^c$, we set
$f^{acd}f^{bcd}=C_A \delta^{ab}$. For example, for the fundamental
$N$~representation of~$G=SU(N)$ for which $\dim(N)=N$, our normalization is
\begin{equation}
   C_A=N,\qquad T_R=\frac{1}{2},\qquad
   C_F=\frac{N^2-1}{2N}.
\label{eq:(A1)}
\end{equation}
We also define $T_F \equiv n_f T_R=n_f/2$, where $n_f$ is 
the number of quark flavors.

The $D$-dimensional Euclidean action of the vectorial gauge theory is given by
\begin{equation}
   S=\int\mathrm{d}^Dx\, \frac{1}{4 g_0^2} F_{\mu\nu}^a(x)F_{\mu\nu}^a(x)
   +\int\mathrm{d}^Dx\,\Bar{\psi}(x) (\Slash{D}+m_0)\psi(x) .
\label{eq:(A2)}
\end{equation}
The field strength is defined by
\begin{equation}
   F_{\mu\nu}(x)
   =\partial_\mu A_\nu(x)-\partial_\nu A_\mu(x)+ [A_\mu(x),A_\nu(x)],
\label{eq:(A3)}
\end{equation}
for $A_\mu(x)=A_\mu^a(x)T^a$ and~$F_{\mu\nu}(x)=F_{\mu\nu}^a(x)T^a$, 
where $g_0$ is the bare gauge coupling and $m_0$ is the bare mass parameter. The covariant derivative on the fermion is
\begin{equation}
   D_\mu=\partial_\mu+ A_\mu,
\label{eq:(A4)}
\end{equation}
and $\Slash{D}\equiv\gamma_\mu D_\mu$, where $\gamma_\mu$ denotes the Hermitian
Dirac matrix.

\section{EMT and renormalization}
\label{app:B}
The EMT is written by bare quantities but is finite. 
This property gives non-trivial information on renormalization of composite operators, 
as explained in Ref.~\cite{Suzuki:2013gza}. 
We revisit this subject with the basis $\hat{\mO}_{\mu \nu}$.  

First, we consider quenched QCD. The EMT is given by
\be
T_{\mu \nu}(x)=\hat{\mathcal{O}}_{1, \mu \nu}(x)-\frac{1}{4} \hat{\mathcal{O}}_{2, \mu \nu}(x) .
\ee
The renormalization of the composite operators is carried out by
\be
\left( \begin{array}{c}
\hat{\mathcal{O}}_{1, \mu \nu} (x) \\
\hat{\mathcal{O}}_{2, \mu \nu} (x)
\end{array} \right) 
=\left( \begin{array}{cc}
Z_{11}(g(\mu)) & Z_{12}(g(\mu)) \\
Z_{21}(g(\mu)) & Z_{22}(g(\mu)) 
\end{array} \right)
\left( \begin{array}{c}
\hat{\mathcal{O}}^R_{1, \mu \nu}(x; \mu) \\
\hat{\mathcal{O}}^R_{2, \mu \nu}(x; \mu)
\end{array} \right) ,
\ee
where $\mu$ denotes the renormalization scale. 
Since the scalar operator is not mixed with the tensor operator, $Z_{21}=0$.
Then, the EMT is written as
\be
T_{\mu \nu}(x)=Z_{11} \hat{\mO}^R_{1, \mu \nu}(x)-\frac{1}{4} (Z_{22}-4 Z_{12}) \hat{\mO}^R_{2, \mu \nu}(x) .
\ee
Noting that this quantity is finite and noting also the property of the $\overline{\rm MS}$ scheme 
that a renormalization factor only has divergent terms, negative powers in $\epsilon$,
beyond LO in $g(\mu)$ [$Z_{ij}=\delta_{ij}+\sum_{n \geq 1, m \geq 1} c_{nm} \epsilon^{-m} g(\mu)^{2n}$], we have
\be
Z_{11}=1, \quad Z_{22}-4 Z_{12}=1 .
\ee 
This leads to the EMT in terms of the renormalized operators as
\be
T_{\mu \nu}(x)=\hat{\mO}^R_{1, \mu \nu}(x)-\frac{1}{4}  \hat{\mO}^R_{2, \mu \nu}(x) .
\ee
Now, let us consider the trace of the EMT:
\be
\delta_{\mu \nu} T_{\mu \nu} (x)
=\lt(1-\frac{D}{4} \rt) \frac{1}{g_0^2} F^a_{\rho \sigma} F^a_{\rho \sigma} (x)
=\frac{\epsilon}{2}  \frac{1}{D} \hat{\mO}^{\rm S}_2(x) .
\ee
From Eqs.~\eqref{g0} and \eqref{betafn}, we can rewrite $\epsilon$ as
\be
\epsilon=-\beta(g) \lt(\frac{1}{g}+\frac{1}{Z_g} \frac{\partial Z_g}{\partial g} \rt)  . \label{rewriteepsilon}
\ee
Then, we have
\be
\delta_{\mu \nu} T_{\mu \nu}(x) =-\frac{\beta(g)}{2 D g} Z_{22}  \lt(1+\frac{g}{Z_g} \frac{\partial Z_g}{\partial g} \rt) \hat{\mO}^{R, {\rm S}}_2(x) .
\ee
This quantity should be finite. Since $\beta(g)/(2 D g)$ is finite as $\epsilon \to 0$, 
\be
Z_{22}  \lt(1+\frac{g}{Z_g} \frac{\partial Z_g}{\partial g} \rt)
\ee
should be finite. From the above property of the $\overline{\rm MS}$ scheme, we have
\be
Z_{22}  \lt(1+\frac{g}{Z_g} \frac{\partial Z_g}{\partial g} \rt)=1 .
\ee
Hence, we obtain
\be
Z_{22}=\frac{1}{1+\frac{g}{Z_g} \frac{\partial Z_g}{\partial g} } \lt(=-\frac{\beta(g)}{\epsilon g} \rt)  .
\ee
To summarize, we obtain
\begin{align}
&Z_{11}=1, \quad Z_{12}=-\frac{1}{4}+\frac{1}{4} Z_{22} ,  \non
& Z_{21}=0, \quad Z_{22}=\frac{1}{1+\frac{g}{Z_g} \frac{\partial Z_g}{\partial g} } . \label{quenchZ}
\end{align}
[$Z_g$ is given by Eq.~\eqref{Zg}.] We also obtain the EMT in terms of the renormalized operators:
\begin{align}
&T_{\mu \nu}(x)=\hat{\mO}^R_{1, \mu \nu}(x;\mu)-\frac{1}{4}  \hat{\mO}^R_{2, \mu \nu}(x;\mu) ,   \non
& T_{\rho \rho}(x)=-\frac{\beta(g(\mu))}{8 g(\mu)} \hat{\mO}_{2}^{R, {\rm S}}(x;\mu) . \label{Srenquench}
\end{align}
From these results, we have another non-trivial relation. By taking the trace of the first equation, 
it should coincide with the second one, which implies that
\be
\hat{\mO}^{R, {\rm S}}_1(x)=\frac{1}{4} \lt(1-\frac{\beta(g)}{2 g} \rt) \hat{\mO}^{R, {\rm S}}_2(x)  . \label{traceofren1}
\ee
From this, we can write the traceless part of the EMT as
\begin{align}
T_{\mu \nu}^{\rm TL}(x)
&=\hat{\mO}^R_{1, \mu \nu}(x)-\frac{\delta_{\mu \nu}}{4} \hat{\mO}^{R, {\rm S}}_1 (x)\non
&=\hat{\mO}^R_{1, \mu \nu}(x)-\frac{1}{4} \lt(1-\frac{\beta(g)}{2 g} \rt) \hat{\mO}^{R}_{2, \mu \nu} (x) ,\label{TLrenquench}
\end{align}
in terms of the renormalized operators.

Next, we consider full QCD. 
In this case, we can obtain limited results but still a few useful relations. 
The EMT is given by
\be
T_{\mu \nu}(x)
=\hat{\mathcal{O}}_{1, \mu \nu}(x)-\frac{1}{4} \hat{\mathcal{O}}_{2, \mu \nu}(x)+\frac{1}{4} \hat{\mathcal{O}}_{3, \mu \nu}(x)  .
\ee
By defining a renormalization matrix in a parallel manner, we deduce that
\be
Z_{1 i}-\frac{1}{4} Z_{2 i}+\frac{1}{4} Z_{3 i}
\ee 
is finite. Using the property of the $\MSb$ scheme, we obtain
\begin{align}
& Z_{1 1}+\frac{1}{4} Z_{3 1}=1 , \non
& Z_{1 2}-\frac{1}{4} Z_{2 2}+\frac{1}{4} Z_{3 2}=-\frac{1}{4} , \non
& Z_{1 3}+\frac{1}{4} Z_{3 3}=\frac{1}{4} ,
\end{align}
where we have used $Z_{21}=Z_{23}=0$.
Then, the EMT is expressed with the renormalized operators as
\be
T_{\mu \nu}(x)=\hat{\mO}^R_{1, \mu \nu}(x;\mu)-\frac{1}{4} \hat{\mO}^R_{2, \mu \nu}(x;\mu)+\frac{1}{4} \hat{\mO}^R_{3, \mu \nu}(x;\mu) . \label{tmunufull}
\ee
Now, let us consider the trace part:
\be
\delta_{\mu \nu} T_{\mu \nu}(x)=\lt(1-\frac{D}{4} \rt) \frac{1}{g_0^2} F^a_{\mu \nu} F^a_{\mu \nu}(x) +\frac{1}{2} \sum_f\Bar{\psi}_f(x) \overleftrightarrow{\Slash{D}} \psi_f(x)
=\frac{\epsilon}{2} \frac{1}{D} \hat{\mO}_2^{\rm S}(x)+\frac{1}{2 D} \hat{\mO}^{\rm S}_4(x) .
\ee
By rewriting it with the renormalized operators, we have
\be
\delta_{\mu \nu} T_{\mu \nu}(x)
=-\frac{\beta(g)}{2 D g} Z_{22}  \lt(1+\frac{g}{Z_g} \frac{\partial Z_g}{\partial g} \rt) \hat{\mO}^{R, {\rm S}}_2(x)
+\frac{1}{2 D}(1+\epsilon Z_{2 4}) \hat{\mO}_4^{R, {\rm S}}(x) . \label{eq:(B21)}
\ee
Here, we rewrote $\epsilon$ in the same way as above [Eq.~\eqref{rewriteepsilon}] and used $Z_{44}=1$
because $\mO_{4, \mu \nu}$, which is proportional to $\mO_{5, \mu \nu}$ via the EOM, is a finite operator.
Each coefficient should be finite in Eq.~\eqref{eq:(B21)}. Then, for the coefficient of $\hat{\mO}^{R, {\rm S}}_2$,
we obtain the same result as the quenched case:
\be
Z_{22}  \lt(1+\frac{g}{Z_g} \frac{\partial Z_g}{\partial g} \rt)=1  ,
\ee
which indicates that
\be
\delta_{\mu \nu} T_{\mu \nu}(x)
=-\frac{\beta(g)}{8 g}  \hat{\mO}^{R, {\rm S}}_2(x)+\frac{1}{2 D}(1+\epsilon Z_{2 4}) \hat{\mO}_4^{R, {\rm S}}(x) .
\ee
As one can see, the coefficient of the gluonic operator can be obtained in this way also in full QCD.
For $\hat{\mO}^{R, {\rm S}}_4$, we deduce that $1+\epsilon Z_{2 4}$ should be finite.
(Thus, $Z_{2 4}$ has only simple poles in $\epsilon$ to all orders.)
However, it seems difficult to fix its finite value by this argument alone.
Here, we refer to another argument giving the trace part:
\begin{align}
\delta_{\mu \nu} T_{\mu \nu}(x)
&=-\frac{\beta(g)}{8 g}  \hat{\mO}^{R, {\rm S}}_2(x)-(1+\gamma_m) \frac{1}{4} \hat{\mO}^{R, {\rm S}}_{5}(x) \non
&=-\frac{\beta(g(\mu))}{8 g(\mu)}  \hat{\mO}^{R, {\rm S}}_2(x;\mu)+\frac{1+\gamma_m(g(\mu))}{8} \hat{\mO}^{R, {\rm S}}_{4}(x;\mu) . \label{tmunufulltr}
\end{align}
In the last line, we used the EOM, $\frac{1}{2} \hat{\mO}^R_{4, \mu \nu}+\hat{\mO}^R_{5, \mu \nu}=0$.
By this, $\epsilon Z_{24}=\gamma_m$ follows.
By taking the trace of Eq.~\eqref{tmunufull} and comparing with the above expression, we obtain
\be
\hat{\mO}^{R, {\rm S}}_1(x)+\frac{1}{4} \hat{\mO}_3^{R, {\rm S}}(x)
=\frac{1}{4} \lt(1-\frac{\beta(g)}{2 g} \rt) \hat{\mO}^{R, {\rm S}}_{2}(x)+\frac{1}{8}(1+\gamma_m) \hat{\mO}_4^{R, {\rm S}} (x) . \label{traceofren2}
\ee
The traceless part of the EMT is given by
\begin{align}
T_{\mu \nu}^{\rm TL}(x)
&=\hat{\mO}^R_{1, \mu \nu}(x;\mu)-\frac{1}{4} \lt(1-\frac{\beta(g(\mu))}{2 g(\mu)} \rt) \hat{\mO}^R_{2, \mu \nu}(x;\mu) \non
&\quad{}+\frac{1}{4} \hat{\mO}^R_{3, \mu \nu}(x;\mu) -\frac{1+\gamma_m(g(\mu))}{8} \hat{\mO}^R_{4, \mu \nu}(x;\mu) . \label{tmunufulltl}
\end{align}

\section{Anomalous dimension matrix for dimension-four operators}
\label{app:C}
In this appendix, we give the anomalous dimension matrix for dimension-four operators.
For the definition, see Eq.~\eqref{gammaDef}.
It has a form 
\be
\gamma(g)=\gamma_0 \frac{g^2}{(4 \pi)^2} +\gamma_1 \lt[ \frac{g^2}{(4 \pi)^2} \rt]^2+\cdots .
\ee

In quenched QCD, the anomalous dimension can be obtained from Eqs.~\eqref{quenchZ} and \eqref{Zg}:
\be
\gamma_0=\left( \begin{array}{cc}
0 & -\frac{\beta_0}{2}  \\
0 & -2 \beta_0 \\
\end{array} \right) , \label{gamma0quench}
\ee
\be
\gamma_1=\left( \begin{array}{cc}
0 & -\beta_1  \\
0 & -4 \beta_1 \\
\end{array} \right) , 
\ee
\be
\gamma_2=\left( \begin{array}{cc}
0 & -\frac{3}{2}\beta_2  \\
0 & -6 \beta_2 \\
\end{array} \right)  .
\ee

In full QCD, the leading order matrix is given by \cite{Harlander:2018zpi}
\be
\gamma_0=
\begin{pmatrix}
\frac{8}{3} T_F & -\frac{11}{6} C_A & -\frac{4}{3} C_F & -\frac{7}{3} C_F \\
0 & -2 \beta_0  & 0 & -12 C_F \\
-\frac{32}{3}  T_F & \frac{8 }{3} T_F & \frac{16}{3}  C_F& -\frac{8 }{3} C_F\\
0 & 0 & 0 & 0
\end{pmatrix} \, . \label{gamma0full}
\ee
The NLO matrix is given by \cite{Harlander:2018zpi}
\be
\gamma_{1,11}=\frac{4}{27} T_F (35 C_A+74 C_F) ,
\ee
\be
\gamma_{1,12}=\frac{2}{27} (-153 C_A^2+56 C_A  T_F+5 C_F T_F) ,
\ee
\be
\gamma_{1,13}=\frac{4}{27}  C_F (-47 C_A+14 C_F+26 T_F) ,
\ee
\be
\gamma_{1,14}=-\frac{1}{27} C_F (812 C_A+85 C_F-44 T_F) ,
\ee
\be
\gamma_{1,21}=0 ,
\ee
\be
\gamma_{1,22}=\frac{8}{3} (-17 C_A^2+10 C_A T_F +6 C_F T_F) ,
\ee
\be
\gamma_{1,23}=0 ,
\ee
\be
\gamma_{1,24}=-\frac{4}{3} C_F (97 C_A+9 C_F-20 T_F) ,
\ee
\be
\gamma_{1,31}=-\frac{16}{27} T_F (35 C_A+74 C_F) ,
\ee
\be
\gamma_{1,32}=\frac{8}{27} T_F (34 C_A+49 C_F) ,
\ee
\be
\gamma_{1,33}=-\frac{16}{27} C_F (-47 C_A+14 C_F+26 T_F) ,
\ee
\be
\gamma_{1,34}=\frac{4}{27} C_F (-61 C_A+4 C_F +136 T_F) ,
\ee
\be
\gamma_{1,41}=\gamma_{1,42}=\gamma_{1,43}=\gamma_{1,44}=0 .
\ee
As noted in footnote~\ref{fn:4}, our definition of the renormalization factor $Z$
is the inverse of that of Ref.~\cite{Harlander:2018zpi}.

\section{Coefficients $c^{(\prime)}_i(t)$} 
\label{app:D}
The coefficients $c^{(\prime)}_i(t)$ are RG invariant and 
the $L(\mu,t) \equiv \log(2 e^{\gamma_E} \mu^2 t)$ dependence of the perturbative series is determined by the RG equation.
For $i=1, 2$, we have
\begin{align}
c_{i}(t)=\frac{1}{g(\mu)^2} \sum_{n=0}^{\infty} k_{i}^{(n)}(L(\mu,t)) \lt[ \frac{g(\mu)^2}{(4 \pi)^2} \rt]^{n} ,
\end{align}
with
\begin{align}
&k_i^{(0)}(L(\mu,t))=k_i^{(0)}(L=0) ,\non
&k_i^{(1)}(L(\mu,t))=k_i^{(1)}(L=0)-\beta_0 k_i^{(0)} L(\mu,t) , \non
&k_i^{(2)}(L(\mu,t))=k_i^{(2)}(L=0)-\beta_1 k_i^{(0)}  L(\mu,t) , \non
&k_i^{(3)}(L(\mu,t))=k_i^{(3)}(L=0)+\lt(\beta_0 k_i^{(2)}(L=0) - \beta_2 k_i^{(0)} \rt)L(\mu,t)-\frac{1}{2}\beta_0 \beta_1 k_i^{(0)} L(\mu,t)^2 , \non
&k_i^{(4)}(L(\mu,t))=k_i^{(4)}(L=0)+\lt(-\beta_3 k_i^{(0)}+\beta_1 k_i^{(2)}(L=0)+2 \beta_0 k_i^{(3)}(L=0) \rt)L(\mu,t) \non
&\qquad{}\qquad{}\qquad{}+\lt\{ \lt(-\frac{1}{2} \beta_1^2 -\beta_0 \beta_2  \rt)k_i^{(0)}+\beta_0^2 k_i^{(2)}(L=0) \rt\}L(\mu,t)^2
-\frac{1}{3} \beta_0^2 \beta_1 k_i^{(0)} L(\mu,t)^3 . \label{eq:(D2)}
\end{align}
For $i=3, 4$, we have
\be
c^{(\prime)}_{i}(t)=\sum_{n=0}^{\infty} k_{i}^{(n)}(L(\mu,t)) \lt[ \frac{g(\mu)^2}{(4 \pi)^2} \rt]^{n}
\ee
with
\begin{align}
&k_i^{(0)}(L(\mu,t))=k_i^{(0)}(L=0) ,\non
&k_i^{(1)}(L(\mu,t))=k_i^{(1)}(L=0) , \non
&k_i^{(2)}(L(\mu,t))=k_i^{(2)}(L=0)+\beta_0 k_i^{(1)}  L(\mu,t) , \non
&k_i^{(3)}(L(\mu,t))=k_i^{(3)}(L=0)+\lt(\beta_1 k_i^{(1)} +2 \beta_0 k_i^{(2)}(L=0) \rt) L(\mu,t)+\beta_0^2 k_i^{(1)} L(\mu,t)^2 .
\end{align}

The explicit NLO result is given by \cite{Suzuki:2013gza, Makino:2014taa}
\be
c_1(t)=\frac{1}{g^2}+\lt(-\frac{7}{3} C_A+\frac{3}{2} T_F -\beta_0 L(\mu,t) \rt)  \frac{1}{(4 \pi)^2} ,
\ee
\be
c_2(t)=\frac{1}{(4 \pi)^2} \lt(\frac{11}{24} C_A-\frac{3}{8} T_F \rt) , 
\ee
\be
c_3(t)=\frac{1}{4}+\frac{g^2}{(4 \pi)^2} C_F \lt(\frac{3}{8}+\log{2}+\frac{3}{4} \log{3} \rt) ,
\ee
\be
c'_4(t)=\frac{1}{8}+\frac{g^2}{(4 \pi)^2} C_F \lt(\frac{11}{16}+\frac{1}{2} \log{2}+\frac{3}{8} \log{3} \rt)  .
\ee

\section{Properties of the matrix $K(\mu; \mu_0)$}
\label{app:E}
We list some relations that the matrix $K$ should satisfy.
Since $K$ describes evolution, it should satisfy
\begin{align}
& K(\mu=\mu_0; \mu_0)=\bold{1}_{4 \times 4} ,\non
& K(\mu; \mu_0) K(\mu_0; \mu)=\bold{1}_{4 \times 4} .
\end{align}
Using the fact that the EMT [Eq.~\eqref{tmunufull}] is renormalization scale independent, we obtain
\begin{align}
&K_{11}+\frac{1}{4} K_{31}=1 ,\non
&K_{12}-\frac{1}{4} K_{22}+\frac{1}{4} K_{32}=-\frac{1}{4} ,  \non
&K_{13}+\frac{1}{4} K_{33}=\frac{1}{4} .  \label{EMTinv}
\end{align}
Here we used $K_{21}=K_{23}=0$ because $\hat{\mO}^R_{2, \mu \nu}$ is an essentially scalar operator
and is not mixed with $\hat{\mO}^R_{1, \mu \nu}$ or $\hat{\mO}^R_{3, \mu \nu}$. 
Also the trace part of the EMT [Eq.~\eqref{tmunufulltr}] is renormalization scale independent, which leads to
\begin{align}
& -\frac{\beta(g(\mu))}{8 g(\mu)} K(\mu; \mu_0)_{22}=-\frac{\beta(g(\mu_0))}{8 g(\mu_0)} , \non
&  -\frac{\beta(g(\mu))}{8 g(\mu)} K(\mu; \mu_0)_{24}=\frac{1}{8} \{\gamma_m(g(\mu_0))-\gamma_m(g(\mu)) \} \, . \label{EMTtrinv}
\end{align}
These are exact relations, not relying on perturbation theory.
One can check that $K(\mu; \mu_0)$ of Eq.~\eqref{Kfull} satisfies all the relations (at appropriate order).

\section{$\zeta^R$ at NLO}
\label{app:F}
We show the NLO result for $\zeta^R$, defined in Eq.~\eqref{zetaRdef}.
For 
\begin{align}
\zeta^R(t; g(\mu), \mu)=
\begin{pmatrix}
g^2  & 0 & 0 & 0 \\
0  & g^2 & 0 &  0\\
0 & 0 & 1 & 0 \\
0 & 0 & 0 & 1\\
\end{pmatrix}
+\frac{g^2}{(4 \pi)^2} {\zeta^R}^{(1)}+\cdots ,
\end{align} 
the NLO matrix ${\zeta^R}^{(1)}$ is given by
\begin{align}
&{\zeta^R}^{(1)}_{11}=C_A g^2 \lt( \frac{7}{3}+\frac{11}{3} L(\mu, t) \rt) ,\non
&{\zeta^R}^{(1)}_{12}=-C_A g^2 \lt(\frac{1}{6}+\frac{11}{12} L(\mu, t) \rt)  , \non
&{\zeta^R}^{(1)}_{13}=-C_F g^2 \lt(\frac{7}{18}+\frac{2}{3} L(\mu,t) \rt) , \non
&{\zeta^R}^{(1)}_{14}=-C_F g^2 \lt(\frac{59}{36}+\frac{7}{6} L(\mu, t) \rt) , \non
&{\zeta^R}^{(1)}_{21}=0 , \non
&{\zeta^R}^{(1)}_{22}=\frac{7 C_A}{2} g^2 , \non
& {\zeta^R}^{(1)}_{23}=0  , \non
& {\zeta^R}^{(1)}_{24}=-C_F g^2 (5+6 L(\mu,t)) , \non
& {\zeta^R}^{(1)}_{31}=-T_F \lt(6+\frac{16}{3} L(\mu,t) \rt) , \non
& {\zeta^R}^{(1)}_{32}=T_F \lt(3+\frac{4}{3} L(\mu,t) \rt) , \non
& {\zeta^R}^{(1)}_{33}=C_F \lt(\frac{1}{18} -4 \log{2}-3 \log{3}+\frac{8}{3} L(\mu,t) \rt) , \non
& {\zeta^R}^{(1)}_{34}=-C_F \lt( \frac{4}{9}+\frac{4}{3} L(\mu, t) \rt) , \non
&  {\zeta^R}^{(1)}_{41}=0 , \non
&  {\zeta^R}^{(1)}_{42}= \frac{5}{3} T_F , \non
&  {\zeta^R}^{(1)}_{43}=0 , \non 
& {\zeta^R}^{(1)}_{44}=C_F \lt( \frac{1}{2}-4 \log{2}-3 \log{3} \rt) .
\end{align}
 
We can confirm validity of this NLO result for instance as follows. 
By rewriting $\Tilde{\mO}_{1, \mu \nu}^{\rm TL}(t,x)
=\Tilde{\mO}_{1, \mu \nu}(t,x)-(1/4) \Tilde{\mO}_{2, \mu \nu}(t,x)$ in terms of $\hat{\mO}^R(x; \mu)$
with the NLO $\zeta^R(t; g(\mu), \mu)$
and then requiring it is traceless, we obtain
\be
\hat{\mO}^{R, {\rm S}}_1=\frac{1}{4} \hat{\mO}^{R, {\rm S}}_{2}+\frac{g^2}{(4 \pi)^2} \lt[\frac{11}{24} C_A \hat{\mO}^{R, {\rm S}}_2+\frac{7}{12} C_F \hat{\mO}^{R, {\rm S}}_4  \rt]+\mathcal{O}(g^4) \, . \label{O1Rtr}
\ee
Similarly, from $\Tilde{\mO}_{3, \mu \nu}^{\rm TL}(t,x)=\Tilde{\mO}_{3, \mu \nu}(t,x)-(1/2)\Tilde{\mO}_{4, \mu \nu}(t,x)$, we obtain
\be
\hat{\mO}_3^{R, {\rm S}}=\frac{1}{2} \hat{\mO}_4^{R, {\rm S}}+\frac{g^2}{(4 \pi)^2} \lt[-\frac{2}{3} T_F \hat{\mO}^{R, {\rm S}}_2+\frac{2}{3} C_F \hat{\mO}^{R, {\rm S}}_4 \rt]+\mathcal{O}(g^4) \, . \label{O3Rtr}
\ee
One can check that Eq.~\eqref{traceofren2} is correctly reproduced from these results.

\section{Higher order correction to the matrix $K$}
\label{app:G}
In this appendix, we clarify how we specified
the parametrical errors of the LO calculations in Eqs.~\eqref{O1TL}--\eqref{O4S}.
For this purpose, we need to know what kind of higher-order corrections 
appear in the matrices $\zeta^R$ and $K$ in Eq.~\eqref{tildeOrewrite}.
Here, we investigate this issue particularly for the matrix $K$,
because its higher-order effect is more difficult to see than the matrix $\zeta^R$
(whose higher-order effect is just given by a higher power in $g(\mu(t))^2$).

We first develop a general argument to detect higher-order correction to the matrix $K$.
The RG equation for $K$ in Eq.~\eqref{Kfirstappear} is given in the form
\be
\mu \frac{d K}{d \mu}= -\gamma(g) K ,
\ee
or
\be
\frac{d K}{d A}
=-\frac{\gamma(A)}{2 \beta_A(A) } K 
=\lt[\frac{\gamma_0}{2 \beta_0 A}+\frac{\beta_0  \gamma_1-\beta_1 \gamma_0}{2 \beta_0^2} +\mathcal{O}(A)\rt] K ,
\ee
with $A \equiv g(\mu)^2/(4\pi)^2$ and $\beta_A(A)\equiv -\sum_{i=0}^{\infty} \beta_i A^{i+2}$.
Here and hereafter, we define $\gamma_0, \gamma_1, \dots$ as 
\be
\gamma(A)=\gamma_0 A+\gamma_1 A^2+\mathcal{O}(A^3) .
\ee
$\gamma_0$ and $\gamma_1$ are not commutative generally.
The formal solution to the above RG equation is given by a path ordered product,
but it is difficult to evaluate explicitly. 
In fact, we can obtain the NLO $K$ matrix as follows. 
Denoting the LO solution by $K_{\rm LO}(A;A_0)$ (with~$A_0=g(\mu_0)^2/(4 \pi)^2$), which satisfies 
\be
\frac{d }{d A}K_{\rm LO}(A;A_0)=\frac{\gamma_0}{2 \beta_0 A} K_{\rm LO}(A;A_0) ,
\ee
and writing the NLO solution as $K_{\rm NLO}=K_{\rm LO} \tilde{K}$,
we have
\begin{align}
\frac{d K_{\rm NLO}}{d A}
&=\frac{d K_{\rm LO}}{d  A} \tilde{K}+ K_{\rm LO} \frac{d \tilde{K}}{d A} \non
&=\lt[\frac{\gamma_0}{2 \beta_0 A}+\frac{\beta_0  \gamma_1-\beta_1 \gamma_0}{2 \beta_0^2} +\mathcal{O}(A) \rt] K_{\rm LO} \tilde{K} ,
\end{align}
and hence,
\be
\frac{d \tilde{K}}{d A}=\lt(K_{\rm LO}^{-1}  \frac{\beta_0  \gamma_1-\beta_1 \gamma_0}{2 \beta_0^2} K_{\rm LO} \rt) \tilde{K} .
\ee
Since $K_{\rm LO}^{-1}  \frac{\beta_0  \gamma_1-\beta_1 \gamma_0}{2 \beta_0^2}  K_{\rm LO}$ can be regarded as a function of $A$ and $A_0$,
we can obtain $\tilde{K}$ as
\be
\tilde{K}(A;A_0)=\exp \lt[\int_{A_0}^{A} dx \lt[K_{\rm LO}^{-1}  \frac{\beta_0  \gamma_1-\beta_1 \gamma_0}{2 \beta_0^2}  K_{\rm LO} \rt](x;A_0)  \rt] .
\ee
Then the NLO $K$ matrix is given by
\begin{align}
K_{\rm NLO}(A;A_0)
&=K_{\rm LO}(A;A_0)  \exp \lt[\int_{A_0}^{A}dx \lt[K_{\rm LO}^{-1}  \frac{\beta_0  \gamma_1-\beta_1 \gamma_0}{2 \beta_0^2}  K_{\rm LO} \rt](x;A_0)  \rt] \non
&=K_{\rm LO}(A;A_0)+\int_{A_0}^{A}dx \, K_{\rm LO}(A;A_0) K_{\rm LO}^{-1}(x;A_0)  \frac{\beta_0  \gamma_1-\beta_1 \gamma_0}{2 \beta_0^2}  K_{\rm LO}(x;A_0)+\cdots . \label{KatNLO}
\end{align}

Now we reveal the behavior of the second term, which can be regarded as the NLO correction term.
For simplicity, we consider two-dimensional operator space.
We denote the eigenvalues of $\gamma_0/(2 \beta_0)$ by $\lambda_1$ and $\lambda_2$ ($\lambda_1 < \lambda_2$).
Then $K_{\rm LO}(A;A_0)$ is given by a linear combination of $\{(A/A_0)^{\lambda_1}, (A/A_0)^{\lambda_2} \}$.
In Eq.~\eqref{KatNLO}, we note that 
\be
K_{\rm LO}(A;A_0) K_{\rm LO}^{-1}(x;A_0) 
=K_{\rm LO}(A;A_0) K_{\rm LO}(A_0;x)=K_{\rm LO}(A;x) ,
\ee
because $K_{\rm LO}$ is an evolution matrix.
Then the integrand is given by
\begin{align}
&K_{\rm LO}(A;A_0) K_{\rm LO}^{-1}(x;A_0)  \frac{\beta_0  \gamma_1-\beta_1 \gamma_0}{2 \beta_0^2}  K_{\rm LO}(x;A_0) \non
&=K_{\rm LO}(A;x) \frac{\beta_0  \gamma_1-\beta_1 \gamma_0}{2 \beta_0^2}  K_{\rm LO}(x;A_0) \non
&=({\text{linear comb. of~~}} \{(A/x)^{\lambda_1}, (A/x)^{\lambda_2} \}) \times
({\text{linear comb. of~~}} \{(x/A_0)^{\lambda_1}, (x/A_0)^{\lambda_2} \}) \non
&=({\text{linear comb. of~~}} \{(A/A_0)^{\lambda_1}, (A/A_0)^{\lambda_2}, 
(A^{\lambda_1}/A_0^{\lambda_2}) x^{\lambda_2-\lambda_1}, (A^{\lambda_2}/A_0^{\lambda_1}) x^{\lambda_1-\lambda_2} \}) .
\end{align}
After the integration, we have
\begin{align}
&\int_{A_0}^{A}dx \, K_{\rm LO}(A;A_0) K_{\rm LO}^{-1}(x;A_0)  \frac{\beta_0  \gamma_1-\beta_1 \gamma_0}{2 \beta_0^2}  K_{\rm LO}(x;A_0) \non
&=({\text{linear comb. of~~}} \{(A/A_0)^{\lambda_1} A_0, (A/A_0)^{\lambda_2} A_0, 
(A/A_0)^{\lambda_1} A, (A/A_0)^{\lambda_2} A \}) . \label{eq:(G11)}
\end{align}
Therefore either order in $A$ or $A_0$ is raised by one in the NLO correction term
compared to~$K_{\rm LO}$.

In the following we concretely study the NLO effects in full QCD.
To simplify calculations, we decompose the four operators $\hat{\mathcal{O}}_1, \dots, \hat{\mathcal{O}}_4$
into scalar parts and traceless tensor parts. Then, anomalous dimension matrices are two-by-two matrices.

Before this, for convenience we introduce $\zeta'^R$: 
\be
\zeta'^R=
\begin{pmatrix}
1/g(\mu(t))^2 & 0& 0& 0\\
0 & 1/g(\mu(t))^2 & 0 &0 \\
0 & 0& 1 & 0 \\
0 & 0 & 0 & 1
\end{pmatrix} 
\zeta^R .
\ee 
This matrix describes the time evolution of 
$\{1/g(\mu(t))^2 \tilde{\mathcal{O}}_1, 1/g(\mu(t))^2 \tilde{\mathcal{O}}_2, \tilde{\mathcal{O}}_3, \tilde{\mathcal{O}}_4\}$
as seen from the definition [cf. Eq.~\eqref{FtoUF}].
The advantage is that the perturbation order of this matrix is organized by the power of $g$.

Let us begin with scalar operators.
Scalar operators are given by linear combinations of $\hat{\mathcal{O}}_2^{R, {\rm S}}$ and $\hat{\mathcal{O}}_4^{R, {\rm S}}$.
Then, we have an RG equation of [cf. Eq.~\eqref{RGRenOp}]
\be
\mu \frac{d}{d \mu} \lt( \begin{array}{c}
\hat{\mathcal{O}}^{R, {\rm S}}_{2} \\ 
\hat{\mathcal{O}}^{R, {\rm S}}_{4}  
\end{array} \rt)
=-\gamma^{\rm S} \lt( \begin{array}{c}
\hat{\mathcal{O}}^{R, {\rm S}}_{2} \\ 
\hat{\mathcal{O}}^{R, {\rm S}}_{4}  
\end{array} \rt) .
\ee
The matrix $\gamma^{\rm S}$ is readily read off from $\gamma_0$ and $\gamma_1$ given in App.~\ref{app:C}. 
The evolution matrix $K^{\rm S}$ at LO, 
satisfying $\frac{d K^{\rm S}_{\rm LO}}{d A}=\frac{\gamma^{\rm S}_0}{2 \beta_0 A} K^{\rm S}_{\rm LO}$,
is obtained as
\be
K^{\rm S}_{\rm LO}
=\begin{pmatrix}
\frac{A_0}{A} & \frac{6 C_F}{\beta_0} \lt(\frac{A_0}{A}-1\rt) \\
0 &1
\end{pmatrix} , \label{eq:(G14)}
\ee
where the eigenvalues of $\gamma_0^{\rm S}/(2 \beta_0)$ are $\lambda_1=-1$ and $\lambda_2=0$.
From the general argument above [in particular from Eq.~\eqref{eq:(G11)}],
we see that the NLO $K$ matrix is given in the form
\be
K^{\rm S}_{\rm NLO}=K^{\rm S}_{\rm LO}+({\text{linear comb. of }} \{A_0^2/A, A_0, A\}) . \label{eq:(G15)}
\ee

The knowledge on the order of the NLO correction enables us to accurately estimate the parametrical error
of the LO calculation.
In the LO calculation, we use $\zeta'^R_{\rm LO}=\{1\}+\mathcal{O}(A)$
and $K^{\rm S}_{\rm LO}=\{A_0/A , 1\}+\mathcal{O}(A_0^2/A,A_0, A)$.
Here we imply that the left-hand side is given by a linear combination of the functions inside $\{ \}$
and possesses the errors shown by $\mathcal{O}(...)$.
We use this notation hereafter.
Then we have
\be
\zeta'^{R, {\rm S}}_{\rm LO} K_{\rm LO}^{\rm S}=\{A_0/A, 1 \}+\mathcal{O}(A_0^2/A, A_0, \dots) , \label{eq:(G16)}
\ee
where $\zeta'^{R, {\rm S}}$ is the two-by-two matrix which relates the flowed operators
$(1/g(\mu(t))^2)\tilde{\mO}_2^{\rm S}(t,x)$ and $\tilde{\mO}_4^{\rm S}(t,x)$ to
the unflowed operators $\hat{\mO}_2^{R, {\rm S}}(x;\mu(t))$ and $\tilde{\mO}_4^{R, {\rm S}}(x;\mu(t))$.\fn{
Explicitly, it is given by
\be
\zeta'^{R, {\rm S}}=
\begin{pmatrix}
\zeta'^{R, {\rm S}}_{22} &\zeta'^{R, {\rm S}}_{24} \\
\zeta'^{R, {\rm S}}_{42} &\zeta'^{R, {\rm S}}_{44}
\end{pmatrix} . \nonumber 
\ee
At LO, this is the unit matrix.}
Then, for $(1/g(\mu(t))^2)\tilde{\mathcal{O}}_2^{\rm S}(t,x)|_{\rm LO}=(\zeta'^{R, {\rm S}}_{\rm LO} K_{\rm LO}^S)_{11} \hat{\mO}^{R, {\rm S}}_2(x;\mu_0)+(\zeta'^{R, {\rm S}}_{\rm LO} K_{\rm LO}^S)_{12} \hat{\mO}^{R, {\rm S}}_4(x;\mu_0)$,
from Eq.~\eqref{eq:(G16)} we obtain
\begin{align}
\frac{1}{g(\mu(t))^2}\tilde{\mathcal{O}}_2^{\rm S}(t,x)
&=\frac{1}{g(\mu(t))^2} 
\lt[g(\mu_0)^2 \lt(\hat{\mathcal{O}}_2^{R, {\rm S}}(x;\mu_0)+ \frac{6 C_F}{\beta_0} \hat{\mathcal{O}}_4^{R, {\rm S}}(x;\mu_0) \rt) +\mathcal{O}(g(\mu_0)^4) \rt] \non
&\quad{}+\mathcal{O}(g(\mu(t))^0) , \label{eq:(G17)}
\end{align}
where the error and neglected higher-order terms are shown by $\mathcal{O}(...)$.
(Here we set $\mu=\mu(t)$.)
This result corresponds to Eq.~\eqref{O2S}.
For $\tilde{\mathcal{O}}_4^{\rm S}(t,x)$, we have
\begin{align}
\tilde{\mathcal{O}}_4^{\rm S}(t,x)
&=[1+\mathcal{O}(A_0)+\mathcal{O}(A_0^2)+\mathcal{O}(A, A A_0, A^2)]\hat{\mathcal{O}}_4^{R, {\rm S}}(x;\mu_0)  \non
&\quad{}+[\mathcal{O}(A_0)+\mathcal{O}(A_0^2)+\mathcal{O}(A, A A_0, A^2)] \hat{\mathcal{O}}_2^{R, {\rm S}}(x;\mu_0) 
\label{eq:(G18)} ,
\end{align}
which corresponds to Eq.~\eqref{O4S}.
In obtaining this result, we have noted the following facts.
First, $\tilde{\mO}_4^{\rm S}(t,x)$ is written in terms of $\hat{\mO}_2^{R, {\rm S}}(x;\mu(t))$ and 
$\hat{\mO}_4^{R, {\rm S}}(x;\mu(t))$ with $\zeta'^R$ as
\be
\tilde{\mathcal{O}}_4^{\rm S}(t,x)=(1+\mathcal{O}(A)) \hat{\mathcal{O}}_4^{R, {\rm S}}(x;\mu(t))
+\mathcal{O}(A)\hat{\mathcal{O}}_2^{R, {\rm S}}(x;\mu(t)) . \label{eq:(G19)}
\ee
Here, the key is that $\hat{\mathcal{O}}_2^{R, {\rm S}}(x;\mu(t))$ is multiplied by a factor of $\mathcal{O}(A)$.
Then in rewriting $\hat{\mO}_2^{R, {\rm S}}(x;\mu(t))$ in terms of 
$\hat{\mO}_2^{R, {\rm S}}(x;\mu_0)$ and $\hat{\mO}_4^{R, {\rm S}}(x;\mu_0)$ with 
$K_{\rm LO}^{\rm S}=\{A_0/A, 1\}+\mathcal{O}(A_0^2/A,A_0,A)$ [cf. Eqs.~\eqref{eq:(G14)}
and \eqref{eq:(G15)}], we do not have $\mO(1/A)$ contributions from this part.
Also one can see that $\mathcal{O}(A)\hat{\mathcal{O}}_2^{R, {\rm S}}(x;\mu(t))$ gives
the $\mO(A_0)$, $\mO(A_0^2)$, and $\mathcal{O}(A, A A_0, A^2)$ terms
in the coefficient of $\hat{\mO}_2^{R, {\rm S}}(x;\mu_0)$ in Eq.~\eqref{eq:(G18)}
and gives $\mO(A_0)$, $\mO(A_0^2)$, and $\mO(A, A A_0, A^2)$ contributions 
to the coefficient of $\hat{\mO}_4^{R, {\rm S}}(x;\mu_0)$ in Eq.~\eqref{eq:(G18)}.
Secondly, in rewriting $\hat{\mO}_4^{R, {\rm S}}(x;\mu(t))$ in terms of
$\hat{\mO}_2^{R, {\rm S}}(x;\mu_0)$ and $\hat{\mO}_4^{R, {\rm S}}(x;\mu_0)$ with $K^{\rm S}$,
the relevant components $K^{\rm S}_{21}$ and $K^{\rm S}_{22}$ are exactly given by $(K^{\rm S})_{21}=0$ and  $(K^{\rm S})_{22}=1$, and thus $\hat{\mO}_4^{R, {\rm S}}(x;\mu(t))=\hat{\mO}_4^{R, {\rm S}}(x;\mu_0)$.
This is because $\hat{\mO}^{R, {\rm S}}_4(x;\mu)$ is proportional to 
$\hat{\mO}^{R, {\rm S}}_5(x;\mu) (:=\mO^{R, {\rm S}}_5(x;\mu))$ due to the EOM, and $\hat{\mO}^{\rm S}_5(x;\mu)$ 
is a finite operator, which does not need renormalization.
Hence, we do not have $\mO(1/A)$ contributions in Eq.~\eqref{eq:(G18)}.
The $\mathcal{O}(A_0)$ and $\mathcal{O}(A_0^2)$ terms 
in Eq.~\eqref{eq:(G18)}, respectively,
can be explicitly obtained with the combination of the NLO $\zeta'^R$ and the LO $K$, 
and that of the NLO $\zeta'^R$ and the NLO $K$.

Although we have obtained the main results for the scalar parts 
in Eqs.~\eqref{eq:(G17)} and~\eqref{eq:(G18)},
we are also able to explicitly obtain the NLO $K$ matrix with Eq.~\eqref{KatNLO}.
It would be interesting to check the validity of such an explicit NLO $K$ matrix.
As a possible check, we consider here the small flow time limit $t \to 0$, where $g(\mu(t)) \to 0$. 
Then we focus on higher order in $A_0$ rather than $A$.
As mentioned above, we can give accurate coefficients up to $A_0^2=[g(\mu_0)^2/(4 \pi)^2]^2$
in rewriting $\tilde{\mO}_2^{\rm S}(t,x)$ and $\tilde{\mO}_4^{\rm S}(t,x)$ 
in terms of $\hat{\mO}^{R, {\rm S}}_2(x;\mu_0)$ and $\hat{\mO}^{R, {\rm S}}_4(x;\mu_0)$
by using the NLO $K$ matrix and the NLO $\zeta'^R$.
Explicitly, we obtain 
\begin{align}
\tilde{\mathcal{O}}_2^{\rm S}(t,x)
&=g(\mu_0)^2 \lt(\hat{\mathcal{O}}_2^{R, {\rm S}}(x;\mu_0)+\frac{6 C_F}{\beta_0} \hat{\mathcal{O}}_4^{R, {\rm S}}(x;\mu_0) \rt) \non
&\quad{}+\frac{g(\mu_0)^4}{(4 \pi)^2 \beta_0}  
\bigg[ \lt(\frac{34}{3} C_A^2-\frac{20}{3} C_A T_F-4 C_F T_F \rt)\hat{\mathcal{O}}_2^{R, {\rm S}}(x;\mu_0) \non
&\qquad{}\qquad{}\qquad{}
+\lt(\frac{97}{3} C_A C_F+3 C_F^2 - \frac{20}{3} C_F T_F \rt)\hat{\mathcal{O}}_4^{R, {\rm S}}(x;\mu_0) \bigg] ,
\end{align}
\begin{align}
\tilde{\mathcal{O}}_4^{\rm S}(t,x)
&=\hat{\mathcal{O}}_4^{R, {\rm S}}(x;\mu_0)+\frac{g(\mu_0)^2}{(4 \pi)^2} \frac{5}{3} T_F \lt( \hat{\mathcal{O}}_2^{R, {\rm S}}(x;\mu_0)+\frac{6 C_F}{\beta_0} \hat{\mathcal{O}}_4^{R, {\rm S}}(x;\mu_0)  \rt)  \non
&\quad{}+\lt[\frac{g(\mu_0)^2}{(4 \pi)^2} \rt]^2  
\frac{5 T_F}{3 \beta_0} \bigg[ \lt(\frac{34}{3} C_A^2-\frac{20}{3} C_A T_F-4 C_F T_F \rt) \hat{\mathcal{O}}_2^{R, {\rm S}}(x;\mu_0) \non
&\qquad{}\qquad{}\qquad{}\qquad{}\qquad{}
+\lt(\frac{97}{3} C_A C_F+3 C_F^2 - \frac{20}{3} C_F T_F \rt)\hat{\mathcal{O}}_4^{R, {\rm S}}(x;\mu_0) \bigg]  .
\end{align}
Here we just showed
the leading order contribution for small $g(\mu(t))$.
We can confirm the validity of these results as follows.
In the expression of $T^{\rm S}(x)$ in terms of the flowed operators, 
$T^{\rm S}(x)=c_2(t) \tilde{\mathcal{O}}_2^{\rm S}(t,x)+c'_4(t) \tilde{\mathcal{O}}_4^{\rm S}(t,x)$,
when one uses the above results to rewrite the flowed operators in terms of the unflowed operators
at $\mu=\mu_0$, the two-loop order expression of 
$T^{\rm S}(x)$ \eqref{tmunufulltr} with the renormalization scale $\mu=\mu_0$ 
can be correctly reproduced with the LO $c_2$, $c'_4$. (Again, since we consider the small $g(\mu(t))$ limit,
it is sufficient to use the LO $c_2$, $c'_4$, whose higher-order result just affects higher powers in $g(\mu(t))$.)

Now let us move on to the traceless parts and do a parallel analysis.
In studying the traceless parts, we have a complication that traceless operators to be considered
should be changed depending on perturbation order.
At LO, the traceless operators are given by
\be
\hat{\mathcal{O}}_{1, \mu \nu}^{R, {\rm TL}}
=\hat{\mathcal{O}}_{1, \mu \nu}^R-\frac{1}{4} \hat{\mathcal{O}}_{2, \mu \nu}^R , \label{LOTLrel1}
\ee
\be
\hat{\mathcal{O}}_{3, \mu \nu}^{R, {\rm TL}}
=\hat{\mathcal{O}}_{3, \mu \nu}^R-\frac{1}{2} \hat{\mathcal{O}}_{4, \mu \nu}^R ,  \label{LOTLrel3}
\ee
while at NLO they are given by [cf. \eqref{O1Rtr} and \eqref{O3Rtr}]
\be
\hat{\mathcal{O}}_{1, \mu \nu}^{R, {\rm TL}}
=\hat{\mathcal{O}}_{1, \mu \nu}^R-\frac{1}{4} \hat{\mathcal{O}}_{2, \mu \nu}^R
-A \lt(\frac{11}{24} C_A \hat{\mathcal{O}}_{2, \mu \nu}^R +\frac{7}{12} C_F \hat{\mathcal{O}}_{4, \mu \nu}^R\rt) , \label{NLOTLrel1}
\ee
\be
\hat{\mathcal{O}}_{3, \mu \nu}^{R, {\rm TL}}
=\hat{\mathcal{O}}_{3, \mu \nu}^R-\frac{1}{2} \hat{\mathcal{O}}_{4, \mu \nu}^R
-A \lt(-\frac{2}{3} T_F \hat{\mathcal{O}}_{2, \mu \nu}^R +\frac{2}{3} C_F \hat{\mathcal{O}}_{4, \mu \nu}^R\rt) . \label{NLOTLrel3}
\ee
When we consider traceless operators at N$^k$LO ($k$-loop), 
we need to know the anomalous dimension matrix for
the four operators $\hat{\mathcal{O}}_{1, \mu \nu}^{R}, \dots, \hat{\mathcal{O}}_{4, \mu \nu}^{R}$
at N$^k$LO ($(k+1)$-loop).
For instance at NLO, when we consider 
\be
\mu \frac{d}{d \mu} \hat{\mathcal{O}}_{1, \mu \nu}^{R, {\rm TL}}
=\mu \frac{d}{d \mu}  \lt[\hat{\mathcal{O}}_{1, \mu \nu}^R-\frac{1}{4} \hat{\mathcal{O}}_{2, \mu \nu}^R
-A \lt(\frac{11}{24} C_A \hat{\mathcal{O}}_{2, \mu \nu}^R +\frac{7}{12} C_F \hat{\mathcal{O}}_{4, \mu \nu}^R\rt) \rt] ,
\ee
the $\mathcal{O}(A)$ term inside the square brackets gives an $\mathcal{O}(A^2)$ term after $\mu d/(d \mu)$ is operated.
Then for consistency we need to know the NLO anomalous dimension matrix.

We define the anomalous dimension matrix for traceless operators as
\be
\mu \frac{d}{d \mu} 
\begin{pmatrix}
\hat{\mathcal{O}}_{1, \mu \nu}^{R,{\rm TL}} \\
\hat{\mathcal{O}}_{3, \mu \nu}^{R,{\rm TL}} 
\end{pmatrix}
=-\gamma^{\rm TL}  \begin{pmatrix}
\hat{\mathcal{O}}_{1, \mu \nu}^{R,{\rm TL}}  \\
\hat{\mathcal{O}}_{3, \mu \nu}^{R,{\rm TL}}  
\end{pmatrix} .
\ee
Writing $\gamma^{\rm TL}= \gamma_0^{\rm TL} A+\gamma_1^{\rm TL} A^2$, we have 
\be
\gamma^{\rm TL}_0=
\begin{pmatrix}
\frac{8 T_F}{3} & -\frac{4 C_F}{3} \\
-\frac{32 T_F}{3} &  \frac{16 C_F}{3} 
\end{pmatrix} ,
\ee
\be
\gamma^{\rm TL}_1=
\begin{pmatrix}
\frac{4}{27} (35 C_A T_F+74 C_F T_F) & -\frac{4}{27} (47 C_A C_F-14 C_F^2 -26 C_F T_F) \\
-\frac{16}{27} (35 C_A T_F+74 C_F T_F) & \frac{16}{27}  (47 C_A C_F-14 C_F^2 -26 C_F T_F) 
\end{pmatrix} .
\ee
The eigenvalues of $\gamma_0/ 2 \beta_0$ are $0$ and 
$\lambda \equiv \frac{\frac{8}{3} (2 C_F+T_F)}{ 2 \beta_0}=\frac{4 (2 C_F+T_F)}{11 C_A-4 T_F}$.
The LO $K$ matrix is given by
\be
K_{\rm LO}^{\rm TL}=\frac{1}{2 C_F+T_F}
\begin{pmatrix}
2 C_F +T_F \lt(\frac{A}{A_0} \rt)^{\lambda} &  \frac{C_F}{2} \lt(1-  \lt( \frac{A}{A_0} \rt)^{\lambda} \rt) \\
4 T_F\lt(1-  \lt( \frac{A}{A_0} \rt)^{\lambda} \rt) &  T_F+2 C_F  \lt( \frac{A}{A_0} \rt)^{\lambda}
\end{pmatrix} .
\ee
From Eq.~\eqref{eq:(G11)}, the NLO $K^{\rm TL}$ takes the form
\be
K^{\rm TL}_{\rm NLO}=K^{\rm TL}_{\rm LO}+({\text{linear comb. of }} \{A_0,  (A/A_0)^{\lambda} A_0,  A, (A/A_0)^{\lambda} A \}) .
\ee
The knowledge on the order of the NLO correction enables us to accurately estimate the parametrical error
of the LO calculation.
We have
\begin{align}
\zeta'^{R,{\rm TL}}_{\rm LO} K_{\rm LO}^{\rm TL}
&=(\{1\}+\mathcal{O}(A)) (\{1, (A/A_0)^{\lambda}\}+\mathcal{O}(A_0,  (A/A_0)^{\lambda} A_0,  A, (A/A_0)^{\lambda} A )) \non
&=\{1,  (A/A_0)^{\lambda}\}+\mathcal{O}(A_0,  (A/A_0)^{\lambda} A_0, \dots)  .
\end{align}
Then we can show the LO results of the traceless flowed operators with explicit parametric errors (and neglected higher-order terms):
\be
\frac{1}{g(\mu)^2}\tilde{\mathcal{O}}^{\rm TL}_{1,\mu \nu}(t,x)
=\frac{2 C_F}{2 C_F+T_F} \lt( \hat{\mathcal{O}}^{R,{\rm TL}} _{1,\mu \nu}(\mu_0)+ \frac{1}{4} \hat{\mathcal{O}}^{R,{\rm TL}} _{3,\mu \nu}(\mu_0) +\mathcal{O}(A_0) \rt)+\mathcal{O}((A/A_0)^{\lambda}) , \label{O1TLLO}
\ee
\be
\tilde{\mathcal{O}}^{\rm TL}_{3,\mu \nu}(t,x)
=\frac{4 T_F}{2 C_F+T_F} \lt( \hat{\mathcal{O}}^{R,{\rm TL}} _{1,\mu \nu}(\mu_0)+ \frac{1}{4} \hat{\mathcal{O}}^{R,{\rm TL}} _{3,\mu \nu}(\mu_0)+\mathcal{O}(A_0) \rt)+\mathcal{O}((A/A_0)^{\lambda}) . \label{O3TLLO}
\ee
These correspond to Eqs.~\eqref{O1TL} and \eqref{O3TL}.
Here we understand $\hat{\mathcal{O}}^{R,{\rm TL}} _{1,3 \mu \nu}(\mu_0)$ as the LO ones 
given by Eqs.~\eqref{LOTLrel1} and \eqref{LOTLrel3}.
They have $\mathcal{O}(A_0)$ errors and this is consistent with the above errors.

Calculating the NLO $K$ matrix, we can explicitly show higher-order results. 
Again we are interested in higher order in $g(\mu_0)$ 
and just show the leading contribution for small $g(\mu(t))$.
With the NLO $K$ matrix (and LO $\zeta'^R$ matrix, which is sufficient for the present purpose), we have
\be
\frac{1}{g(\mu(t))^2}\tilde{\mathcal{O}}^{\rm TL}_{1,\mu \nu}(t,x)
=\frac{2 C_F}{2 C_F+T_F} \lt( \hat{\mathcal{O}}^{R,{\rm TL}} _{1,\mu \nu}(x;\mu_0)+ \frac{1}{4} \hat{\mathcal{O}}^{R,{\rm TL}} _{3,\mu \nu}(x; \mu_0) \rt) ,
\ee
\be
\tilde{\mathcal{O}}^{\rm TL}_{3,\mu \nu}(t,x)
=\frac{4 T_F}{2 C_F+T_F} \lt( \hat{\mathcal{O}}^{R,{\rm TL}} _{1,\mu \nu}(x; \mu_0)+ \frac{1}{4} \hat{\mathcal{O}}^{R,{\rm TL}} _{3,\mu \nu}(x; \mu_0) \rt) .
\ee
Now $\hat{\mathcal{O}}^{R,{\rm TL}} _{1,3 \mu \nu}(\mu_0)$ are the ones at NLO [Eqs.~\eqref{NLOTLrel1} and \eqref{NLOTLrel3}].
Although there is no apparent difference from the LO calculations \eqref{O1TLLO} and \eqref{O3TLLO},
the $\mathcal{O}(A_0)$ terms are now fixed and turn out to be zero.
(The expected error is now $\mathcal{O}(A_0^2)$). 
One can confirm that the one-loop expression of $T^{\rm TL}$ \eqref{tmunufulltl}
with the renormalization scale $\mu=\mu_0$ can be
correctly reproduced from $T_{\mu \nu}^{\rm TL}(x)=c_1(t) \Tilde{\mathcal{O}}^{\rm TL}_{1, \mu \nu}(t, x)+c_3(t) \Tilde{\mathcal{O}}^{\rm TL}_{3, \mu \nu}(t,x)$ with the LO $c_1$, $c_3$ and the above results. 

\section{Results for thermodynamics quantities with other $t \to 0$ extrapolation functions}
\label{app:H}
In this appendix, for reference, we present the results obtained with other $t\to 0$ extrapolation functions.
Here we use a linear function in $t$ for the entropy density and trace anomaly,
and use a linear function with the anomalous dimension, $g(\mu(t))^{\lambda_3} t$, given in Eq.~\eqref{TLdim6}, for the entropy density. 

\begin{figure}[tbhp]
\begin{minipage}{0.5\hsize}
\begin{center}
\vspace{1mm}
\includegraphics[width=7.25cm]{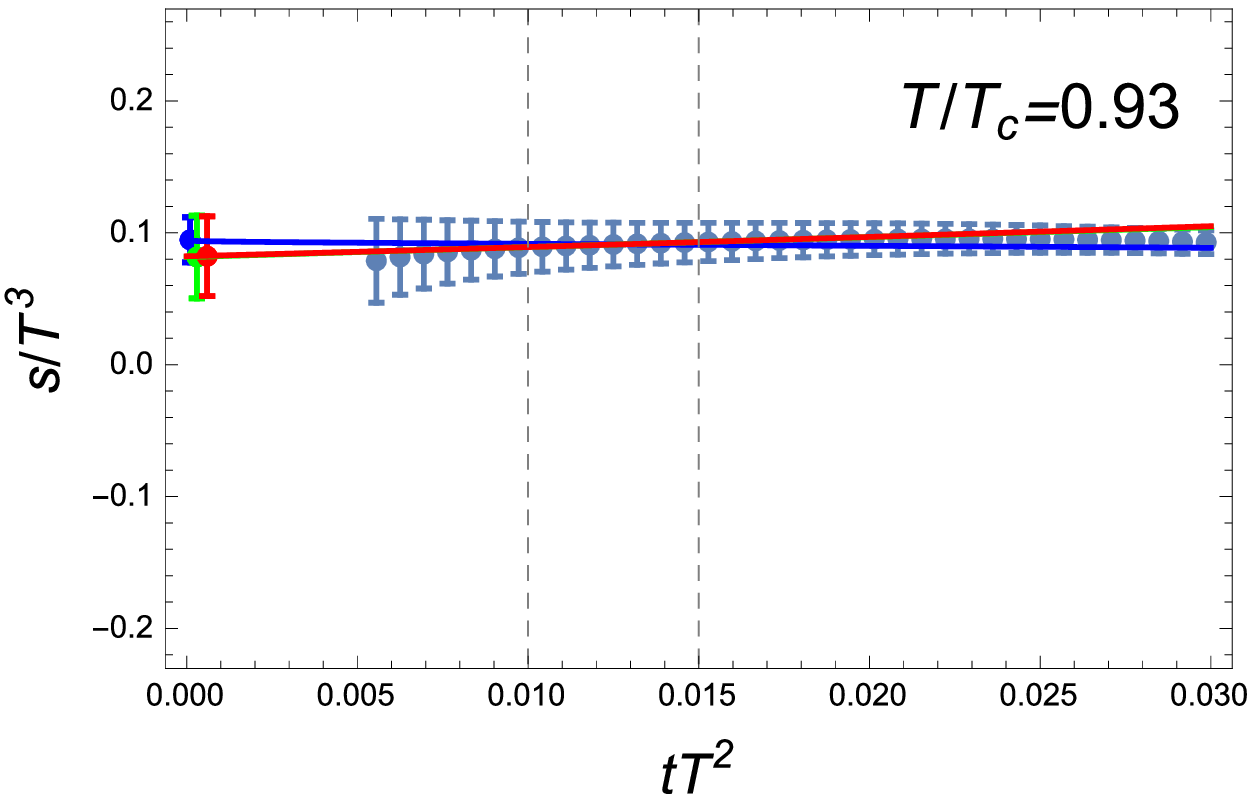}
\end{center}
\end{minipage}
\begin{minipage}{0.5\hsize}
\begin{center}
\includegraphics[width=7cm]{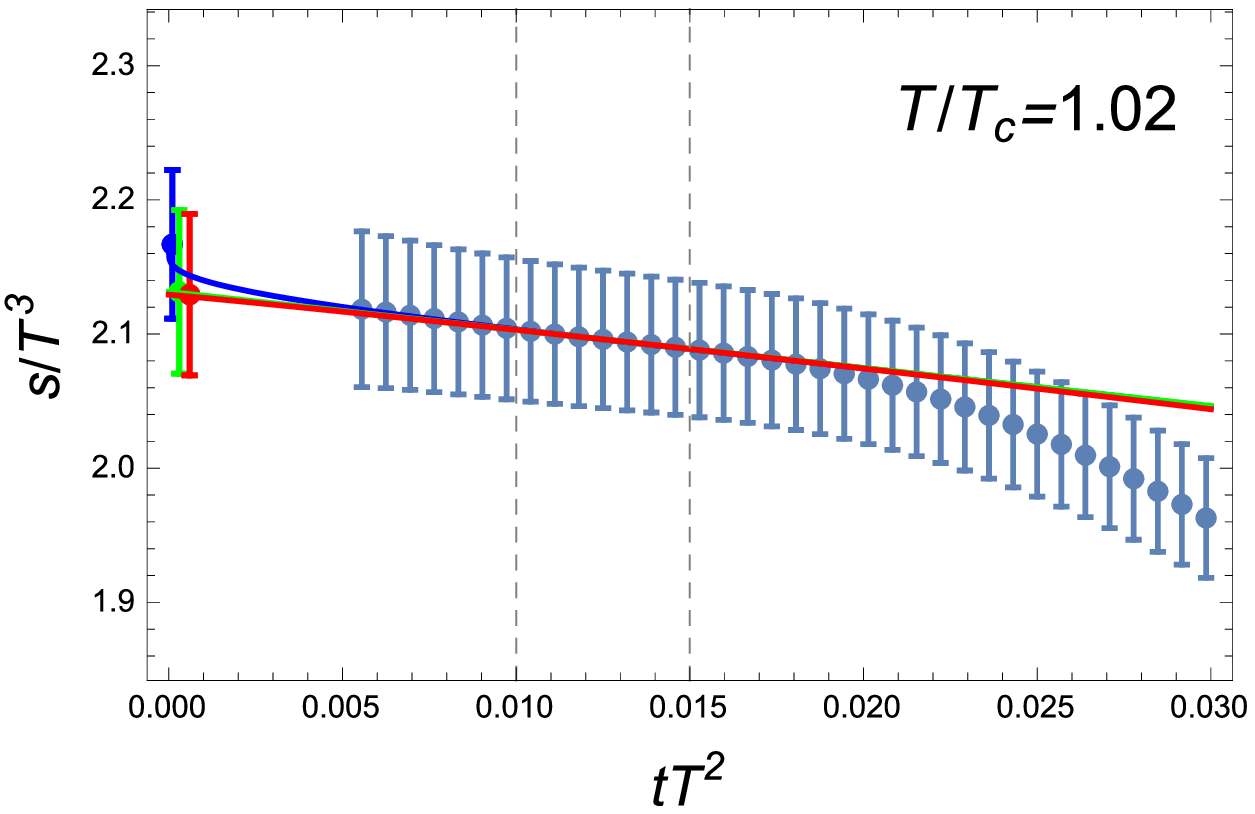}
\end{center}
\end{minipage}
\begin{minipage}{0.5\hsize}
\begin{center}
\vspace{1mm}
\includegraphics[width=7.05cm]{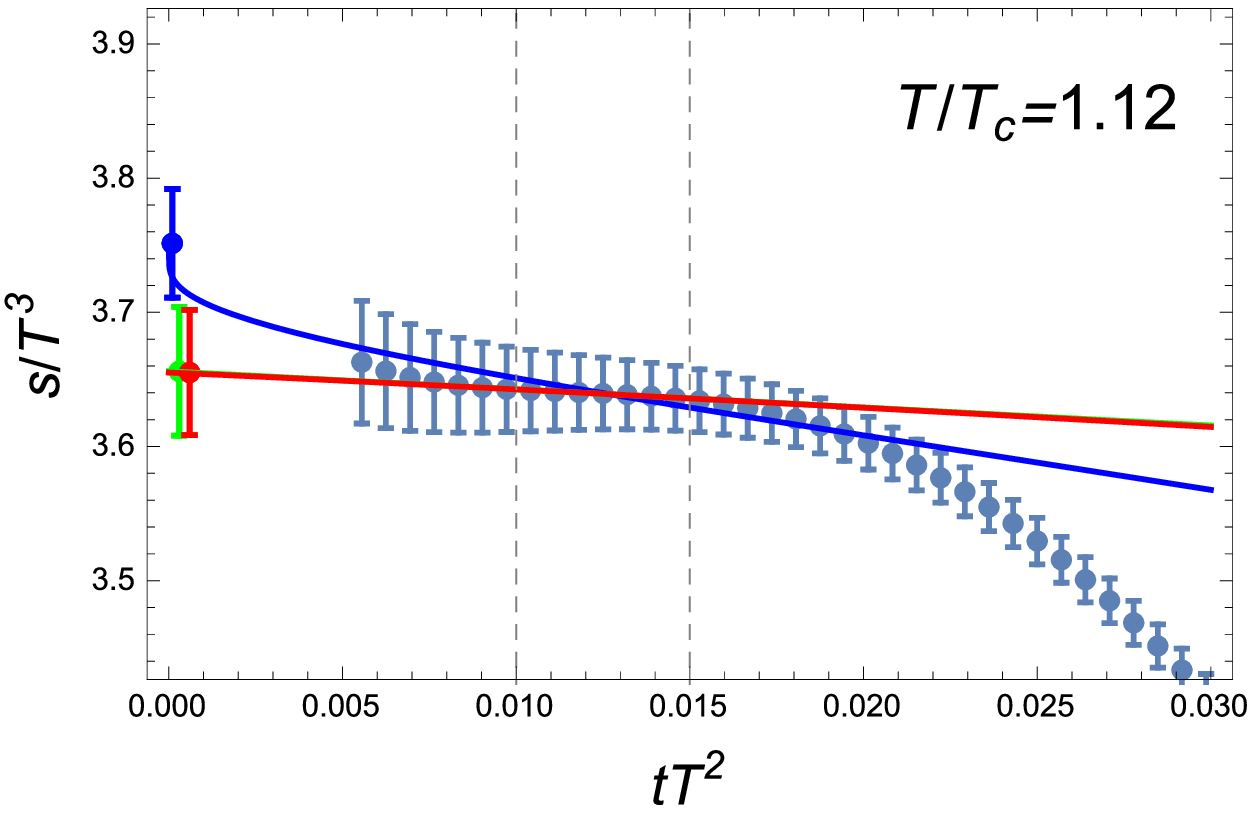}
\end{center}
\end{minipage}
\begin{minipage}{0.5\hsize}
\begin{center}
\includegraphics[width=7cm]{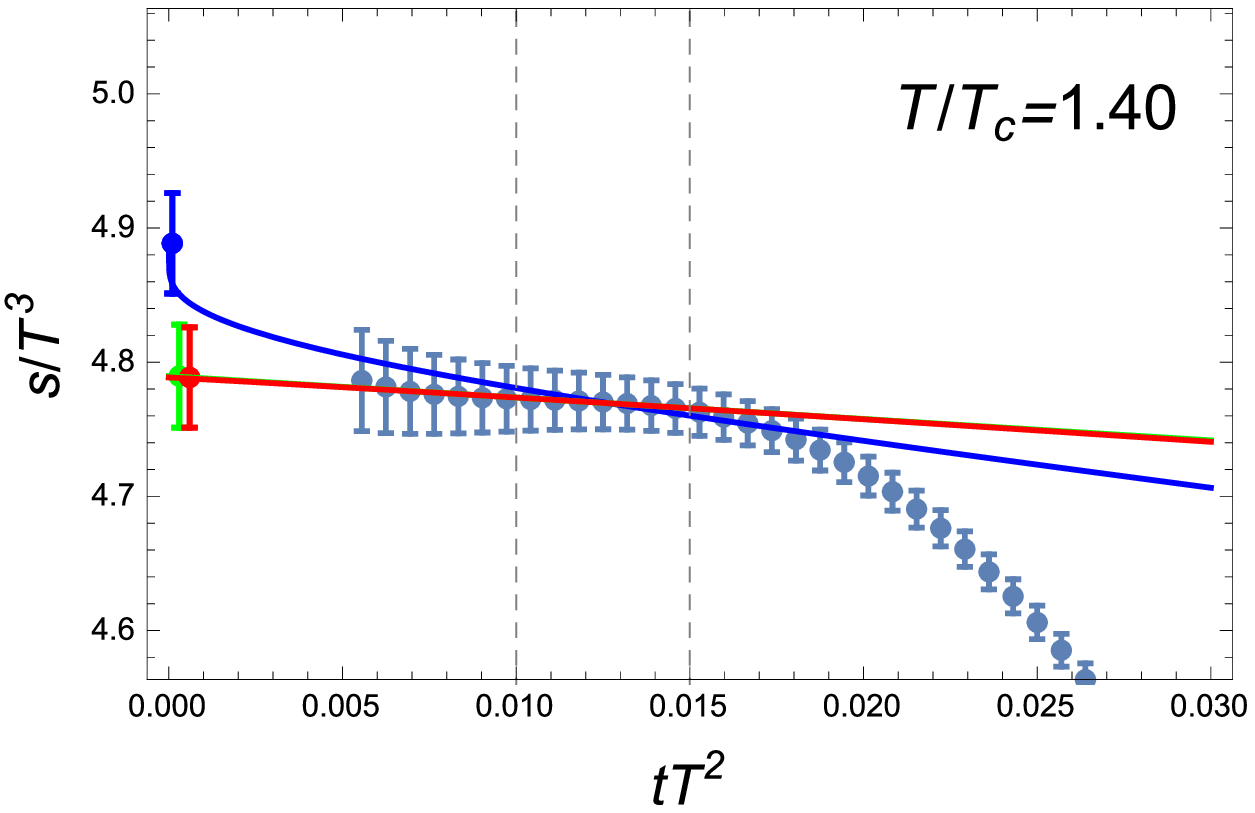}
\end{center}
\end{minipage}
\begin{minipage}{0.5\hsize}
\begin{center}
\vspace{1mm}
\includegraphics[width=7.05cm]{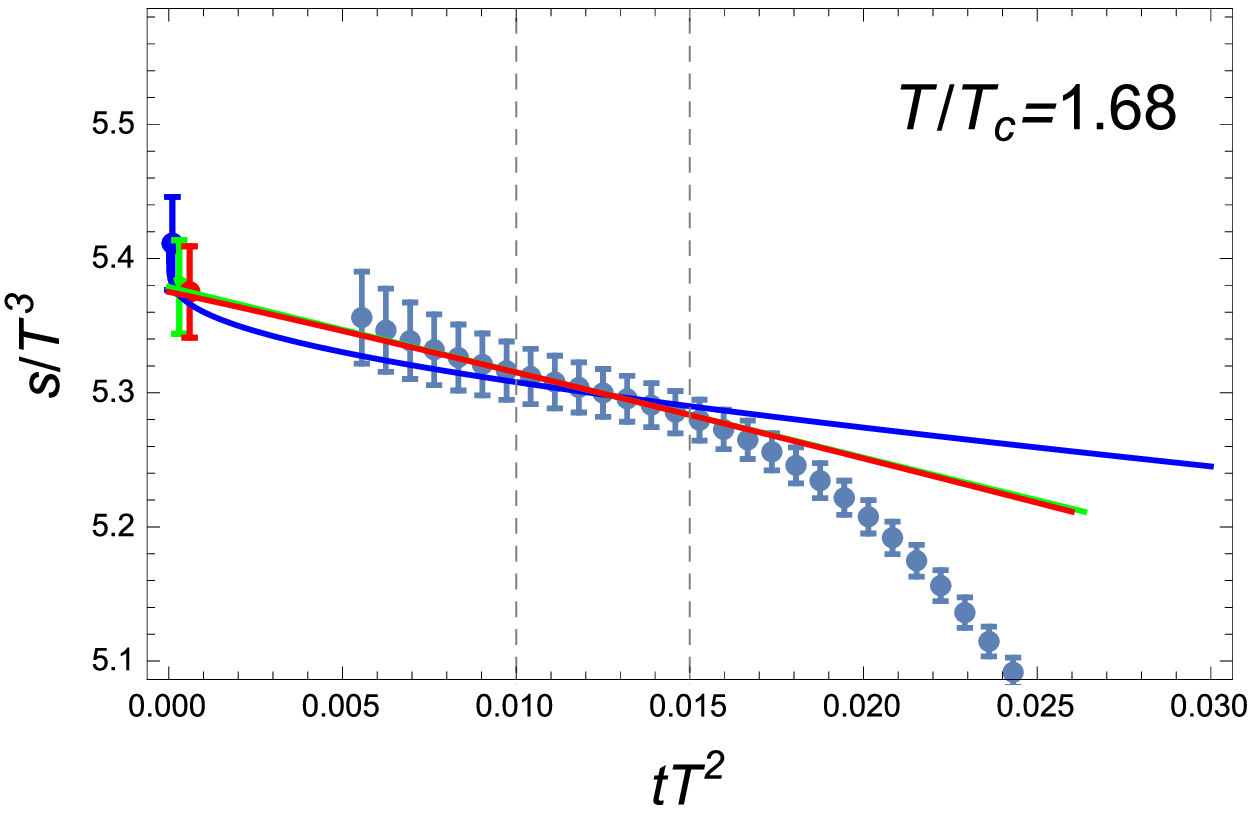}
\end{center}
\end{minipage}
\begin{minipage}{0.5\hsize}
\begin{center}
\includegraphics[width=7cm]{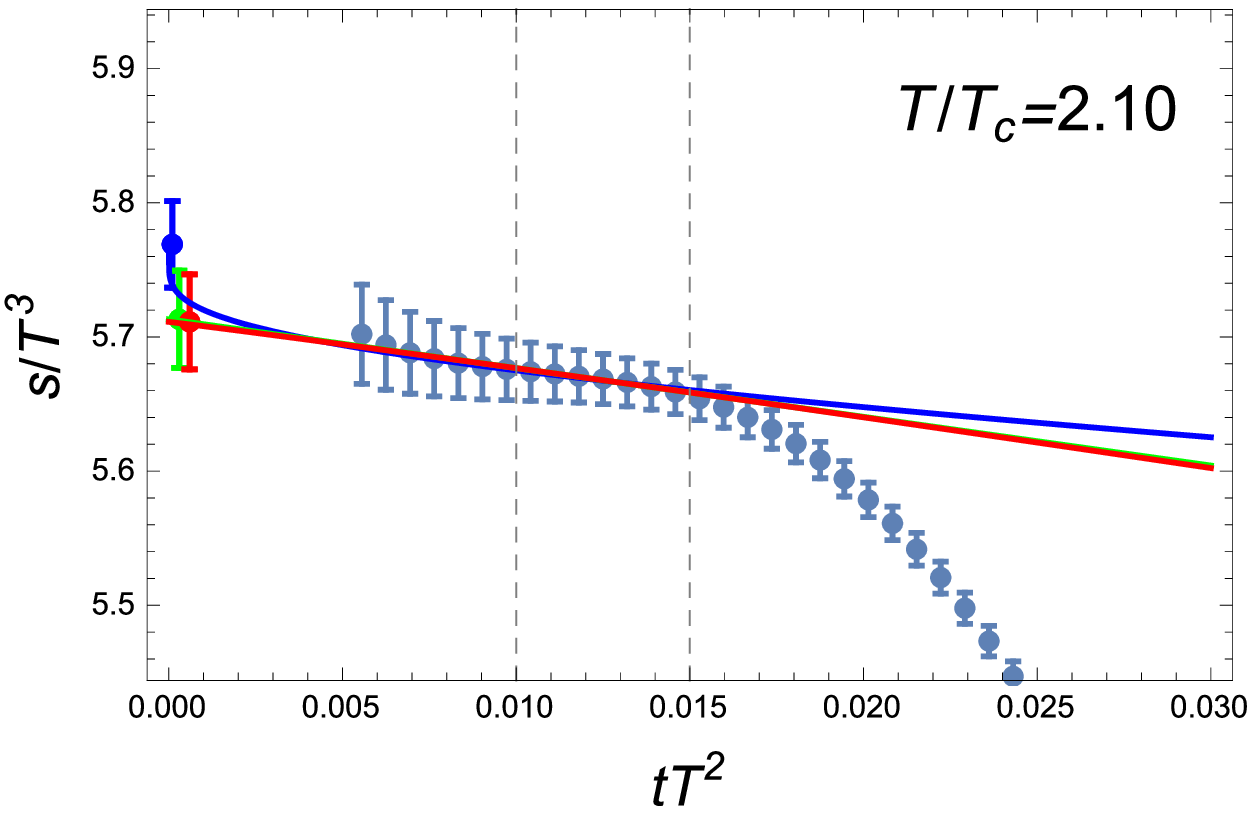}
\end{center}
\end{minipage}
\begin{minipage}{0.5\hsize}
\begin{center}
\vspace{1mm}
\includegraphics[width=7.05cm]{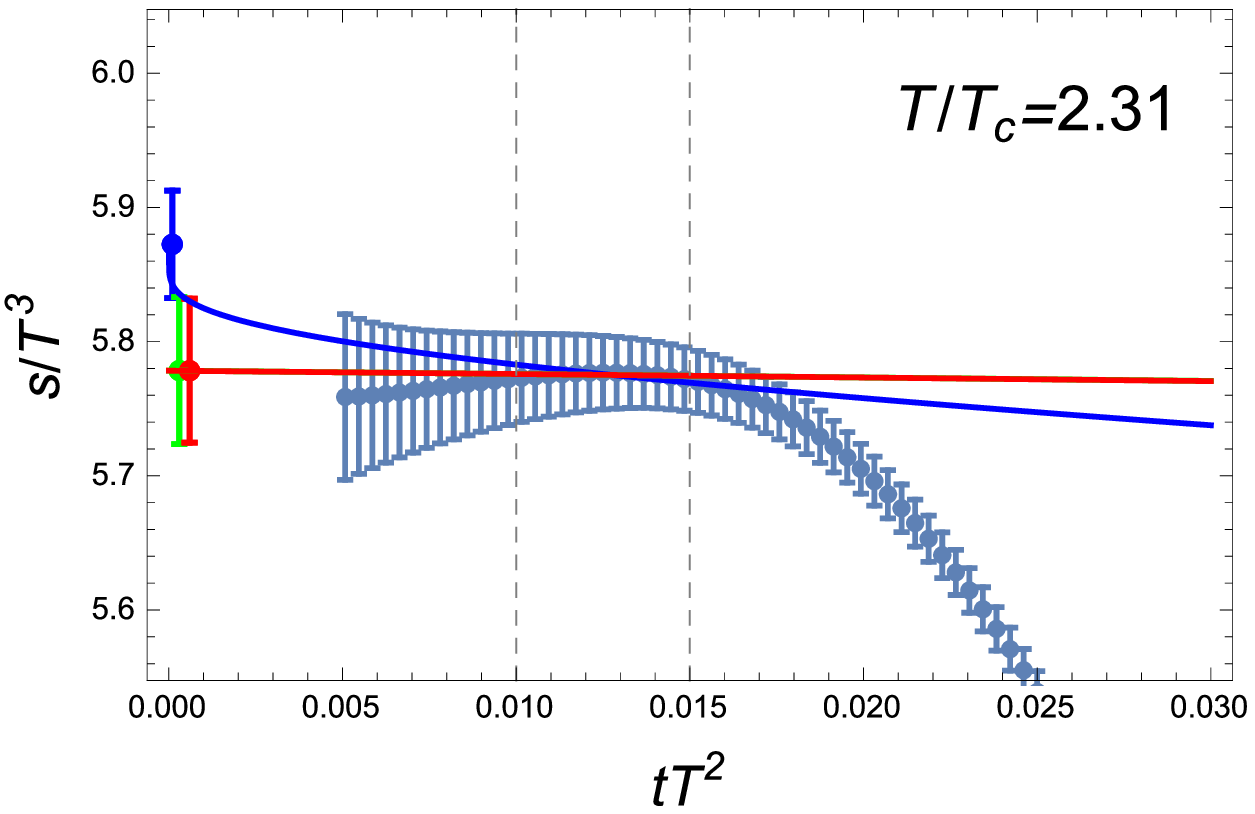}
\end{center}
\end{minipage}
\begin{minipage}{0.5\hsize}
\begin{center}
\includegraphics[width=7cm]{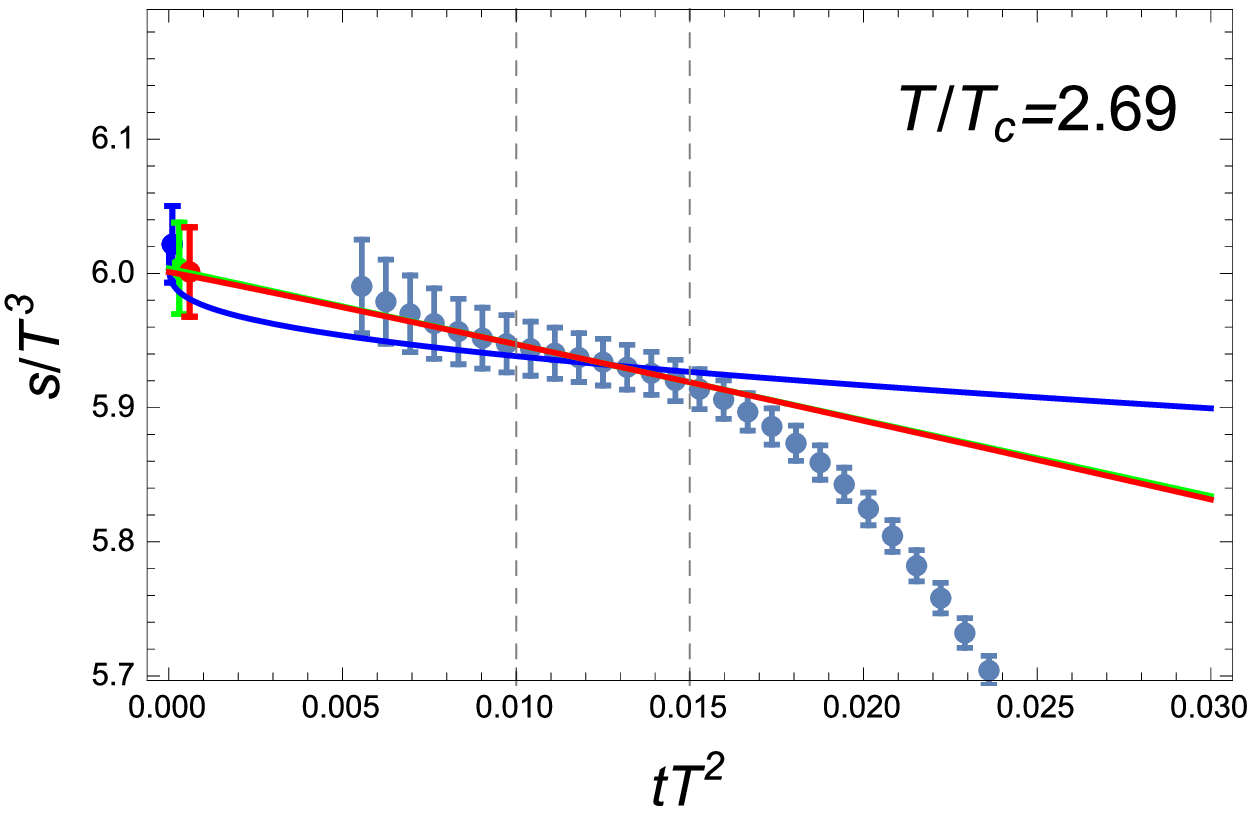}
\end{center}
\end{minipage}
\caption{NLO analysis of the entropy density with different $t \to 0$ extrapolations. 
The blue lines correspond to our main extrapolation function Eq.~\eqref{resquenchTL},
green ones to linear functions, and red ones to linear functions with the anomalous dimension [Eq.~\eqref{TLdim6}].
The gray dashed lines show the fit range used in the $t \to 0$ extrapolation.}
\label{fig:NLOentropyOtherExtrpls}
\end{figure}

The results are summarized in Tables~\ref{tab:OtherExtrplsforEntropy} and \ref{tab:OtherExtrplsforTA}.
One can see that the perturbative extrapolation functions, Eqs.~\eqref{resquenchTL} and \eqref{resquenchS}, 
give smaller differences in the NLO and N$^2$LO results than the other linear-type functions.
We can also see from Table~\ref{tab:OtherExtrplsforEntropy} that the numerical impact of
the inclusion of the  anomalous dimension of dimension-six operators is not significant.
This can also be seen in Fig.~\ref{fig:NLOentropyOtherExtrpls},
where we show the $t \to 0$ extrapolation analyses with different extrapolation functions. 
\begin{table}[htbp]
\begin{center}
\begin{tabular}{c|c|c|c}
\hline\hline
   $s/T^3$ (NLO) \\
\hline
 $T/T_c$ & Perturbative function & Linear & Linear with anomalous dim.  \\ \hline
0.93 & $0.095 (17)$ & $0.082(31)$ & $0.082(30)$ \\
1.02 & $2.167(55)$ & $2.131(61)$  & $2.129(60)$  \\ 
1.12 & $3.751(41)$ & $3.656(48)$ & $3.655(47)$  \\
1.40 & $4.889(37)$ & $4.790(38)$ & $4.789(37)$ \\
1.68 & $5.411(35)$ & $5.379(35)$ & $5.375(34)$ \\
2.10 & $5.769(32)$ & $5.713(36)$ & $5.711(35)$\\
2.31 & $5.873(40)$ & $5.778(55)$ & $5.778(54)$ \\
2.69 & $6.022(29)$ & $6.004(34)$ & $6.001(33)$\\
\hline 
$s/T^3$ (N$^2$LO) \\ \hline
$T/T_c$ & Perturbative function & Linear & Linear with anomalous dim.  \\ \hline
0.93 & $0.095 (17)$ & $0.083(32)$ & $0.084(31)$ \\
1.02 & $2.164(55)$ & $2.167(62)$  & $2.168(61)$  \\ 
1.12 & $3.748(37)$ & $3.715(49)$ & $3.717(48)$  \\
1.40 & $4.885(32)$ & $4.860(39)$ & $4.862(38)$ \\
1.68 & $5.408(28)$ & $5.451(35)$ & $5.450(35)$ \\
2.10 & $5.766(25)$ & $5.782(37)$ & $5.782(36)$\\
2.31 & $5.870(34)$ & $5.844(56)$ & $5.846(54)$ \\
2.69 & $6.020(22)$ & $6.068(35)$ & $6.066(34)$\\
\hline
\end{tabular}
\caption{Comparison of the results for the entropy density obtained with
different $t \to 0$ extrapolation functions. 
``Perturbative function" means Eq.~\eqref{resquenchTL} with $k=1$ or $k=2$,
``Linear" means a linear function in $t$, and ``Linear with anomalous dim." means Eq.~\eqref{TLdim6}.
We only show statistical errors.}
\label{tab:OtherExtrplsforEntropy}
\end{center}
\end{table}

\begin{table}[htbp]
\begin{center}
\begin{tabular}{c|c|c}
\hline\hline
   $\Delta/T^4$ (NLO) \\
\hline
 $T/T_c$ & Perturbative function & Linear  \\ \hline
0.93 & $0.081 (20)$ & $0.066(32)$  \\
1.02 & $1.942(53)$ & $1.946(54)$    \\ 
1.12 & $2.562(35)$ & $2.567(32)$  \\
1.40 & $1.770(22)$ & $1.779(22)$ \\
1.68 & $1.180(14)$ & $1.203(18)$  \\
\hline 
$\Delta/T^4$ (N$^2$LO) \\ \hline
$T/T_c$ & Perturbative function & Linear \\ \hline
0.93 & $0.081(20)$ & $0.066(31)$  \\
1.02 & $1.941(52)$ & $1.933(54)$    \\ 
1.12 & $2.561(30)$ & $2.550(31)$  \\
1.40 & $1.770(18)$ & $1.769(22)$ \\
1.68 & $1.180(12)$ & $1.196(18)$  \\
\hline
\end{tabular}
\caption{Comparison of the results for the trace anomaly obtained with
different $t \to 0$ extrapolation functions. ``Perturbative function" means
Eq.~\eqref{resquenchS} with $k=1$ or $k=2$, and
``Linear" means a linear function in $t$.
We only show statistical errors.}
\label{tab:OtherExtrplsforTA}
\end{center}
\end{table}

\clearpage


\begin{thebibliography}{00}

\bibitem{Suzuki:2016ytc}
H.~Suzuki, ``{Energy--momentum tensor on the lattice: recent developments},''
  \href{http://dx.doi.org/10.22323/1.256.0002}{{\em PoS} {\bfseries
  LATTICE2016} (2017) 002}, \href{http://arxiv.org/abs/1612.00210}{{\ttfamily
  arXiv:1612.00210 [hep-lat]}}.

\bibitem{Suzuki:2013gza}
H.~Suzuki, ``{Energy--momentum tensor from the Yang--Mills gradient flow},''
  \href{http://dx.doi.org/10.1093/ptep/ptt059, 10.1093/ptep/ptv094}{{\em PTEP}
  {\bfseries 2013} (2013) 083B03},
  \href{http://arxiv.org/abs/1304.0533}{{\ttfamily arXiv:1304.0533 [hep-lat]}}.
[Erratum: PTEP2015,079201(2015)].

\bibitem{Makino:2014taa}
H.~Makino and H.~Suzuki, ``{Lattice energy--momentum tensor from the
  Yang--Mills gradient flow---inclusion of fermion fields},''
  \href{http://dx.doi.org/10.1093/ptep/ptu070, 10.1093/ptep/ptv095}{{\em PTEP}
  {\bfseries 2014} (2014) 063B02},
  \href{http://arxiv.org/abs/1403.4772}{{\ttfamily arXiv:1403.4772 [hep-lat]}}.
[Erratum: PTEP2015,079202(2015)].

\bibitem{Narayanan:2006rf}
R.~Narayanan and H.~Neuberger, ``{Infinite N phase transitions in continuum
  Wilson loop operators},''
  \href{http://dx.doi.org/10.1088/1126-6708/2006/03/064}{{\em JHEP} {\bfseries
  03} (2006) 064}, \href{http://arxiv.org/abs/hep-th/0601210}{{\ttfamily
  arXiv:hep-th/0601210}}.

\bibitem{Luscher:2009eq}
M.~L\"{u}scher, ``{Trivializing maps, the Wilson flow and the HMC algorithm},''
  \href{http://dx.doi.org/10.1007/s00220-009-0953-7}{{\em Commun. Math. Phys.}
  {\bfseries 293} (2010) 899--919},
  \href{http://arxiv.org/abs/0907.5491}{{\ttfamily arXiv:0907.5491 [hep-lat]}}.

\bibitem{Luscher:2010iy}
M.~L{\"u}scher, ``{Properties and uses of the Wilson flow in lattice QCD},''
  \href{http://dx.doi.org/10.1007/JHEP08(2010)071,
  10.1007/JHEP03(2014)092}{{\em JHEP} {\bfseries 08} (2010) 071},
  \href{http://arxiv.org/abs/1006.4518}{{\ttfamily arXiv:1006.4518 [hep-lat]}}.
[Erratum: JHEP03,092(2014)].

\bibitem{Luscher:2011bx}
M.~L\"{u}scher and P.~Weisz, ``{Perturbative analysis of the gradient flow in
  non-abelian gauge theories},''
  \href{http://dx.doi.org/10.1007/JHEP02(2011)051}{{\em JHEP} {\bfseries 02}
  (2011) 051},
\href{http://arxiv.org/abs/1101.0963}{{\ttfamily arXiv:1101.0963 [hep-th]}}.

\bibitem{Luscher:2013cpa}
M.~L\"{u}scher, ``{Chiral symmetry and the Yang--Mills gradient flow},''
  \href{http://dx.doi.org/10.1007/JHEP04(2013)123}{{\em JHEP} {\bfseries 04}
  (2013) 123}, \href{http://arxiv.org/abs/1302.5246}{{\ttfamily arXiv:1302.5246
  [hep-lat]}}.

\bibitem{Harlander:2018zpi}
R.~V. Harlander, Y.~Kluth, and F.~Lange, ``{The two-loop energy--momentum
  tensor within the gradient-flow formalism},''
  \href{http://dx.doi.org/10.1140/epjc/s10052-018-6415-7}{{\em Eur. Phys. J.}
  {\bfseries C78} no.~11, (2018) 944},
\href{http://arxiv.org/abs/1808.09837}{{\ttfamily arXiv:1808.09837 [hep-lat]}}.

\bibitem{Artz:2019bpr}
J.~Artz, R.~V. Harlander, F.~Lange, T.~Neumann, and M.~Prausa, ``{Results and
  techniques for higher order calculations within the gradient-flow
  formalism},'' \href{http://dx.doi.org/10.1007/JHEP06(2019)121}{{\em JHEP}
  {\bfseries 06} (2019) 121}, \href{http://arxiv.org/abs/1905.00882}{{\ttfamily
  arXiv:1905.00882 [hep-lat]}}. [Erratum: JHEP 10, 032 (2019)].
  
\bibitem{Asakawa:2013laa}
{\bfseries FlowQCD}, M.~Asakawa, T.~Hatsuda, E.~Itou, M.~Kitazawa, and
  H.~Suzuki, ``{Thermodynamics of SU(3) gauge theory from gradient flow on the
  lattice},'' \href{http://dx.doi.org/10.1103/PhysRevD.90.011501}{{\em Phys.
  Rev. D} {\bfseries 90} no.~1, (2014) 011501},
  \href{http://arxiv.org/abs/1312.7492}{{\ttfamily arXiv:1312.7492 [hep-lat]}}.
  [Erratum: Phys.Rev.D 92, 059902 (2015)].

\bibitem{Taniguchi:2016ofw}
Y.~Taniguchi, S.~Ejiri, R.~Iwami, K.~Kanaya, M.~Kitazawa, H.~Suzuki, T.~Umeda,
  and N.~Wakabayashi, ``{Exploring $N_{f}$ = 2+1 QCD thermodynamics from the
  gradient flow},'' \href{http://dx.doi.org/10.1103/PhysRevD.96.014509}{{\em
  Phys. Rev. D} {\bfseries 96} no.~1, (2017) 014509},
  \href{http://arxiv.org/abs/1609.01417}{{\ttfamily arXiv:1609.01417
  [hep-lat]}}. [Erratum: Phys.Rev.D 99, 059904 (2019)].

\bibitem{Kitazawa:2016dsl}
M.~Kitazawa, T.~Iritani, M.~Asakawa, T.~Hatsuda, and H.~Suzuki, ``{Equation of
  State for SU(3) Gauge Theory via the Energy--Momentum Tensor under Gradient
  Flow},'' \href{http://dx.doi.org/10.1103/PhysRevD.94.114512}{{\em Phys. Rev.
  D} {\bfseries 94} no.~11, (2016) 114512},
  \href{http://arxiv.org/abs/1610.07810}{{\ttfamily arXiv:1610.07810
  [hep-lat]}}.

\bibitem{Ejiri:2017wgd}
S.~Ejiri, R.~Iwami, M.~Shirogane, N.~Wakabayashi, K.~Kanaya, M.~Kitazawa,
  H.~Suzuki, Y.~Taniguchi, and T.~Umeda, ``{Determination of latent heat at the
  finite temperature phase transition of SU(3) gauge theory},''
  \href{http://dx.doi.org/10.22323/1.256.0058}{{\em PoS} {\bfseries
  LATTICE2016} (2017) 058}, \href{http://arxiv.org/abs/1701.08570}{{\ttfamily
  arXiv:1701.08570 [hep-lat]}}.

\bibitem{Kitazawa:2017qab}
M.~Kitazawa, T.~Iritani, M.~Asakawa, and T.~Hatsuda, ``{Correlations of the
  energy--momentum tensor via gradient flow in SU(3) Yang-Mills theory at finite
  temperature},'' \href{http://dx.doi.org/10.1103/PhysRevD.96.111502}{{\em
  Phys. Rev. D} {\bfseries 96} no.~11, (2017) 111502},
  \href{http://arxiv.org/abs/1708.01415}{{\ttfamily arXiv:1708.01415
  [hep-lat]}}.

\bibitem{Kanaya:2017cpp}
{\bfseries WHOT-QCD}, K.~Kanaya, S.~Ejiri, R.~Iwami, M.~Kitazawa, H.~Suzuki,
  Y.~Taniguchi, and T.~Umeda, ``{Equation of state in (2+1)-flavor QCD at
  physical point with improved Wilson fermion action using gradient flow},''
  \href{http://dx.doi.org/10.1051/epjconf/201817507023}{{\em EPJ Web Conf.}
  {\bfseries 175} (2018) 07023},
  \href{http://arxiv.org/abs/1710.10015}{{\ttfamily arXiv:1710.10015
  [hep-lat]}}.

\bibitem{Taniguchi:2017ibr}
{\bfseries WHOT-QCD}, Y.~Taniguchi, S.~Ejiri, K.~Kanaya, M.~Kitazawa,
  A.~Suzuki, H.~Suzuki, and T.~Umeda, ``{Energy--momentum tensor correlation
  function in $N_f = 2 + 1$ full QCD at finite temperature},''
  \href{http://dx.doi.org/10.1051/epjconf/201817507013}{{\em EPJ Web Conf.}
  {\bfseries 175} (2018) 07013},
  \href{http://arxiv.org/abs/1711.02262}{{\ttfamily arXiv:1711.02262
  [hep-lat]}}.

\bibitem{Yanagihara:2018qqg}
R.~Yanagihara, T.~Iritani, M.~Kitazawa, M.~Asakawa, and T.~Hatsuda,
  ``{Distribution of Stress Tensor around Static Quark--Anti-Quark from
  Yang-Mills Gradient Flow},''
  \href{http://dx.doi.org/10.1016/j.physletb.2018.09.067}{{\em Phys. Lett. B}
  {\bfseries 789} (2019) 210--214},
  \href{http://arxiv.org/abs/1803.05656}{{\ttfamily arXiv:1803.05656
  [hep-lat]}}.

\bibitem{Hirakida:2018uoy}
T.~Hirakida, E.~Itou, and H.~Kouno, ``{Thermodynamics for pure SU(2) gauge
  theory using gradient flow},''
  \href{http://dx.doi.org/10.1093/ptep/ptz003}{{\em PTEP} {\bfseries 2019}
  no.~3, (2019) 033B01}, \href{http://arxiv.org/abs/1805.07106}{{\ttfamily
  arXiv:1805.07106 [hep-lat]}}.

\bibitem{Shirogane:2018zbp}
M.~Shirogane, S.~Ejiri, R.~Iwami, K.~Kanaya, M.~Kitazawa, H.~Suzuki,
  Y.~Taniguchi, and T.~Umeda, ``{Equation of state near the first order phase
  transition point of SU(3) gauge theory using gradient flow},''
  \href{http://dx.doi.org/10.22323/1.334.0164}{{\em PoS} {\bfseries
  LATTICE2018} (2018) 164}, \href{http://arxiv.org/abs/1811.04220}{{\ttfamily
  arXiv:1811.04220 [hep-lat]}}.

\bibitem{Iritani:2018idk}
T.~Iritani, M.~Kitazawa, H.~Suzuki, and H.~Takaura, ``{Thermodynamics in
  quenched QCD: energy\textendash{}momentum tensor with two-loop order
  coefficients in the gradient-flow formalism},''
  \href{http://dx.doi.org/10.1093/ptep/ptz001}{{\em PTEP} {\bfseries 2019}
  no.~2, (2019) 023B02}, \href{http://arxiv.org/abs/1812.06444}{{\ttfamily
  arXiv:1812.06444 [hep-lat]}}.

\bibitem{Taniguchi:2019eid}
Y.~Taniguchi, A.~Baba, A.~Suzuki, S.~Ejiri, K.~Kanaya, M.~Kitazawa, T.~Shimojo,
  H.~Suzuki, and T.~Umeda, ``{Study of energy--momentum tensor correlation
  function in $N_f=2+1$ full QCD for QGP viscosities},''
  \href{http://dx.doi.org/10.22323/1.334.0166}{{\em PoS} {\bfseries
  LATTICE2018} (2019) 166}, \href{http://arxiv.org/abs/1901.01666}{{\ttfamily
  arXiv:1901.01666 [hep-lat]}}.

\bibitem{Kitazawa:2019otp}
M.~Kitazawa, S.~Mogliacci, I.~Kolb\'e, and W.~A. Horowitz, ``{Anisotropic
  pressure induced by finite-size effects in SU(3) Yang-Mills theory},''
  \href{http://dx.doi.org/10.1103/PhysRevD.99.094507}{{\em Phys. Rev. D}
  {\bfseries 99} no.~9, (2019) 094507},
  \href{http://arxiv.org/abs/1904.00241}{{\ttfamily arXiv:1904.00241
  [hep-lat]}}.

\bibitem{Kanaya:2019okb}
K.~Kanaya, A.~Baba, A.~Suzuki, S.~Ejiri, M.~Kitazawa, H.~Suzuki, Y.~Taniguchi,
  and T.~Umeda, ``{Study of 2+1 flavor finite-temperature QCD using improved
  Wilson quarks at the physical point with the gradient flow},''
  \href{http://dx.doi.org/10.22323/1.363.0088}{{\em PoS} {\bfseries
  LATTICE2019} (2019) 088}, \href{http://arxiv.org/abs/1910.13036}{{\ttfamily
  arXiv:1910.13036 [hep-lat]}}.

\bibitem{Taniguchi:2020mgg}
{\bfseries WHOT-QCD}, Y.~Taniguchi, S.~Ejiri, K.~Kanaya, M.~Kitazawa,
  H.~Suzuki, and T.~Umeda, ``{$N_f$ = 2+1 QCD thermodynamics with gradient flow
  using two-loop matching coefficients},''
  \href{http://dx.doi.org/10.1103/PhysRevD.102.014510}{{\em Phys. Rev. D}
  {\bfseries 102} no.~1, (2020) 014510},
  \href{http://arxiv.org/abs/2005.00251}{{\ttfamily arXiv:2005.00251
  [hep-lat]}}. [Erratum: Phys.Rev.D 102, 059903 (2020)].

\bibitem{Yanagihara:2020tvs}
R.~Yanagihara, M.~Kitazawa, M.~Asakawa, and T.~Hatsuda, ``{Distribution of
  Energy--Momentum Tensor around a Static Quark in the Deconfined Phase of SU(3)
  Yang-Mills Theory},''
  \href{http://dx.doi.org/10.1103/PhysRevD.102.114522}{{\em Phys. Rev. D}
  {\bfseries 102} no.~11, (2020) 114522},
  \href{http://arxiv.org/abs/2010.13465}{{\ttfamily arXiv:2010.13465
  [hep-lat]}}.

\bibitem{Shirogane:2020muc}
M.~Shirogane, S.~Ejiri, R.~Iwami, K.~Kanaya, M.~Kitazawa, H.~Suzuki, Y.~Taniguchi and T.~Umeda,
``{Latent heat and pressure gap at the first-order deconfining phase transition of SU(3) Yang-Mills theory using the small flow-time expansion method},''
\href{https://doi.org/10.1093/ptep/ptaa184}{{\em PTEP}
  {\bfseries 2021} (2021) 013B08},
  \href{http://arxiv.org/abs/2011.10292}{{\ttfamily arXiv:2011.10292 [hep-lat]}}.

\bibitem{Kim:2015ywa}
H.~Kim and S.~H. Lee, ``{Renormalization of dimension 6 gluon operators},''
  \href{http://dx.doi.org/10.1016/j.physletb.2015.07.028}{{\em Phys. Lett. B}
  {\bfseries 748} (2015) 352--355},
  \href{http://arxiv.org/abs/1503.02280}{{\ttfamily arXiv:1503.02280
  [hep-ph]}}.

\bibitem{Boyd:1996bx}
G.~Boyd, J.~Engels, F.~Karsch, E.~Laermann, C.~Legeland, M.~Lutgemeier, and
  B.~Petersson, ``{Thermodynamics of SU(3) lattice gauge theory},''
  \href{http://dx.doi.org/10.1016/0550-3213(96)00170-8}{{\em Nucl. Phys. B}
  {\bfseries 469} (1996) 419--444},
  \href{http://arxiv.org/abs/hep-lat/9602007}{{\ttfamily
  arXiv:hep-lat/9602007}}.

\bibitem{Okamoto:1999hi}
{\bfseries CP-PACS}, M.~Okamoto {\em et~al.}, ``{Equation of state for pure
  SU(3) gauge theory with renormalization group improved action},''
  \href{http://dx.doi.org/10.1103/PhysRevD.60.094510}{{\em Phys. Rev. D}
  {\bfseries 60} (1999) 094510},
  \href{http://arxiv.org/abs/hep-lat/9905005}{{\ttfamily
  arXiv:hep-lat/9905005}}.

\bibitem{Borsanyi:2012ve}
S.~Borsanyi, G.~Endrodi, Z.~Fodor, S.~D. Katz, and K.~K. Szabo, ``{Precision
  SU(3) lattice thermodynamics for a large temperature range},''
  \href{http://dx.doi.org/10.1007/JHEP07(2012)056}{{\em JHEP} {\bfseries 07}
  (2012) 056}, \href{http://arxiv.org/abs/1204.6184}{{\ttfamily arXiv:1204.6184
  [hep-lat]}}.

\bibitem{Borsanyi:2013bia}
S.~Borsanyi, Z.~Fodor, C.~Hoelbling, S.~D. Katz, S.~Krieg, and K.~K. Szabo,
  ``{Full result for the QCD equation of state with 2+1 flavors},''
  \href{http://dx.doi.org/10.1016/j.physletb.2014.01.007}{{\em Phys. Lett. B}
  {\bfseries 730} (2014) 99--104},
  \href{http://arxiv.org/abs/1309.5258}{{\ttfamily arXiv:1309.5258 [hep-lat]}}.

\bibitem{Bazavov:2014pvz}
{\bfseries HotQCD}, A.~Bazavov {\em et~al.}, ``{Equation of state in ( 2+1
  )-flavor QCD},'' \href{http://dx.doi.org/10.1103/PhysRevD.90.094503}{{\em
  Phys. Rev. D} {\bfseries 90} (2014) 094503},
  \href{http://arxiv.org/abs/1407.6387}{{\ttfamily arXiv:1407.6387 [hep-lat]}}.

\bibitem{Shirogane:2016zbf}
M.~Shirogane, S.~Ejiri, R.~Iwami, K.~Kanaya, and M.~Kitazawa, ``{Latent heat at
  the first order phase transition point of SU(3) gauge theory},''
  \href{http://dx.doi.org/10.1103/PhysRevD.94.014506}{{\em Phys. Rev. D}
  {\bfseries 94} no.~1, (2016) 014506},
  \href{http://arxiv.org/abs/1605.02997}{{\ttfamily arXiv:1605.02997
  [hep-lat]}}.

\bibitem{Giusti:2014ila}
L.~Giusti and M.~Pepe, ``{Equation of state of a relativistic theory from a
  moving frame},'' \href{http://dx.doi.org/10.1103/PhysRevLett.113.031601}{{\em
  Phys. Rev. Lett.} {\bfseries 113} (2014) 031601},
  \href{http://arxiv.org/abs/1403.0360}{{\ttfamily arXiv:1403.0360 [hep-lat]}}.

\bibitem{Giusti:2016iqr}
L.~Giusti and M.~Pepe, ``{Equation of state of the SU(3) Yang\textendash{}Mills
  theory: A precise determination from a moving frame},''
  \href{http://dx.doi.org/10.1016/j.physletb.2017.04.001}{{\em Phys. Lett. B}
  {\bfseries 769} (2017) 385--390},
  \href{http://arxiv.org/abs/1612.00265}{{\ttfamily arXiv:1612.00265
  [hep-lat]}}.

\bibitem{Caselle:2018kap}
M.~Caselle, A.~Nada, and M.~Panero, ``{QCD thermodynamics from lattice
  calculations with nonequilibrium methods: The SU(3) equation of state},''
  \href{http://dx.doi.org/10.1103/PhysRevD.98.054513}{{\em Phys. Rev. D}
  {\bfseries 98} no.~5, (2018) 054513},
  \href{http://arxiv.org/abs/1801.03110}{{\ttfamily arXiv:1801.03110
  [hep-lat]}}.

\bibitem{Aoki:2019cca}
{\bfseries Flavour Lattice Averaging Group}, S.~Aoki {\em et~al.}, ``{FLAG
  Review 2019: Flavour Lattice Averaging Group (FLAG)},''
  \href{http://dx.doi.org/10.1140/epjc/s10052-019-7354-7}{{\em Eur. Phys. J. C}
  {\bfseries 80} no.~2, (2020) 113},
  \href{http://arxiv.org/abs/1902.08191}{{\ttfamily arXiv:1902.08191
  [hep-lat]}}.



\end{thebibliography}



\end{document}